\documentclass[12pt, leqno]{article}
\usepackage{amsmath,setspace,multirow}
\usepackage[font=small,labelfont=bf]{caption}
\usepackage[top=1.2in, bottom=1.2in, left=1.18in, right=1.18in]{geometry}

\usepackage[round]{natbib}
\usepackage{color,soul}

\DeclareCaptionStyle{italic}[justification=centering]{labelfont={bf},textfont={it},labelsep=colon}
\usepackage{graphicx,psfrag,epsf,textcomp,epstopdf, amsthm, mathabx}
\usepackage{enumerate, dsfont}
\usepackage{natbib}
\usepackage{url,xcolor}
\usepackage{booktabs, subfig, bm, paralist,mathpazo,tikz,todonotes,longtable,microtype,algorithm}

\usepackage[pdftex,colorlinks=true]{hyperref}
\definecolor{darkblue}{rgb}{0,0,.6}
\hypersetup{citecolor=darkblue,linkcolor=darkblue,urlcolor=darkblue}
\definecolor{DarkRed}{rgb}{.7,0,.4}

\usepackage{comment}

\newcommand{\blind}{0}

\graphicspath{{plots/}}

\newsavebox\CBox

 \newtheorem{@definition}{\sc Definition}[section]

\begin{document}

\def\spacingset#1{\renewcommand{\baselinestretch}{#1}\small\normalsize} \spacingset{1}

\if0\blind
{
  \title{\bf Robust bootstrap prediction intervals for univariate and multivariate autoregressive time series models}}
    \author{
Ufuk Beyaztas\footnote{Corresponding author: Department of Economics and Finance, Piri Reis University University, 34940 Istanbul, Turkey; Email: ubeyaztas@pirireis.edu.tr, ORCID: 0000-0002-5208-4950} \\
Department of Economics and Finance \\
Piri Reis University University \\
\\
Han Lin Shang \\
Department of Actuarial Studies and Business Analytics \\
Macquarie University
}
  \maketitle
 \fi

\if1\blind
{
  \bigskip
  \bigskip
  \bigskip
  \begin{center}
    {\LARGE\bf Robust bootstrap prediction intervals for univariate and multivariate autoregressive time series models}}
\end{center}
  \medskip
} \fi

\maketitle

\begin{abstract}
The bootstrap procedure has emerged as a general framework to construct prediction intervals for future observations in autoregressive time series models. Such models with outlying data points are standard in real data applications, especially in the field of econometrics. These outlying data points tend to produce high forecast errors, which reduce the forecasting performances of the existing bootstrap prediction intervals calculated based on non-robust estimators. In the univariate and multivariate autoregressive time series, we propose a robust bootstrap algorithm for constructing prediction intervals and forecast regions. The proposed procedure is based on the weighted likelihood estimates and weighted residuals. Its finite sample properties are examined via a series of Monte Carlo studies and two empirical data examples.
\end{abstract}


\newpage
\spacingset{1.56}

\section{Introduction} \label{sec:intro}

Forecasting future observations in autoregressive (AR) time series models is of great interest. Forecasts are usually obtained in the form of \textit{point} and/or \textit{interval} forecasts. A point forecast is an estimate of the unknown future observation conditional on the observed dataset. The point forecast provides a partial information since it does not provide any information about the degree of uncertainty associated with the forecasts. On the other hand, an interval forecast (or prediction interval) can provide reasonable inferences taking into account uncertainty associated with each point forecast \citep[e.g.,][]{Chatfield1993, Kim2001, JoreMitchellVahey2010}. Classical techniques for calculating prediction intervals require strong assumptions, such as Gaussian forecast errors, known lag order, and model parameters. However, these assumptions are generally unknown in practice. They may also be affected by any departure from assumptions that leads to unreliable results. Alternative computational methods to construct prediction intervals without considering distributional assumptions include the use of well-known resampling-based procedures, such as the bootstrap.

Bootstrapping prediction intervals of the AR models have received much attention in the literature. Several bootstrap methodologies have been proposed to improve the empirical coverage accuracy of prediction intervals. Most relevant studies are based on the bootstrap procedure of \cite{ThombsSchucany1990}. In the context of univariate AR models, \cite{Caoetal1997} proposed a computationally efficient bootstrap method. \cite{Masarotto1990} and \cite{Grigoletto1998} proposed alternative bootstrap techniques to obtain prediction intervals in the AR models when the true lag-order $p$ is unknown. \cite{Kabaila1993} suggested bootstrap prediction intervals considering both unconditional inference and inference conditional on the last $p$ observed values. \cite{Breidtetal1995} introduced bootstrap prediction intervals for non-Gaussian autoregressions. \cite{Pascualetal2001} studied the effect of parameter estimation on prediction intervals. \cite{ClementTaylor2001} improved the empirical coverage accuracy of the bootstrap prediction intervals by applying parameter estimation bias-correction methods. \cite{AlonsoPenaRomo2002} proposed the AR-sieve bootstrap procedure to construct nonparametric prediction intervals. \cite{Kim2004} used asymptotically mean-unbiased parameter estimation to construct prediction intervals. \cite{ClementsKim2007} proposed to bootstrap prediction intervals for highly persistent AR time series; \cite{Mukhop2010} proposed a modified version of the AR-sieve bootstrap to obtain prediction intervals. \cite{PanPolitis2016} used forward and backward bootstrap methods with predictive and fitted residuals when constructing prediction intervals. 

In the context of multivariate AR (VAR) models, bootstrap forecast densities/regions were constructed by \cite{Kim1999}, who extended the univariate bootstrap procedure of \cite{ThombsSchucany1990} to the VAR case. \cite{Kim2001, Kim2004} proposed bias-corrected bootstrap forecast regions by using the bootstrap-after-bootstrap approach; and \cite{FresoliRuisPascual2015} introduced a bootstrap method based on a forwarding representation for non-Gaussian unrestricted VAR models.

In the aforementioned literature, the bootstrap analog of AR forecasts is calculated using maximum likelihood, Yule-Walker, or ordinary least squares (OLS) estimators. It is well known that these estimators can be affected by the presence of the \textit{outlier}s, which is common in real datasets. In such cases, the estimates may be severely biased, and this can lead to poor finite sample properties of the bootstrap prediction intervals. On the other hand, robust estimation procedures can be irresponsive to outliers and produce consistent results. In the context of AR-type models, several robust estimation procedures have been proposed. For example, for univariate AR models, \cite{denby} proposed a class of generalized M-estimates to obtain robust estimates of the parameter of a first-order AR model. \cite{collomb} proposed a nonparametric method to estimate the AR model and predict one-step-ahead observations. \cite{connor} proposed a robust recurrent neural network-based learning algorithm to obtain robust predictions for time series. \cite{sejling} proposed two recursive robust estimation algorithms for the estimation of AR models. \cite{politisR} suggested an algorithm for the robust fitting of AR models. \cite{maronna} discussed the effects of outliers on estimation methods in the AR models and proposed several robust procedures, including:
\begin{inparaenum}
\item[1)] a robust Yule-Walker (Rob-YW) method that uses a robust estimation of the autocorrelation function with the Durbin-Levinson algorithm, 
\item[2)] a robust regression (Rob-Reg) where the model parameters are estimated by the MM-type regression estimator proposed by \cite{yohai87}, 
\item[3)] a robust filter algorithm (Rob-Flt) where the model parameters are computed by minimizing a robust scale of prediction residuals, and 
\item[4)] a generalized M estimator based robust regression (Rob-GM). 
\end{inparaenum}
These methods are also discussed by \cite{durre}. 

For the VAR models, \cite{Croux} suggested a robust estimation for the VAR models (Rob-VAR) using the multivariate least trimmed squares estimator of \cite{agullo}. \cite{wds} proposed a robust procedure based on weighted least squares for estimating model parameters in the VAR models. \cite{muleryohai} introduced a new class of robust estimators for the VAR models by extending MM-estimators to the multivariate case.

The aforementioned methods produce robust parameter estimates and more reliable point forecasts in the presence of outliers compared with classical estimation methods. However, bootstrap prediction intervals based on these estimation methods may still not provide efficient inferences for future values, as they produce high forecasting errors corresponding to the outlying observations. Thus, an estimation method that is robust both in the estimation of model parameters and in producing forecasting errors is needed. We examine the impacts of outliers on bootstrapping prediction intervals in the univariate and multivariate AR models. Further, we propose a sufficiently robust bootstrap method for constructing prediction intervals. The proposed method is based on the \textit{minimum density power divergence} estimator of \cite{Basu1998}. Compared with classical techniques, this estimator produces more robust estimates in the presence of outliers or when model assumptions do not hold. Also, it provides weighted residuals used to calculate bootstrap pseudo datasets and forecasts. The residuals obtained by the minimum density power divergence estimator are adjusted by a weight function that down-weights the high errors corresponding to the outliers. In turn, the proposed bootstrap method provides more reliable prediction intervals than those obtained by classical and other robust methods.

In time series, several types of outliers, such as the \textit{additive outlier} (AO), \textit{innovative outlier} (IO), \textit{level shift outlier}, and \textit{transitory change outlier} have been proposed \citep[see, e.g.,][]{Fox1972}. In this study, we restrict our focus to AOs and IOs since they are the most troublesome and commonly studied outlier types in time series. Several Monte Carlo experiments under different data-generating and outlier-generating scenarios and empirical data examples are used to evaluate the finite sample properties of the proposed bootstrap prediction intervals. According to the numerical results in Section~\ref{sec:results}, the proposed robust procedure has competitive coverage performance with narrower interval length in comparison with the OLS and other robust estimation-based bootstrap prediction intervals. Our findings also show that the proposed method has a competitive performance compared with other methods when no contamination is present in the data.

The rest of the paper is organized as follows. In Section \ref{sec:methodology}, we present detailed information about outlier types, the weighted likelihood estimation methodology, and bootstrap procedures and their proposed robust counterparts. Extensive Monte Carlo simulations are performed to examine the finite sample properties of the proposed method in Section~\ref{sec:results}. Empirical data examples are analyzed, and the results are presented in Section~\ref{sec:real}. Section~\ref{sec:conclusion} concludes the paper, along with some idea on how the methodology presented can be further extended.

\section{Methodology} \label{sec:methodology}

Forecasting future observations in the VAR models is similar to forecasting future observations in the univariate AR models. Thus, in this section, we discuss bootstrapping prediction intervals for the VAR models only. Bootstrapping prediction intervals of the AR models can be considered as a particular case of the VAR, where there is only a univariate observed time series. In Section~\ref{sec:outlier}, we first summarize the multivariate AO and IO types. Bootstrap prediction intervals of the VAR models and its robust counterparts are presented in Section \ref{sec:boot}.

\subsection{Outliers} \label{sec:outlier}

Let us consider an $N$ dimensional multivariate time series $\left\lbrace Y_t \right\rbrace$. It follows the VAR($p$) model with the AR order $p$ if
\begin{equation} \label{eq:VAR}
Y_t = \Phi_0 + \Phi_1 Y_{t-1} \cdots + \Phi_p Y_{t-p} + \epsilon_t, \qquad t = 1, \cdots, T,
\end{equation}
where $Y_t = \left( Y_{1t}, \cdots, Y_{Nt} \right)^{\prime} \in \mathds{R}^N$ are the random $N$ vectors of the time series, $\Phi = \left(\Phi_0, \Phi_1, \cdots, \Phi_p \right)$ is the vector of coefficients, which is a $N \times \left[ (p+1) N \right]$ matrix containing the separate $N \times N$ matrices column-wise as blocks, and $\left\lbrace \epsilon_t \right\rbrace_{t \geq 1}$ is a normally distributed multivariate white noise with non-singular covariance matrix $\Sigma_{\epsilon}$. Using the lag operator notation, model~\eqref{eq:VAR} is written as follows:
\begin{equation*}
\phi(B)Y_t = C + \epsilon_t,
\end{equation*}
where $\phi(B) = I - \phi_1B - \cdots - \phi_pB^p$ is matrix a polynomial of order $p$ and $C$ is an $N$ dimensional constant vector.

Following \cite{Tsay2000}, a multivariate time series with a multivariate outlier, $\left\lbrace X_t \right\rbrace$, can be written as follows:
\begin{equation*}
X_t = Y_t + \eta(B) \Delta \mathds{1}_t^{(s)},
\end{equation*}
where $\Delta = \left( \Delta_1, \cdots, \Delta_N \right)^\top$ denotes the size of the outlier and $\mathds{1}_t^{(s)}$ is the binary indicator variable that characterizes the outlier at time point $s$. The outlier type (AO/IO) is defined based on the matrix polynomial $\eta(B)$. If an AO exists at time $s$, then $\eta(B) = I$. On the other hand, if an IO is present at time $s$, then $\eta(B) = \left( \phi(B) \right)^{-1}$.

An IO that affects innovation $\epsilon_t$ can be considered an extraordinary shock influencing the future values of the series $(Y_t, Y_{t+1}, \cdots)$. In contrast, an AO at time point $s$ affects only the corresponding data point $Y_t$, not future values of the series, and can have serious effects on the estimated model parameters, residuals, and accuracy of forecasts. Generally, an AO has a much greater effect on forecasting accuracy than an IO, because an IO at time $s$ affects only the corresponding residual $\epsilon_s$. Still, an AO, which is far from the bulk of the data, causes a large residual and affects future residuals. For more information about the AOs, IOs, and their effects on the model parameters, consult \cite{Fox1972}, \cite{ChangTiaoChen1988} and \cite{McTsai} .

\subsection{Bootstrapping prediction intervals} \label{sec:boot}

Let us consider the VAR($p$) model given in~\eqref{eq:VAR}. Let $\widehat{Y}_t$ and $Y_{T+h}$ for $t = 1, \cdots, T$ and $h = 1, 2, \cdots$ be the fitted values of $Y_t$ obtained by a suitable method such as OLS and $h$-step-ahead unobservable future values, respectively. Then, the point forecast of $Y_{T+h}$ is obtained as follows:
\begin{equation*}
\widehat{Y}_{T+h|T} = \widehat{\Phi}_0 + \widehat{\Phi}_1 \widehat{Y}_{T+h-1|T} + \cdots + \widehat{\Phi}_p \widehat{Y}_{T+h-p|T},
\end{equation*}
where $\widehat{\Phi} = \left(\widehat{\Phi}_0, \widehat{\Phi}_1, \cdots, \widehat{\Phi}_p \right)$ is the estimated parameter matrix of $\Phi$ and $\widehat{Y}_{T+j|T} = Y_{T+j}$ for $j \leq 0$. Define the forecast error covariance matrix by 
\begin{equation*}
\Sigma(h) = \sum_{k=0}^{h-1} \Psi_k \Sigma_{\epsilon} \Psi_k^{\prime},
\end{equation*} 
where $\Psi_k = \sum_{j = 1}^{p-1} \Psi_{k-j} \Phi_j$ with $\Psi_0 = \mathbf{I}_N$ and $\Phi_k = \bm{0}$ for $k > p$ \citep[see, e.g.,][]{Luktepohl1991, Luktepohl2005}. For the VAR models, the prediction interval for a given horizon is obtained in the form of an ellipsoid and/or Bonferroni cube \citep[see, e.g.,][]{Luktepohl1991, Kim1999}. Under the assumption of Gaussian distributed errors, the prediction ellipsoid is obtained by the $h$-step-ahead forecast density, which may be estimated as:
\begin{equation*}\label{Eq:asymp}
Y_{T+h} \sim N \left( \widehat{Y}_{T+h|T}, \widehat{\Sigma}(h) \right),
\end{equation*}
where $\widehat{\Sigma}(h)$ is the estimated forecast error covariance matrix. The calculation of this prediction ellipsoid becomes complicated when the forecast horizon $h \geq 2$. Alternatively, a Bonferroni cube, which provides better forecast regions by taking into account the asymmetry of the distribution, is suggested by \cite{Luktepohl1991}. Thus, we calculate only the Bonferroni cube in our numerical analyses. 

As noted earlier, this procedure may not be sufficient to obtain accurate prediction intervals since it is based on the normality assumption, conformity to which is generally unknown in practice. Therefore, it may be affected by the departure from normality and lead to unreliable results. Alternatively, the bootstrap method may be used to construct a prediction interval for $Y_{T+h}$ without considering any distributional assumption. The underlying idea behind this method is to approximate the distribution of future observations by drawing resamples conditional on the observed data. Early studies on the construction of bootstrap forecast regions in the VAR models were conducted by \cite{Kim1999}, who extended the idea of \cite{ThombsSchucany1990} to the VAR case. \cite{Kim2001, Kim2004} proposed a bootstrap-after-bootstrap approach to obtain bias-corrected bootstrap forecast regions. Later, \cite{FresoliRuisPascual2015} proposed a bootstrap procedure based on the forward recursion to decrease the complexity of the bootstrap method of \cite{Kim1999}. In this study, we consider the technique proposed by \cite{FresoliRuisPascual2015} (abbreviated as ``FRP'' hereafter), whose algorithm is given below.

\begin{itemize}
\item[Step 1.] For a realization of $N$-dimensional multivariate time series $\left\lbrace Y_t \right\rbrace_{t=1,\dots,T}$, determine the lag-order parameter $p$ by Akaike Information Criterion (AIC) and calculate the OLS estimator $\widehat{\Phi}$ of the parameter matrix $\Phi$.
\item[Step 2.] Compute the residuals $\widehat{\epsilon}_t$, for $t = 1, \cdots, T$. Let $\widehat{F}_{\epsilon}$ denote the empirical distribution function of the centered and re-scaled residuals, $\widetilde{\widehat{\epsilon}} = \left( \widehat{\epsilon} - \bar{\widehat{\epsilon}} \right) / \hat{\sigma}$, where $ \bar{\widehat{\epsilon}} = \frac{1}{T} \sum_{t = 1}^T \widehat{\epsilon}_t$ and $\hat{\sigma}$ is the calculated standard deviation of the residuals, respectively.
\item[Step 3.] Generate the bootstrap observations $\left\lbrace Y^*_1, \cdots, Y^*_T \right\rbrace$ via the following recursion:
\begin{equation}
Y^*_t = \widehat{\Phi}_0 + \sum_{i=1}^p \widehat{\Phi}_i Y^*_{t-i} + \epsilon_t^*, \qquad t = 1, \cdots, T, \nonumber
\end{equation}
where $\left\lbrace \epsilon_t^* \right\rbrace_{t=1}^T$ is an independent and identically distributed (iid) sample from $\widehat{F}_{\epsilon}$ and $Y^*_t = Y_t$ for $t = 1, \cdots, p$.
\item[Step 4.] Compute the OLS estimator $\widehat{\Phi}^* = \left(\widehat{\Phi}^*_0, \widehat{\Phi}^*_1, \cdots, \widehat{\Phi}^*_p \right)$ for the bootstrap observations $\left\lbrace Y^*_1, \cdots, Y^*_T \right\rbrace$.
\item[Step 5.] Conditional on the last $p$ original observations, obtain the $h$-step-ahead bootstrap point forecast of $Y_{T+h}$, given by $Y^*_{T+h \vert T}$ as follows:
\begin{equation*}
Y_{T+h|T}^* = \widehat{\Phi}^*_0 + \widehat{\Phi}^*_1 \widehat{Y}^*_{T+h-1|T} + \cdots, \widehat{\Phi}^*_p \widehat{Y}^*_{T+h-p|T} + \epsilon^*_{T+h},\qquad h\geq 1
\end{equation*}
where $\widehat{Y}^*_{T+h|T} = Y_{T+h}$ for $h \leq 0$ and $\epsilon^*_{T+h}$ is randomly drawn from $\widehat{F}_{\epsilon}$. \nonumber
\item[Step 6.] Steps 3-5 are repeated $B$ times to obtain bootstrap replicates $\left\lbrace \widehat{Y}^{*,1}_{T+h|T}, \cdots, \widehat{Y}^{*,B}_{T+h|T} \right\rbrace$ for each $h$.
\end{itemize}
Then, a $100(1-\gamma)\%$ Bonferroni cube is obtained as follows \citep[see, e.g.,][]{Kim1999}
\begin{equation*}
C_{T+h} = \left\lbrace Y_{T+h}|Y_{T+h} \in \bigtimes_{i=1}^N \left\lbrace \left[ q_i^*(\tau), q_i^*(1-\tau)\right] \right\rbrace  \right\rbrace,
\end{equation*}
where $\tau = 1-\gamma/(2N)$, $q_i^*(\tau)$ is the $\tau$\textsuperscript{th} quantile of the generated bootstrap distribution of the $i$\textsuperscript{th} element of $\widehat{Y}_{T+h|T}^*$, and $\bigtimes$ denotes cartesian product.

\subsubsection{Proposed bootstrap procedure}\label{sec:prop}

Let us consider the univariate AR($p$) model and let $\bm{\epsilon} = \left\lbrace \epsilon_1, \cdots, \epsilon_T \right\rbrace$ be a set of realizations of a normally distributed white noise sequence. Then, the joint probability density of $\bm{\epsilon}$ is defined as follows:
\begin{equation*} \label{eq:jdf}
P \left(\bm{\epsilon} \vert \Phi, \sigma^2 \right) = \left( 2\pi\sigma^2 \right)^{-\frac{T}{2}} \exp\left\lbrace -\frac{1}{2\sigma^2} \sum_{t=1}^T \epsilon^2_t  \right\rbrace.
\end{equation*}
Now, let $\mathbf{y}_T = \left\lbrace y_1, \cdots, y_T \right\rbrace$ and $\epsilon_t \left( \Phi \right) = \epsilon_t \left( \Phi \vert \mathbf{y}_*, \bm{\epsilon}_*, \mathbf{y}_T \right) = y_t - \phi_0 - \sum_{i = 1}^p \phi_i y_{t-i}$, respectively, be a realization of a time series and residuals, where $\mathbf{y}_* = \left\lbrace y_{1-p}, \cdots, y_0 \right\rbrace$ and $\bm{\epsilon}_* = \left\lbrace \epsilon_{1-p}, \cdots, \epsilon_0 \right\rbrace$ are the vectors of initial values. Then, the conditional log-likelihood is obtained as follows: 
\begin{eqnarray*}
\ell \left( \bm{\epsilon}_t\left( \Phi \right); \sigma^2 \right) &=& \ln~\mathcal{L} \left( \epsilon_t\left( \Phi \right); \sigma^2 \right), \nonumber \\
&=& -\frac{T}{2} \ln \left( 2\pi \sigma^2 \right) -\frac{1}{2\sigma^2} \sum_{t=1}^T \epsilon_t^2 \left( \Phi \vert \mathbf{y}_*, \bm{\epsilon}_*, \mathbf{y}_T \right). \label{eq:cll}
\end{eqnarray*}

The conditional maximum likelihood (CML) estimators of $\Phi$ and $\sigma^2$ are obtained by the solution of the score functions 
\begin{align*}
u_{\Phi}\left( \epsilon_t\left( \Phi \right); \sigma^2 \right) &= \frac{\partial}{\partial \Phi}~\ln~\mathcal{L} \left( \epsilon_t\left( \Phi \right); \sigma^2 \right)\\
u_{\sigma}\left( \epsilon_t\left( \Phi \right); \sigma^2 \right) &= \frac{\partial}{\partial \sigma}~\ln~\mathcal{L} \left( \epsilon_t\left( \Phi \right); \sigma^2 \right). 
\end{align*}
It is well known that the CML method works well under the assumption of a \textit{correctly specified} model and no outliers present in the data; however, when the data are contaminated by outliers, it produces biased estimates for the model parameters, and the errors computed corresponding to the outliers will be very large \citep[see, e.g.,][]{McTsai}. Thus, the bootstrap prediction intervals based on CML may not provide reliable inferences in the presence of outliers.

Our proposal to construct a robust bootstrap prediction interval/forecast region for the AR models is to replace the CML (or OLS) estimators and residuals by weighted likelihood estimators and weighted residuals in the algorithm given in Section \ref{sec:boot}. The weighted likelihood methodology is suggested by \cite{Markatou1996} and \cite{Markatou1998} to improve the robustness of the estimates, and has been studied in a wide variety of statistical applications; see, for example, \cite{Claudio2001}, \cite{Claudio2002a, Claudio2002b}, and \cite{Claudio2010}. It provides efficient estimators with high breakdown points when the errors are iid. Let $f^*(\cdot,\cdot)$ and $m^*(\cdot,\cdot)$ be a non-parametric kernel density estimator and smoothed model density \citep[see, e.g.,][]{Markatou1998, Claudio2010}, respectively, defined as follows:
\begin{align*}
f^* \left( \epsilon_t\left( \Phi \right), \widehat{F}_{\epsilon} \left(\Phi\right) \right) &= \int k \left(  \epsilon_t\left( \Phi \right); r,g \right) d \widehat{F}_{\epsilon} \left(r; \Phi\right)  \qquad \forall t = 1, \cdots, T, \\
m^* \left( \epsilon_t\left( \Phi \right), \sigma^2 \right) &= \int k \left(  \epsilon_t\left( \Phi \right); r,g \right) dM\left( r; \sigma^2 \right),
\end{align*}
where $\widehat{F}_{\epsilon} \left(\Phi\right)$, $M\left(\sigma^2 \right)$ and $k \left( \epsilon_t\left( \Phi \right); r,g \right)$, respectively, denote the empirical cumulative distribution function based on the residuals $\epsilon_t\left( \Phi \right)$, the normal distribution function with zero mean and variance $\sigma^2$, and a kernel density with bandwidth $g$. Note that both empirical cumulative distribution and normal distribution function are functions of $r$. The Pearson residual $\delta_t$, a measure of the discrepancy between $f^*(\cdot,\cdot)$ and $m^*(\cdot,\cdot)$, and the weight function $\omega\left( \delta_t \right)$ are defined as in~\eqref{eq:pearson} and~\eqref{eq:weights}, respectively:
\begin{align}
\delta_t = \delta \left( \epsilon_t\left( \Phi \right); M\left(\sigma^2 \right), \widehat{F}_{\epsilon} \left(\Phi\right) \right) &= \frac{f^* \left( \epsilon_t\left( \Phi \right), \widehat{F}_{\epsilon} \left(\Phi\right) \right)}{m^* \left( \epsilon_t\left( \Phi \right); \sigma^2 \right)}-1 \qquad \forall t = 1, \cdots, T, \label{eq:pearson} \\
\omega\left( \epsilon_t\left( \Phi \right); M\left(\sigma^2 \right), v \right) = \omega\left( \delta_t \right) &= \min \left\lbrace 1, \frac{\left[ A\left( \delta_t \right) + 1 \right]^+}{\delta_t + 1}\right\rbrace,  \label{eq:weights}
\end{align}
where $ \left[ \cdot \right]^+ $ is the positive part and $ A\left( \cdot \right) $ denotes the residual adjustment function (RAF) of \cite{Lindsay} (e.g., Hellinger RAF $A(\delta) = 2 \left[ \left( \delta + 1 \right)^{1/2} - 1 \right]$). Then, the conditional weighted likelihood estimating equations for $\Phi$ and $\sigma$ are defined as in \eqref{eq:eqphi} and \eqref{eq:eqsigma}, respectively:
\begin{eqnarray}
\frac{1}{T-p} \sum_{t=p+1}^{T} \omega\left( \epsilon_t\left( \Phi \right); M\left(\sigma^2 \right), \widehat{F}_{\epsilon} \left(\Phi\right) \right) u_{\Phi}\left( \epsilon_t\left( \Phi \right); \sigma^2 \right) = 0, \label{eq:eqphi} \\
\frac{1}{T-p} \sum_{t=p+1}^{T} \omega\left( \epsilon_t\left( \Phi \right); M\left(\sigma^2 \right), \widehat{F}_{\epsilon} \left(\Phi\right) \right) u_{\sigma}\left( \epsilon_t\left( \Phi \right); \sigma^2 \right) = 0. \label{eq:eqsigma}
\end{eqnarray}

The weighted likelihood methodology defined above uses weighted score equations to estimate the model parameters. Considering the Pearson residuals in~\eqref{eq:pearson}, if the model assumptions hold and no outliers are present in the data, then $\delta_t$ converges to 0, and so the weights converge to 1. In contrast, if the data are contaminated by outliers (i.e., if the observed frequency $f^*(\cdot,\cdot)$ is relatively large compared with the predicted model $m^*(\cdot,\cdot)$), the Pearson residuals will be large and the outlying observations will receive small weights. Since the weighted likelihood estimating equations given in \eqref{eq:eqphi} and \eqref{eq:eqsigma} are defined on the weighted residuals, it will be robust against the outlying points.

As is mentioned above, the proposed robust bootstrap prediction interval/forecast region for the AR models is similar to the algorithm given in Section~\ref{sec:boot} except using weighted likelihood estimators and weighted residuals. Let $\widehat{\Phi}^{\omega} = \left(\widehat{\Phi}_{0}^{\omega}, \widehat{\Phi}_{1}^{\omega}, \cdots, \widehat{\Phi}_{p}^{\omega} \right)$, $\widehat{\Phi}^{\omega,*} = \left(\widehat{\Phi}^{\omega,*}_0, \widehat{\Phi}^{\omega,*}_1, \cdots, \widehat{\Phi}^{\omega,*}_p \right)$, and $\widehat{\epsilon}^{\omega} = \omega \left(Y_t - \widehat{\Phi}_{0}^{\omega} - \sum_{i=1}^p \widehat{\Phi}_{i}^{\omega} Y_{t-i} \right)$ denote the weighted likelihood counterparts of $\widehat{\Phi}$, $\widehat{\Phi}^{*}$ , and $\widehat{\epsilon}$, respectively. Then, the weighted likelihood based $h$-step-ahead bootstrap point forecast of $Y_{T+h}$, $Y^*_{T+h \vert T}$, is obtained by replacing $\left[ \widehat{\Phi}, \widehat{\Phi}^{*}, \widehat{\epsilon} \right]$ with $\left[ \widehat{\Phi}^{\omega}, \widehat{\Phi}^{\omega,*}, \widehat{\epsilon}^{\omega} \right]$ in the algorith given in Section~\ref{sec:boot}.

Bootstrap prediction intervals based on the CML is very sensitive to outliers, and the estimated residuals corresponding to the outlying points will be large when it is used to estimate the AR model. Thus, the bootstrap prediction intervals based on the CML may not provide reliable results for the unobserved future sequence of the time series when the time series is contaminated by outliers. In contrast, bootstrap prediction intervals based on the weighted likelihood methodology are expected to have improved forecasting accuracy since it down-weights the high forecast errors produced by the outliers.

\section{Numerical results} \label{sec:results}

Under different scenarios and parameter settings, we conduct several Monte Carlo experiments to investigate the finite sample performance of our proposed weighted-likelihood based bootstrap prediction interval. For the univariate AR model, we compare our results with OLS-based bootstrap prediction intervals and four robust estimation-based bootstrap prediction intervals; Rob-YW, Rob-Reg, Rob-Flt, and Rob-GM. For the VAR model, the performance of the proposed bootstrap is compared with FRP and the multivariate least trimmed squares based bootstrap (Rob-VAR) method. For the univariate model, the comparison is made through the empirical coverage probability and length of bootstrap prediction intervals. The empirical coverage probability and volume of the bootstrap Bonferroni cubes, as well as the squared errors between the Bonferroni cubes obtained by the bootstrap methods and the empirical Bonferroni cube, are used as performance metrics for the VAR case. 

Throughout the experiments, first, the synthetic datasets $y_t$ (or $Y_t$) from the AR/VAR processes with sizes $T = 100$ and $T = 250$, along with $R = 5000$ future values $y_{T+h}$ (or $Y_{T+h}$), for each $h = 1, \cdots, 10$,  are created. Then, $B = 1999$ bootstrap resamples are performed to obtain the bootstrap prediction intervals and Bonferroni forecast cubes. The coverage probabilities are calculated as the proportion of the generated $R = 5000$ future observations falling within the constructed bootstrap prediction intervals/Bonferroni forecast cube. For each model and sample size, this procedure is repeated for $MC = 1000$ times to calculate the average values of the performance metrics. The significance level $\gamma$ is set to 0.05 to obtain 95\% bootstrap prediction intervals and/or Bonferroni cubes for future $h = 1, \cdots, 10$ observations. The following method is used to calculate performance metrics.
\begin{itemize}
\item[Step 1.] Generate a univariate AR($p$)/bivariate VAR($p$) series, and for each $h = 1, \cdots, 10$, simulate $R = 5000$ future values based on the generated AR($p$)/VAR($p$) series.

\item[Step 2.] Calculate bootstrap future values $\widehat{y}_{T+h|T}^{*,b}$ (or $\widehat{Y}_{T+h|T}^{*,b}$) for $h = 1, \cdots, 10$ and $b = 1, \cdots, B$. Then, compute the average  coverage probability of the bootstrap prediction interval for $\widehat{y}_{T+h|T}$ as:
\begin{equation}
\frac{1}{MC} \frac{1}{R} \sum_{j=1}^{MC} \sum_{r=1}^R \left\lbrace y_{T+h}^{j,r} \vert y_{T+h}^{j,r} \in \left( q^{*,j}(\gamma/2), q^{*,j}(1 - \gamma/2 \right) \right\rbrace. \nonumber
\end{equation}
Also, estimate the average coverage probabilities of bootstrap forecast regions for $\widehat{Y}_{T+h|T}^*$ as:
\begin{equation}
\frac{1}{MC} \frac{1}{R} \sum_{j=1}^{MC} \sum_{r=1}^R \left\lbrace Y_{T+h}^{j,r}|Y_{T+h}^{j,r} \in \bigtimes_{i=1}^2 \left\lbrace \left[ q_i^{*,j}(\gamma/(2N)), q_i^{*,j}(1-\gamma/(2N))\right] \right\rbrace  \right\rbrace.  \nonumber
\end{equation}

\item[Step 3.] Calculate the prediction interval length as:
\begin{equation}
\frac{1}{MC} \sum_{j=1}^{MC} \left(  q^{*,j} \left( 1 - \gamma/2 \right) -  q^{*,j} \left( \gamma/2 \right) \right) , \nonumber
\end{equation}
and estimate the volume of the bootstrap Bonferroni cube using the following equation:
\begin{equation}
\frac{1}{MC} \sum_{j=1}^{MC} \left[ \left( q_{1}^{*,j}(1-\gamma/(2N)) - q_{1}^{*,j}(\gamma/(2N)) \right) \times \left( q_{2}^{*,j}(1-\gamma/(2N)) - q_{2}^{*,j}(\gamma/(2N)) \right) \right]. \nonumber
\end{equation}

\item[Step 4.] Calculate the squared error between the bootstrap and empirical Bonferroni cubes as follows:
\begin{align}
\frac{1}{MC} \sum_{j=1}^{MC} \left[\left( q_{1}^j(1-\gamma/(2N)) - q_{1}^{*,j}(1-\gamma/(2N))\right)^2 + \left( q_{1}^j(\gamma/(2N)) - q_{1}^{*,j}(\gamma/(2N))\right)^2 \right. \nonumber \\
\qquad + \left. \left( q_{2}^j(1-\gamma/(2N)) - q_{2}^{*,j}(1-\gamma/(2N))\right)^2 + \left( q_{2}^j(\gamma/(2N)) - q_{2}^{*,j}(\gamma/(2N))\right)^2 \right]. \nonumber
\end{align}
\end{itemize}
Since our conclusions do not vary significantly with different sample sizes, to save space, we present only the results for $T = 100$; the results obtained when $T = 250$ can be obtained from the corresponding author on request. 

\subsection{Numerical results for the univariate AR model}

\begin{figure}[!htbp]
  \centering
  \includegraphics[width=5.9cm]{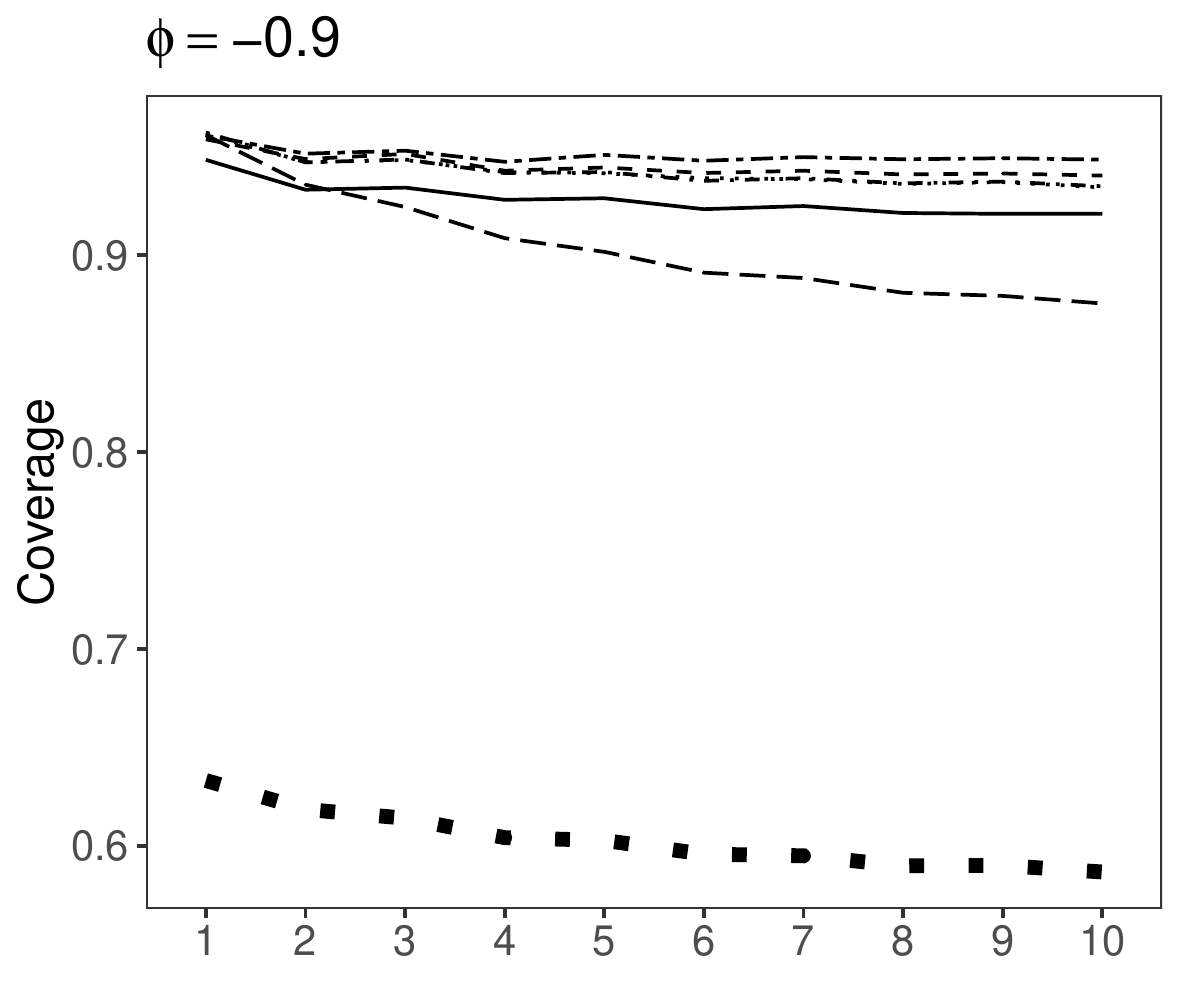}
  \includegraphics[width=5.9cm]{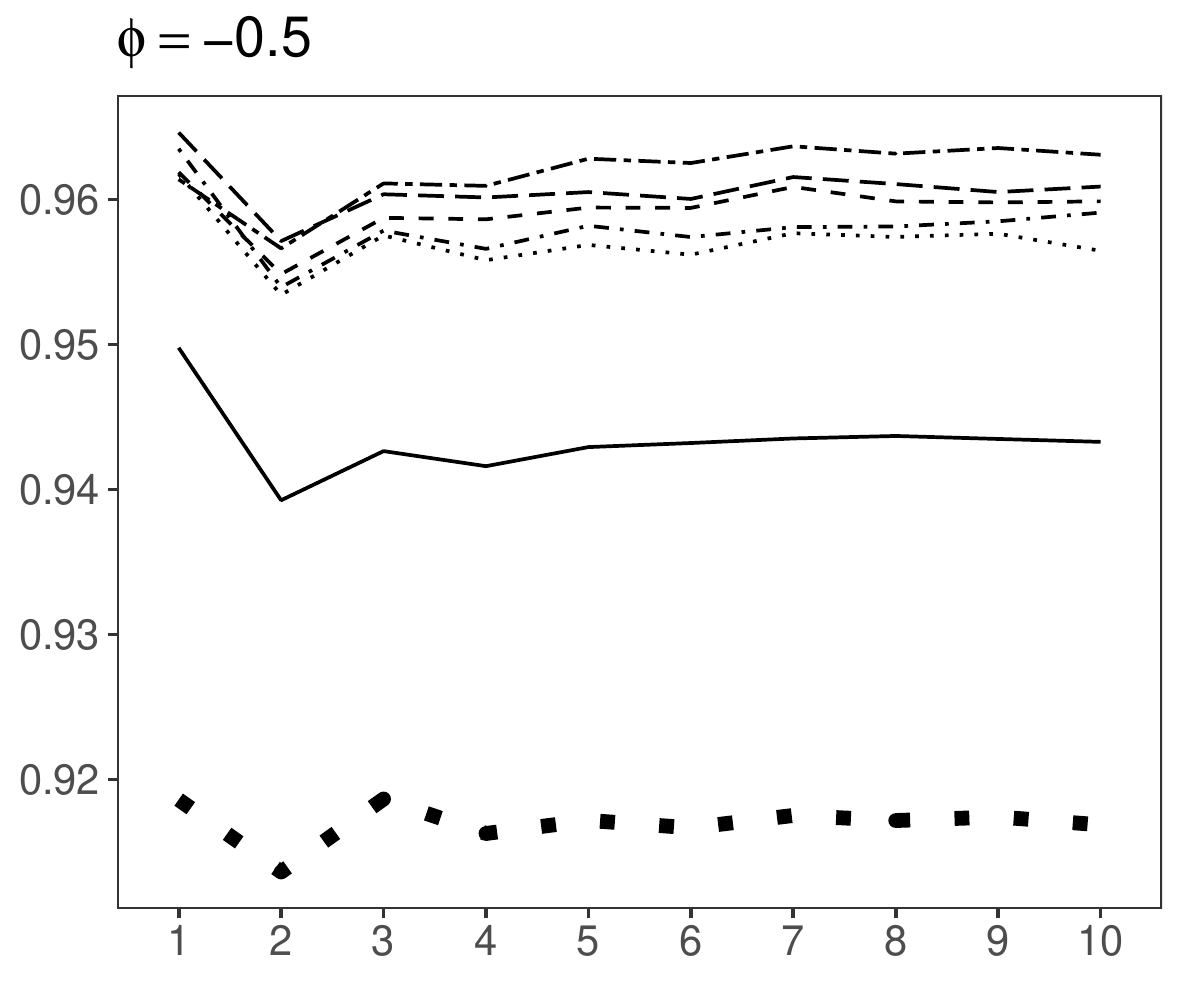}
  \includegraphics[width=5.9cm]{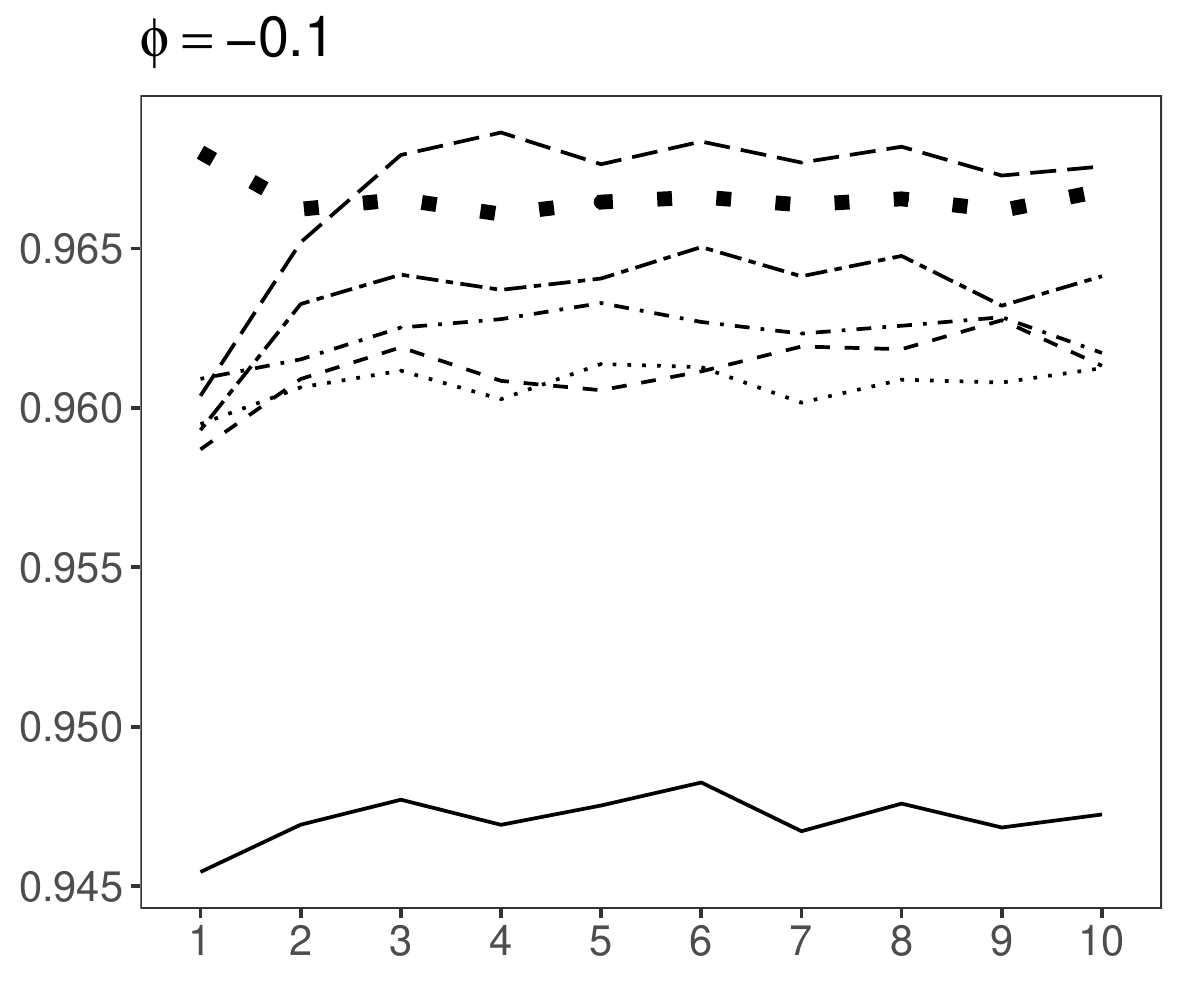}
\\
  \includegraphics[width=5.9cm]{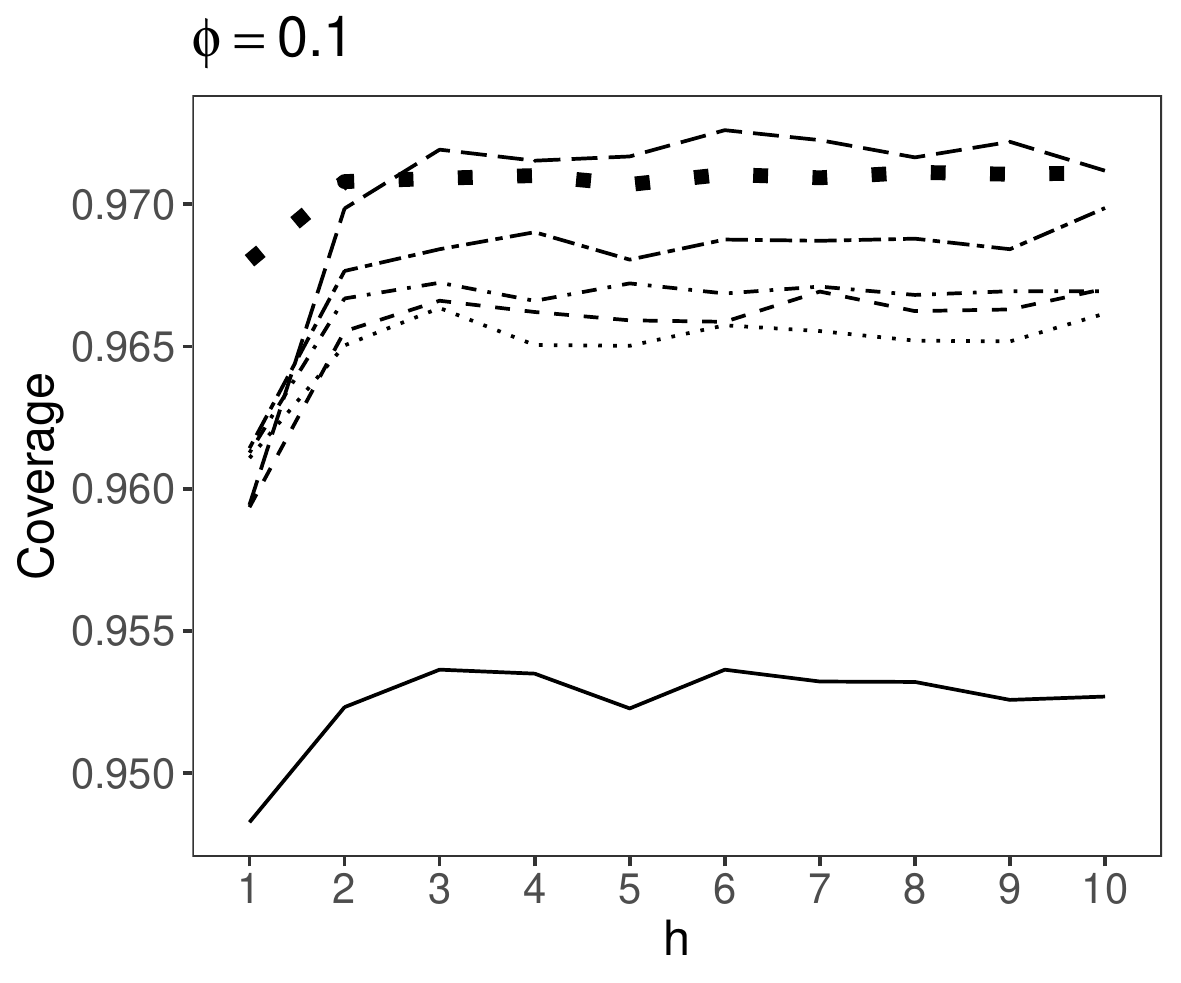}
  \includegraphics[width=5.9cm]{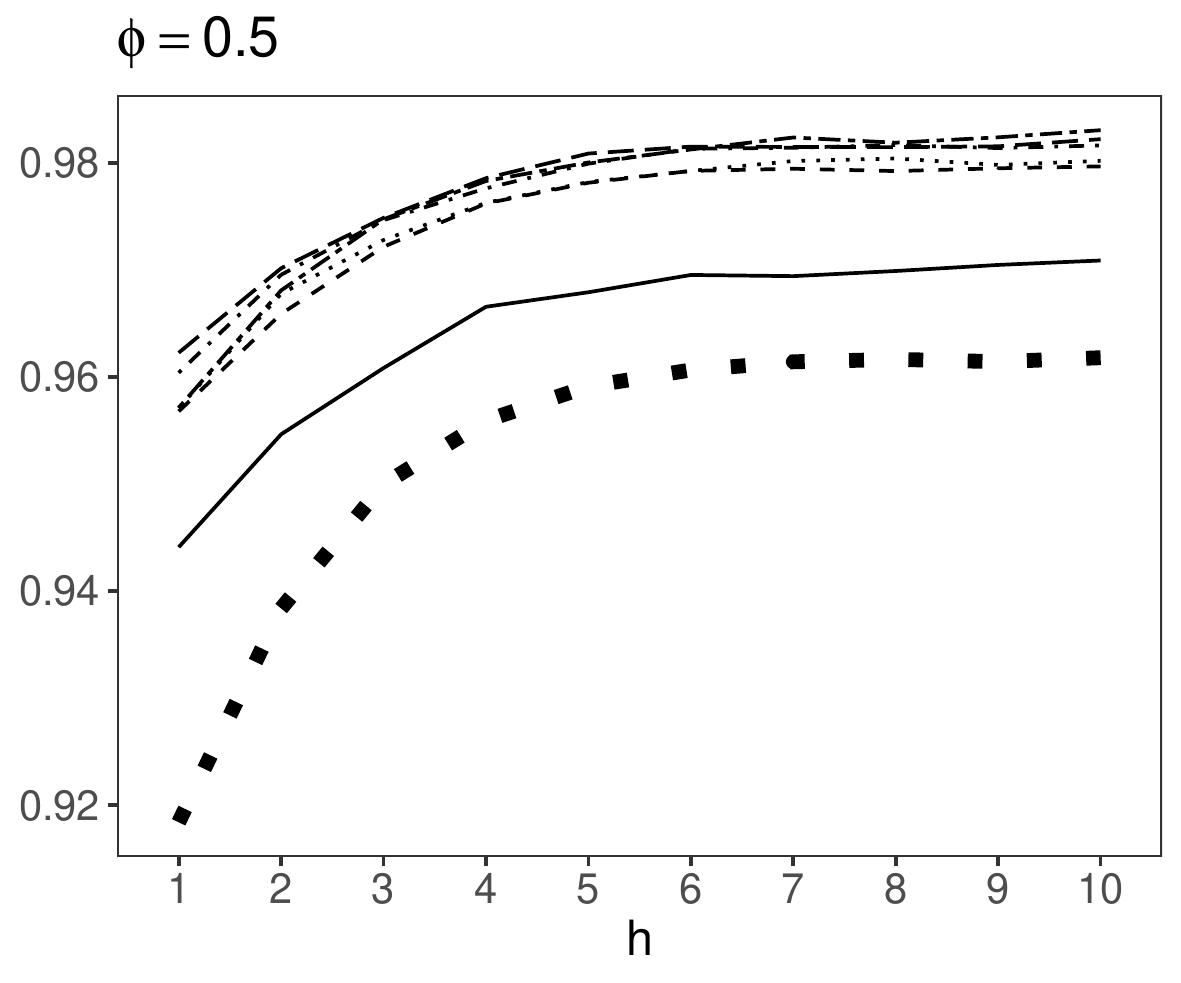}
  \includegraphics[width=5.9cm]{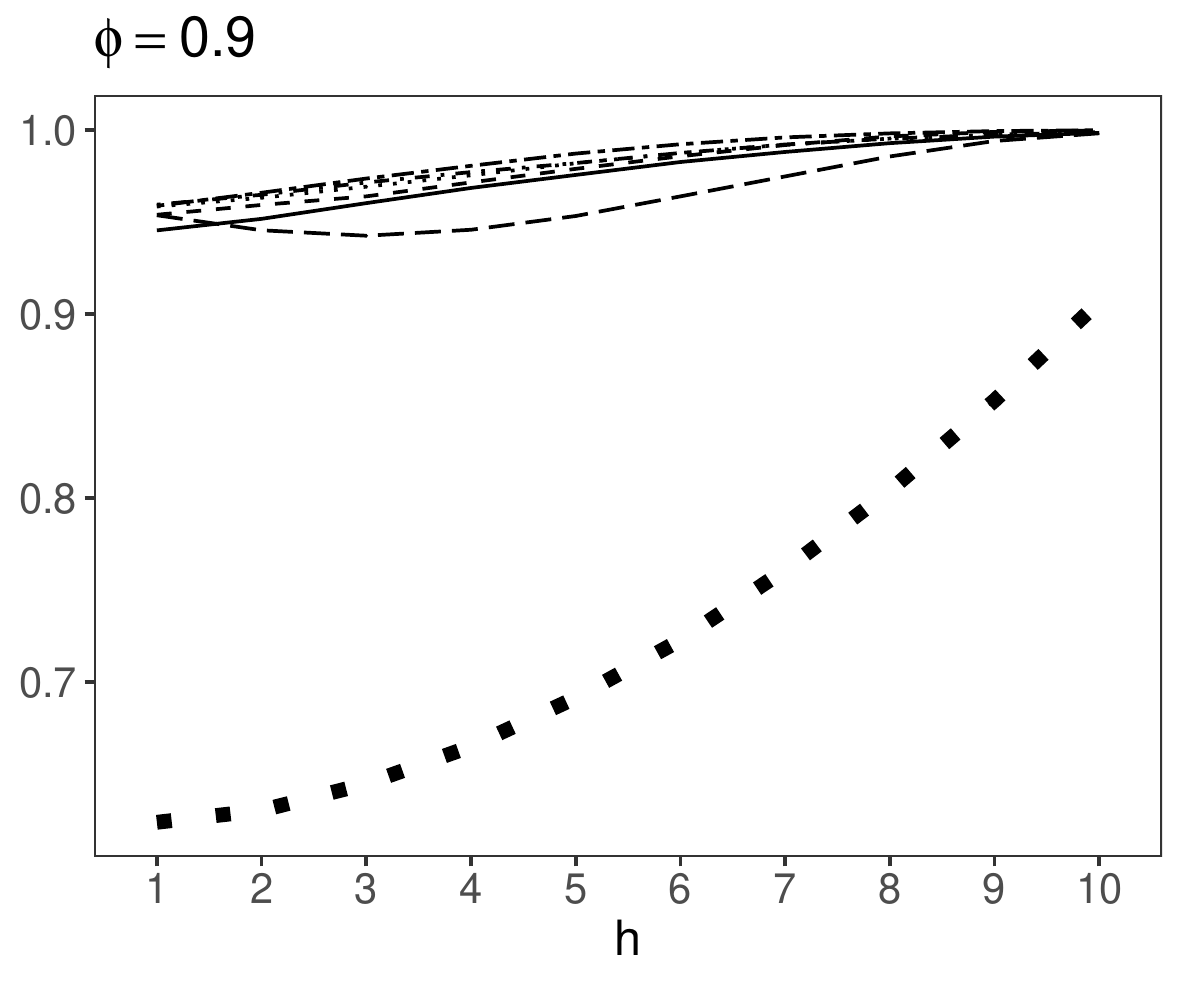}  
  \caption{\small{Monte Carlo averages of the coverages of prediction intervals based on the: OLS (dotted line), weighted likelihood (solid line), Rob-YW (dashed line), Rob-Reg (dot-dashed line), Rob-Flt (long dashed line), and Rob-GM (two dashed line) methods for the AR(1) model when no deliberate outlying data point is inserted into the dataset. Note that square dot points denote the average coverage probabilities when the quantiles of the normal distribution are used.}}
  \label{fig:1}
\end{figure}

\begin{figure}[!htbp]
  \centering
  \includegraphics[width=5.9cm]{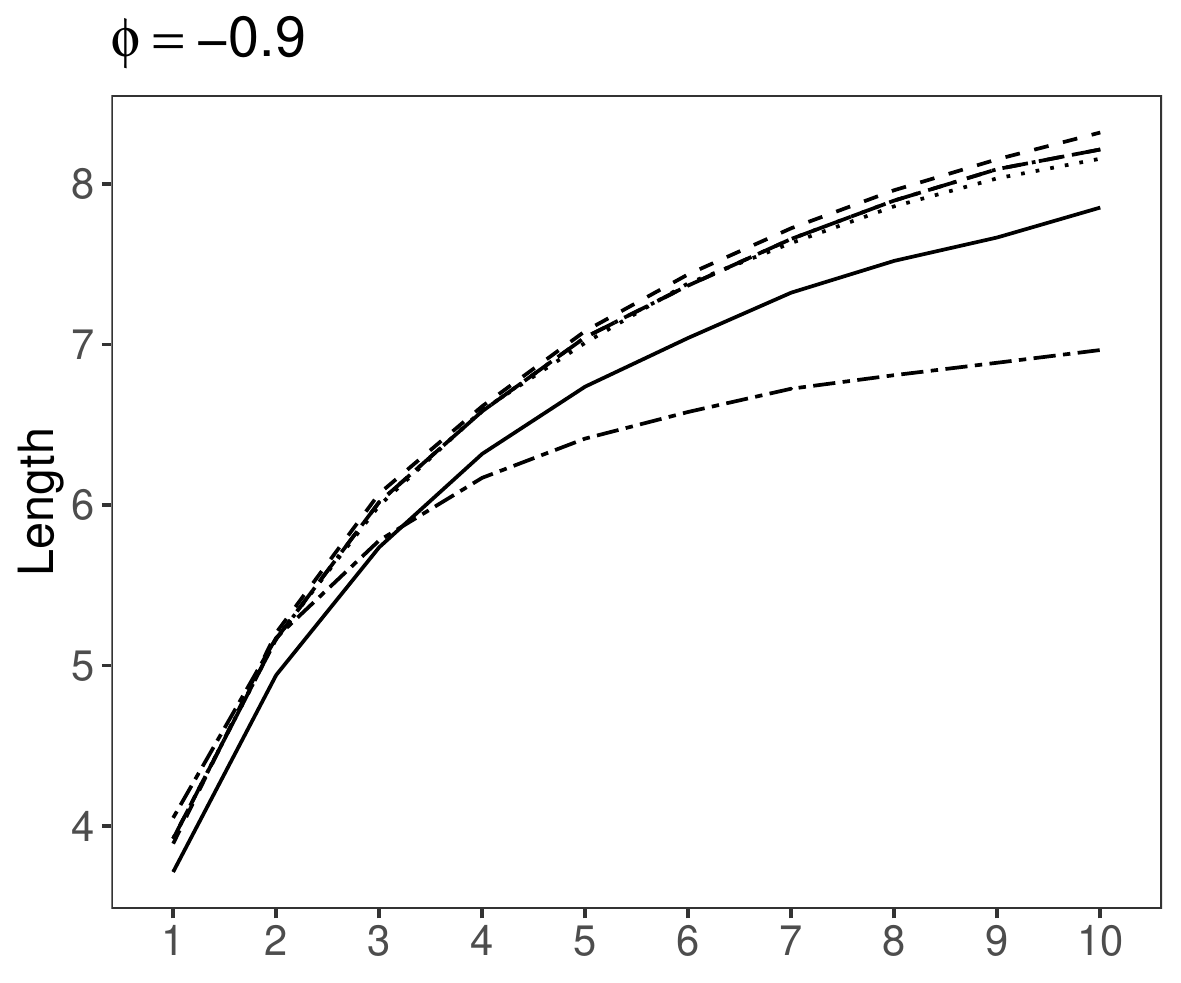}
  \includegraphics[width=5.9cm]{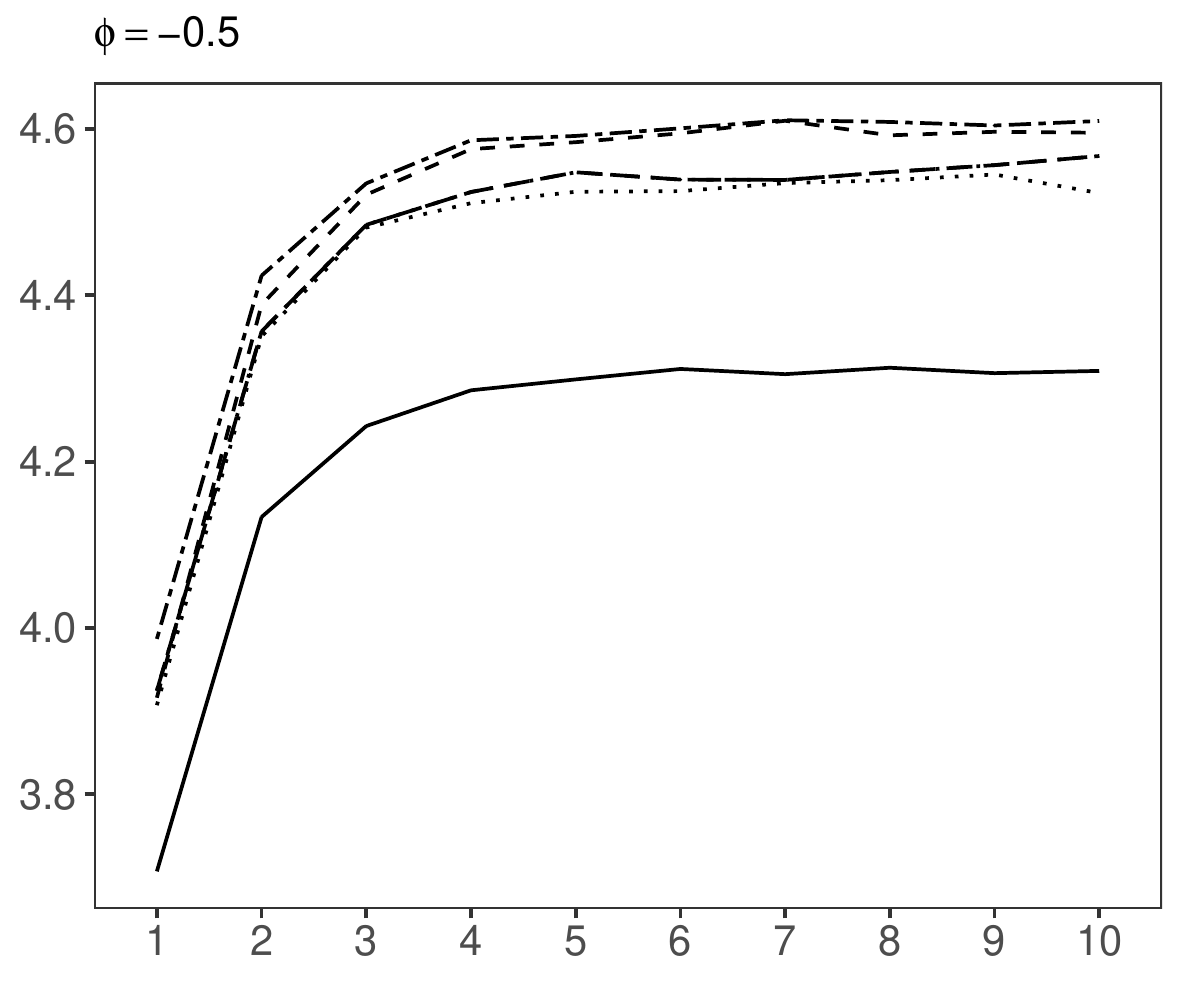}
  \includegraphics[width=5.9cm]{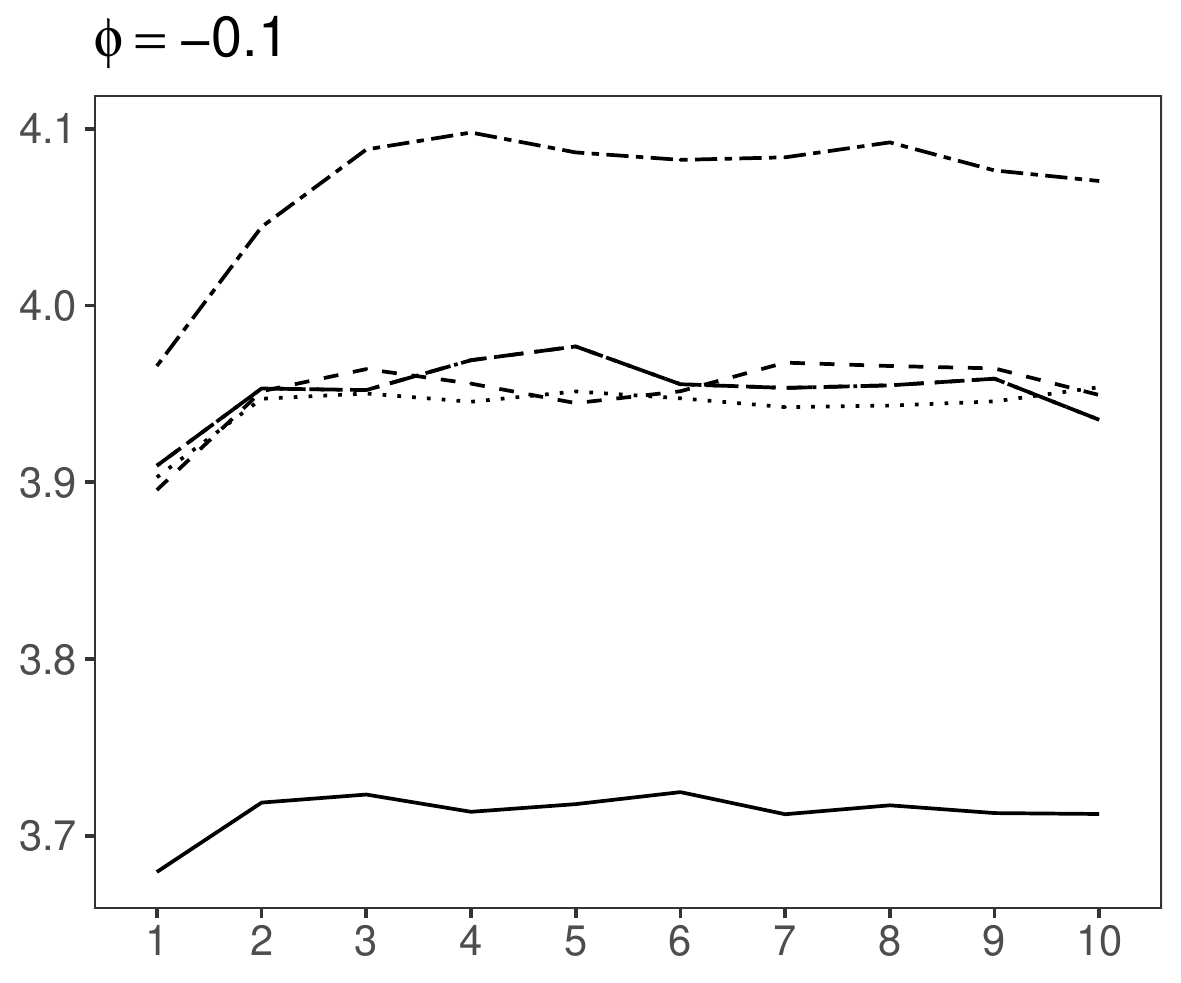}
\\
  \includegraphics[width=5.9cm]{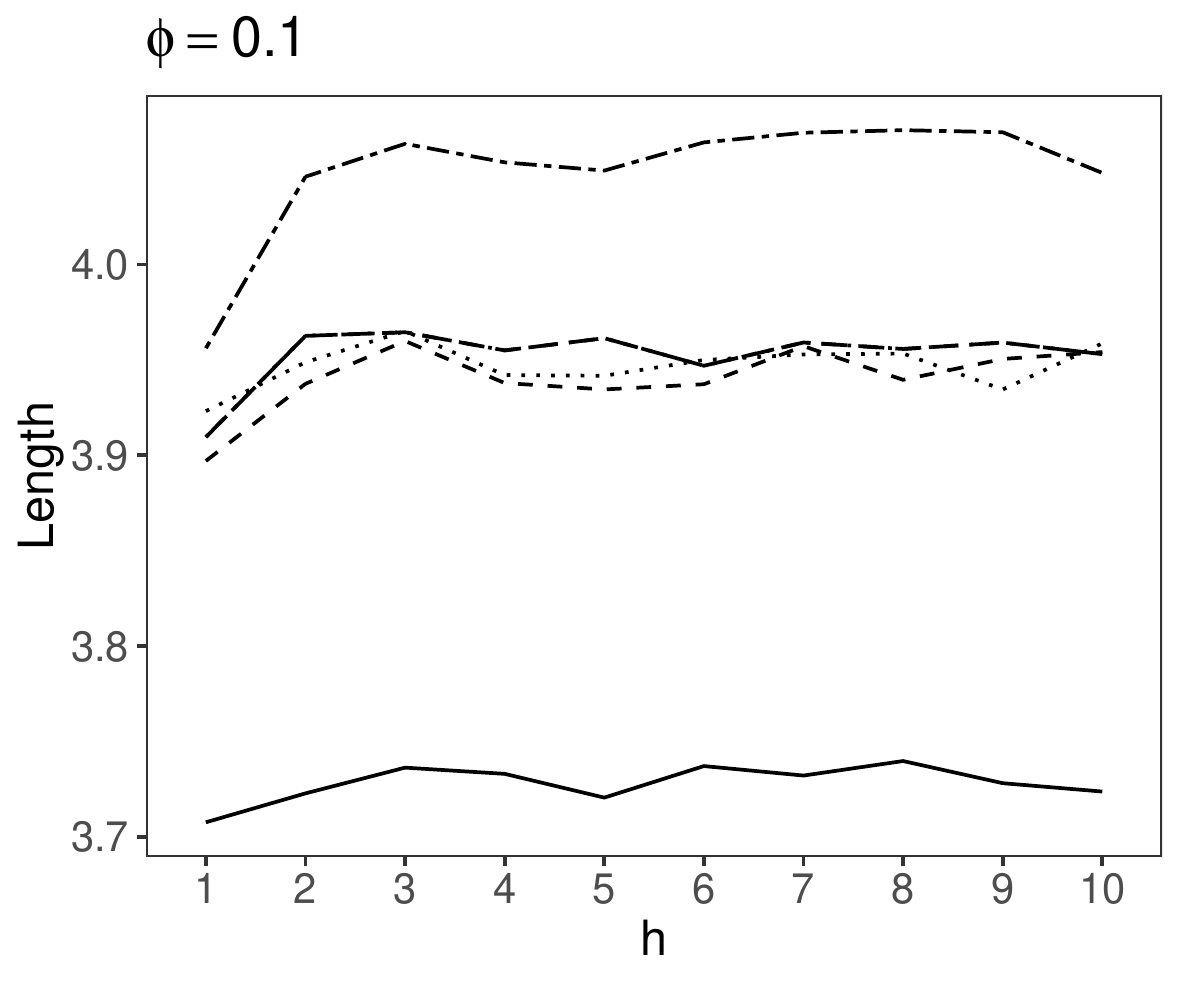}
  \includegraphics[width=5.9cm]{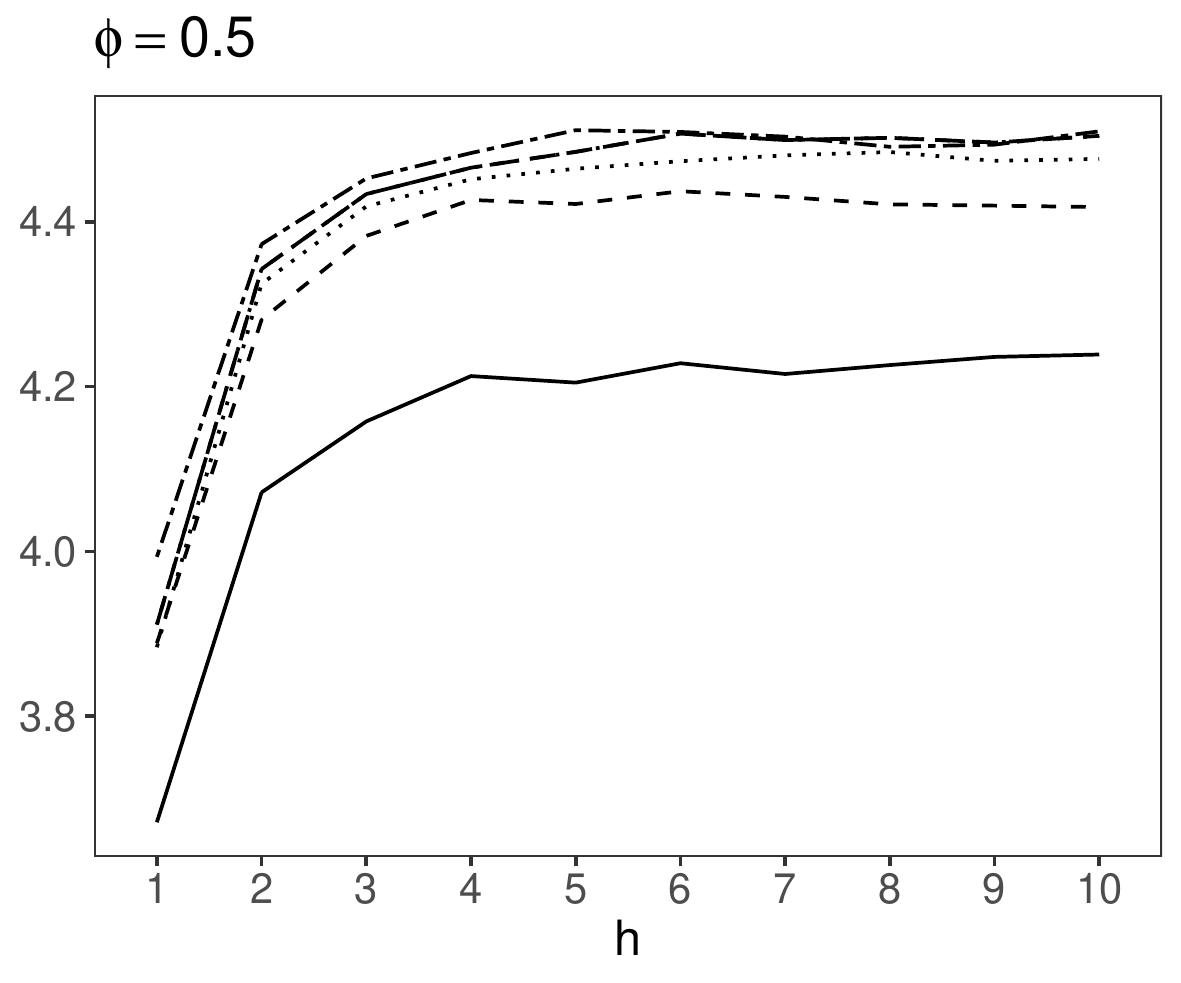}
  \includegraphics[width=5.9cm]{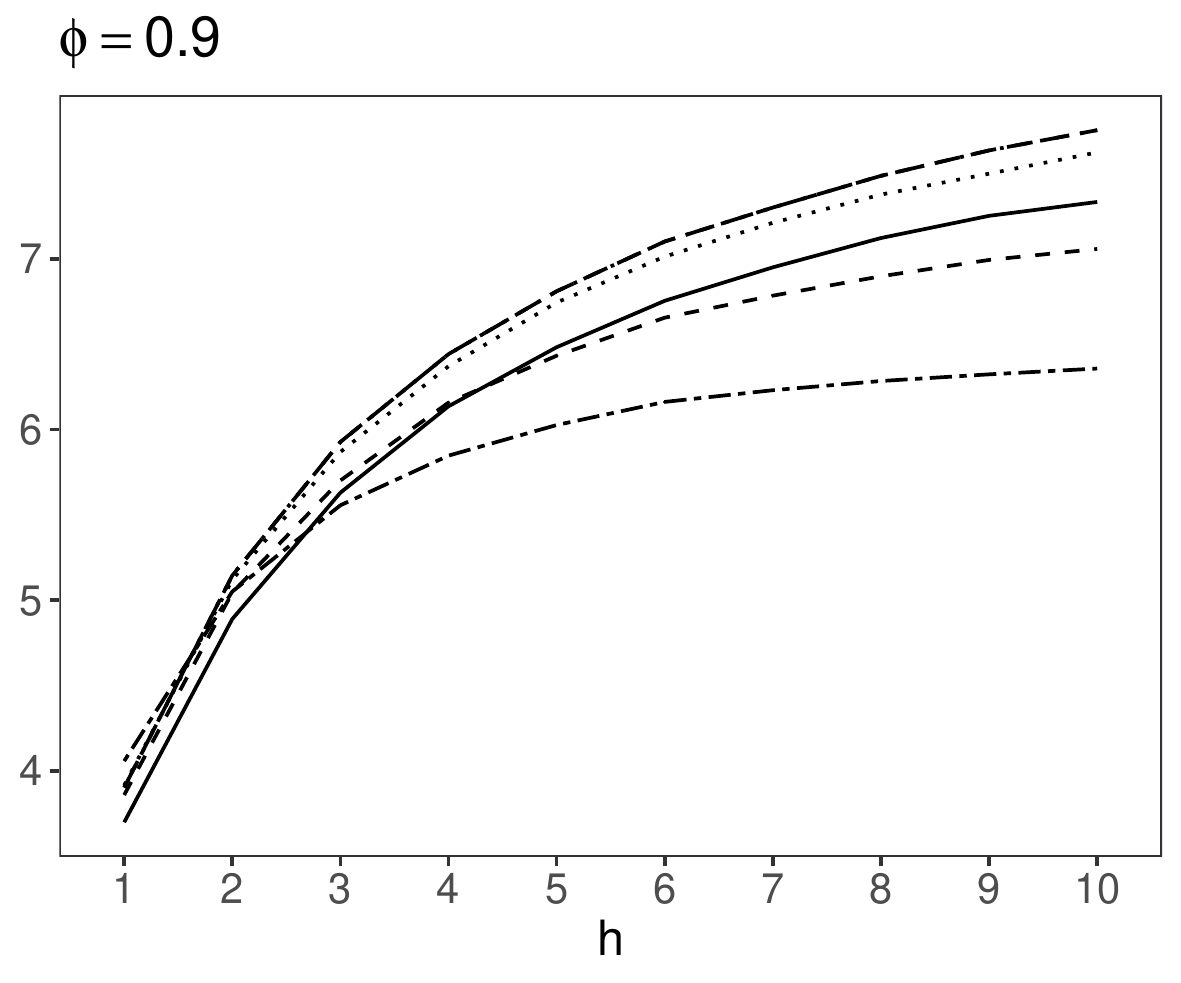}  
  \caption{\small{Monte Carlo averages of the lengths of prediction intervals based on the: OLS (dotted line), weighted likelihood (solid line), Rob-YW (dashed line), Rob-Reg (dot-dashed line), Rob-Flt (long dashed line), and Rob-GM (two dashed line) methods for the AR(1) model when no deliberate outlying data point is inserted into the dataset.}}
  \label{fig:2}
\end{figure}

\begin{figure}[!htbp]
  \centering
  \includegraphics[width=5.7cm]{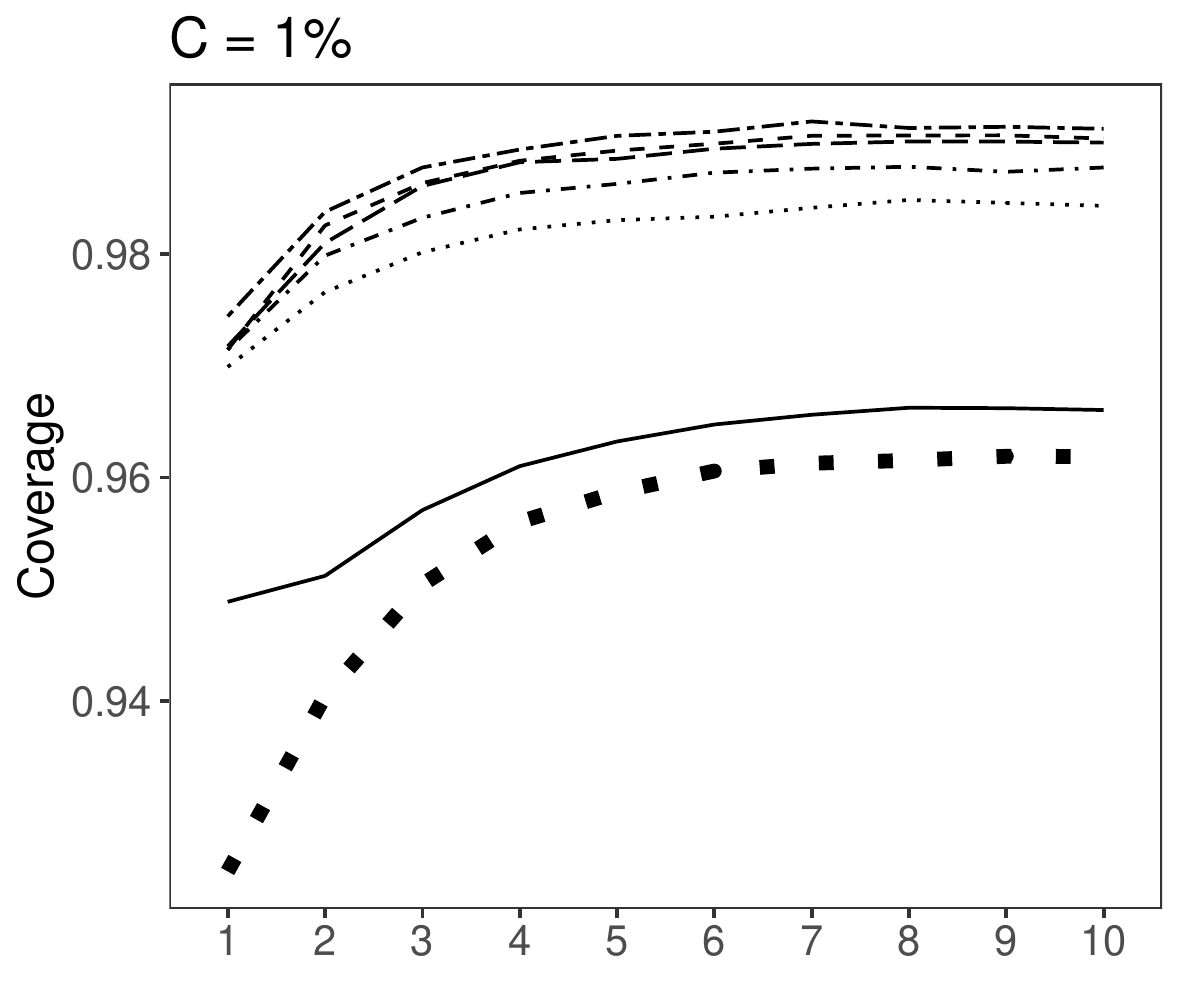}
  \includegraphics[width=5.7cm]{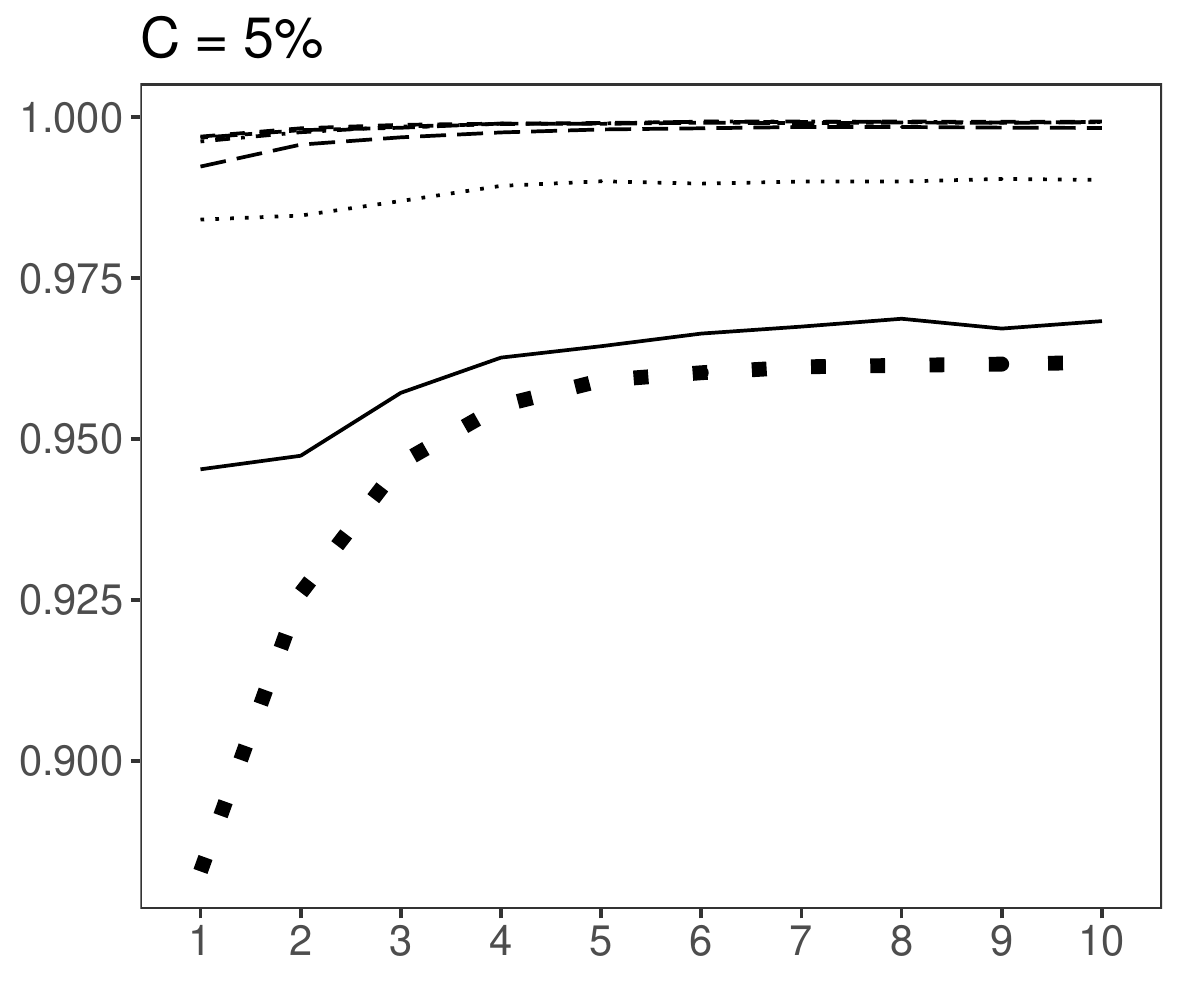}
  \includegraphics[width=5.7cm]{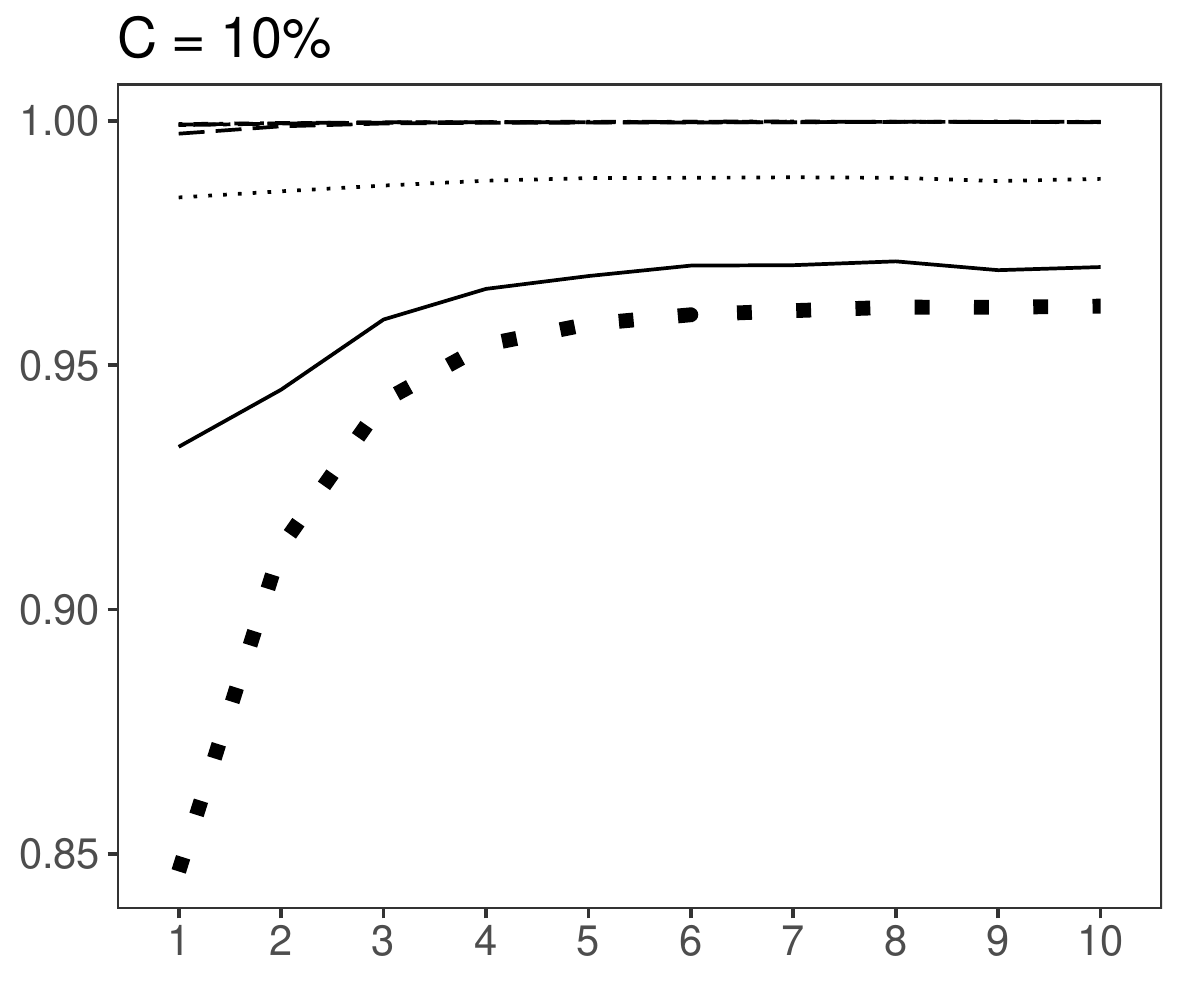}
\\
  \includegraphics[width=5.7cm]{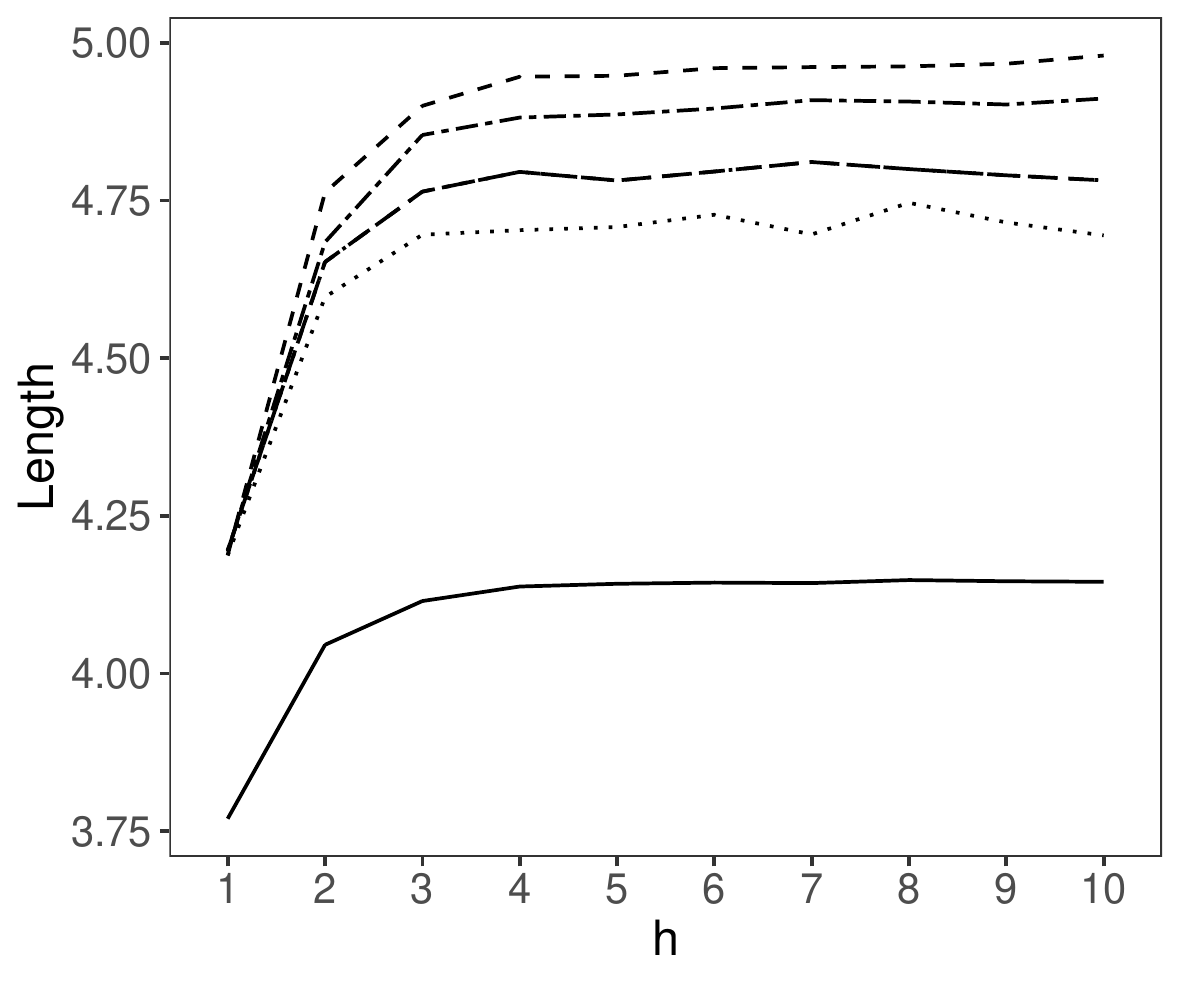}
  \includegraphics[width=5.7cm]{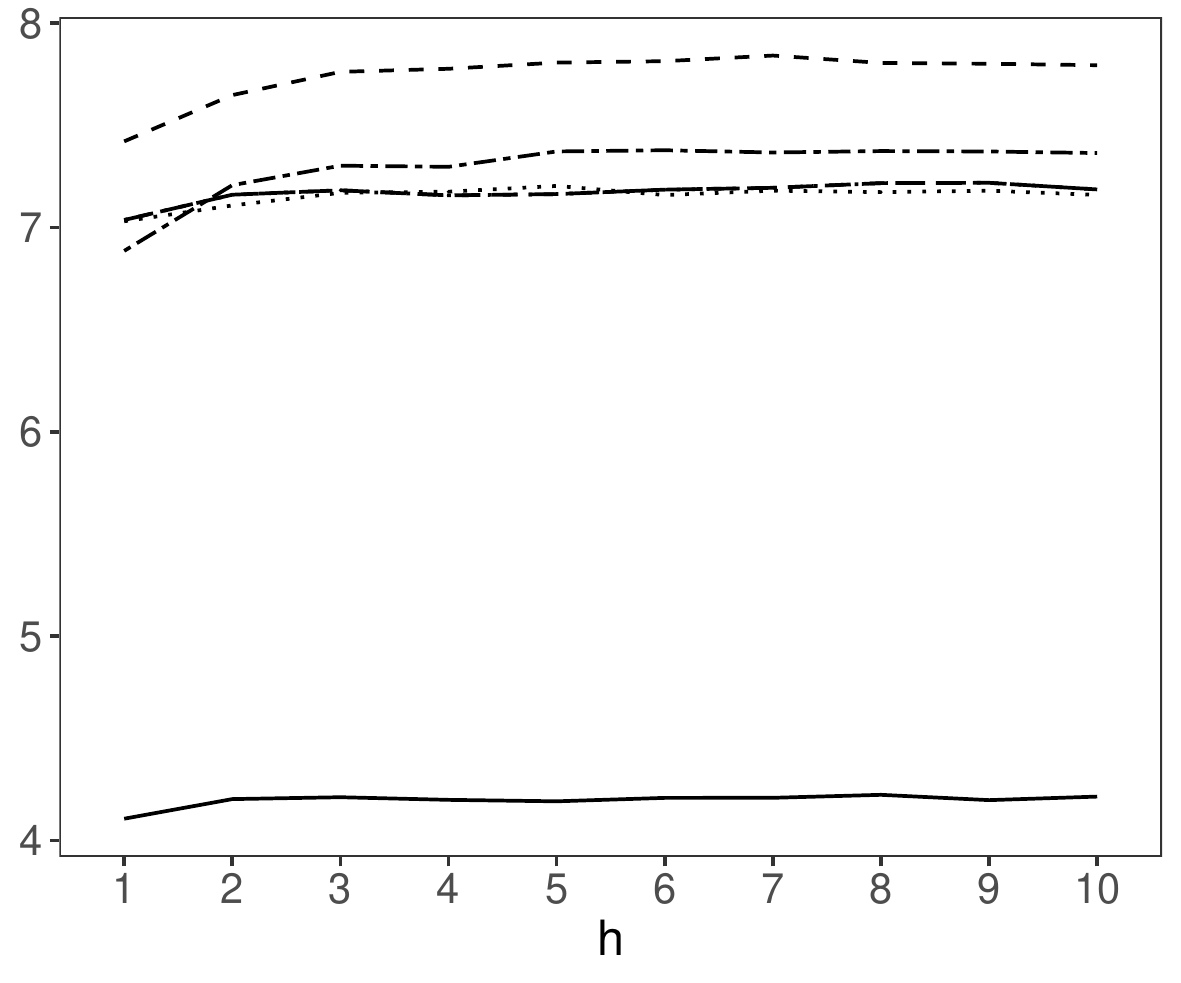}
  \includegraphics[width=5.7cm]{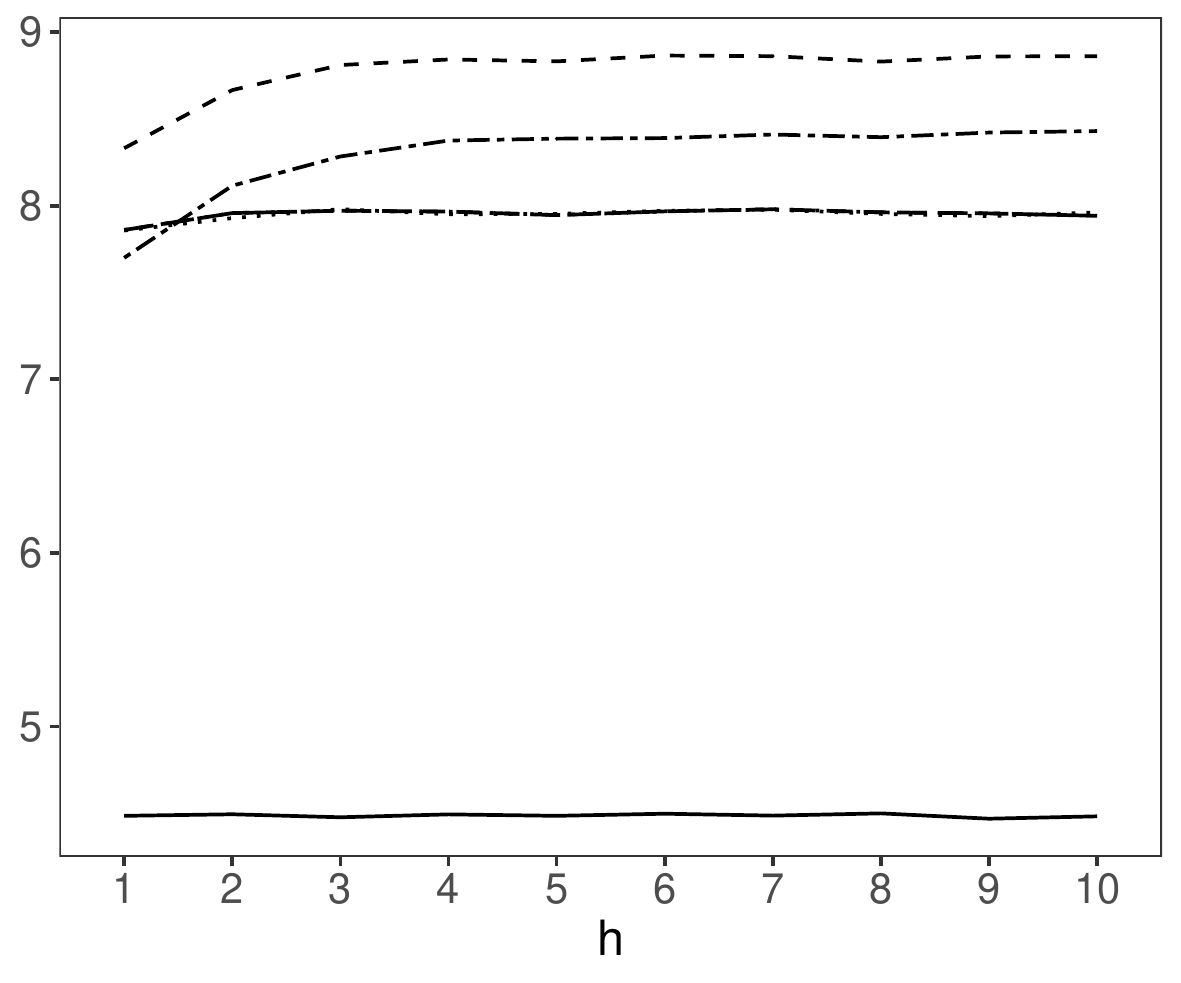}
  \caption{\small{Monte Carlo averages of the coverages (first row) and lengths (second row) of prediction intervals based on the: OLS (dotted line), weighted likelihood (solid line), Rob-YW (dashed line), Rob-Reg (dot-dashed line), Rob-Flt (long dashed line), and Rob-GM (two dashed line) methods for the AR(1) model when $C = \left[ 1\%~\text{(first column)}, 5\%~\text{(second column)}, 10\%~\text{(third column)} \right]$ of the generated training data are contaminated by deliberately inserted AOs.}}
  \label{fig:3}
\end{figure}

\begin{figure}[!htbp]
  \centering
  \includegraphics[width=5.7cm]{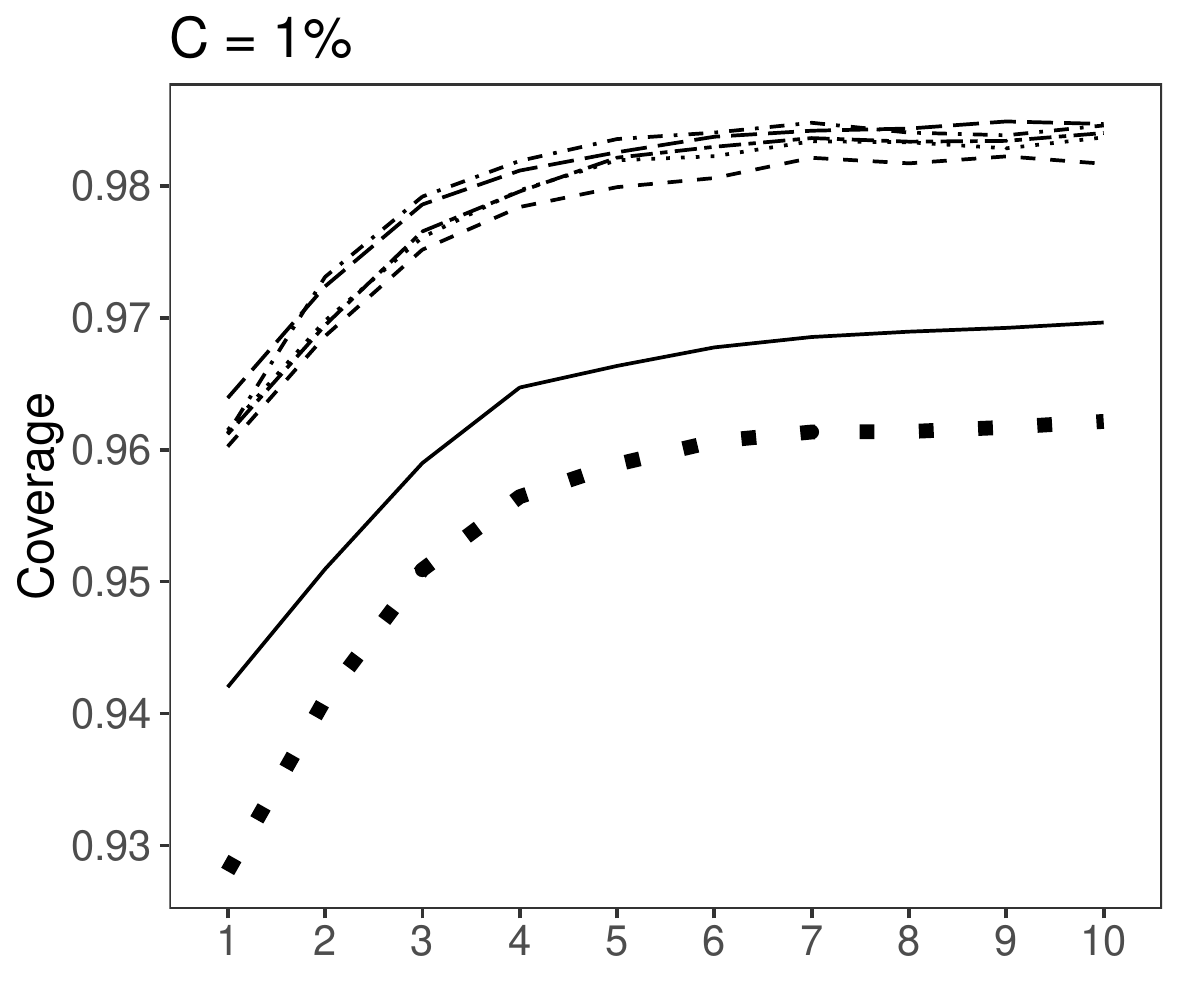}
  \includegraphics[width=5.7cm]{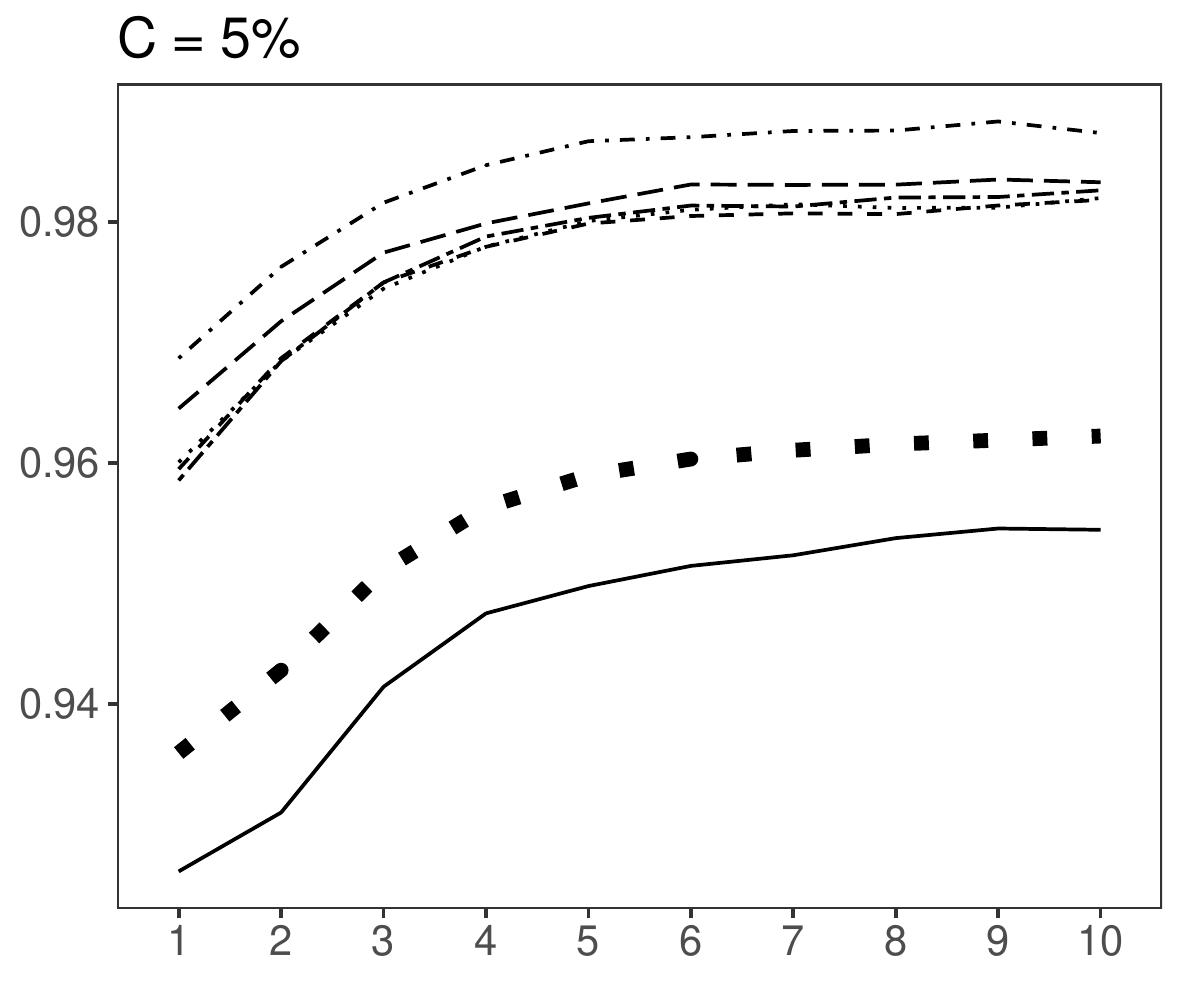}
  \includegraphics[width=5.7cm]{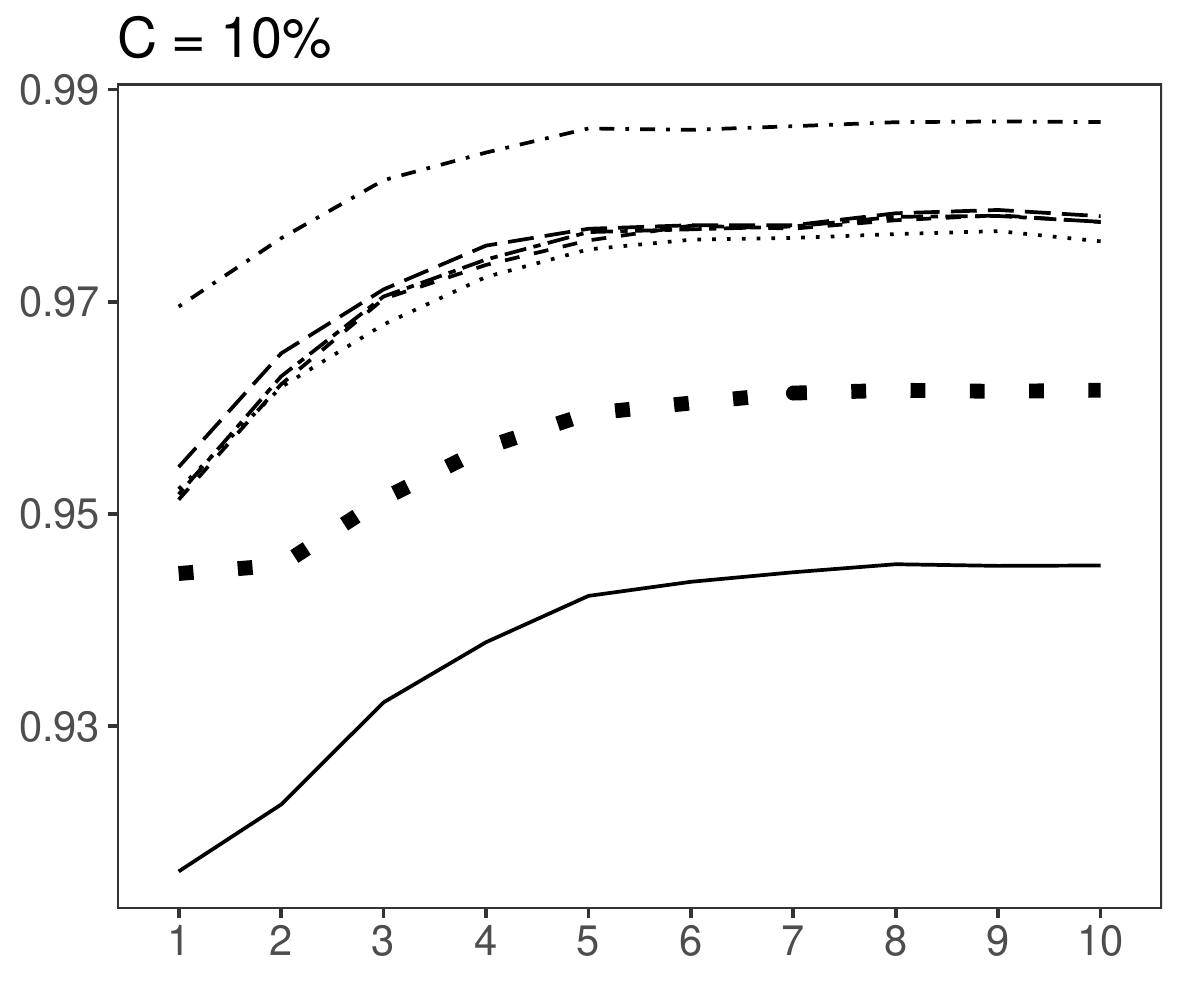}
\\
  \includegraphics[width=5.7cm]{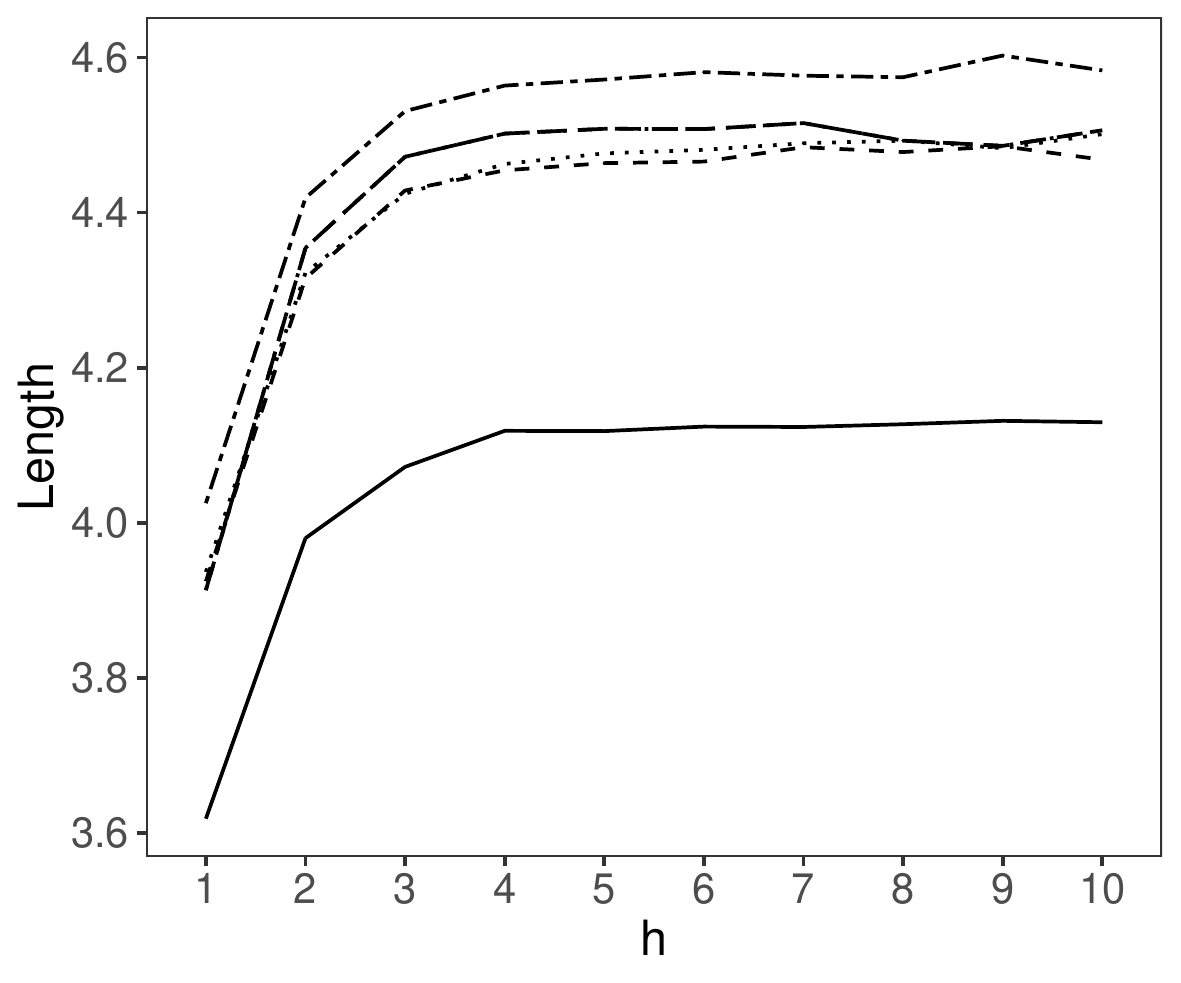}
  \includegraphics[width=5.7cm]{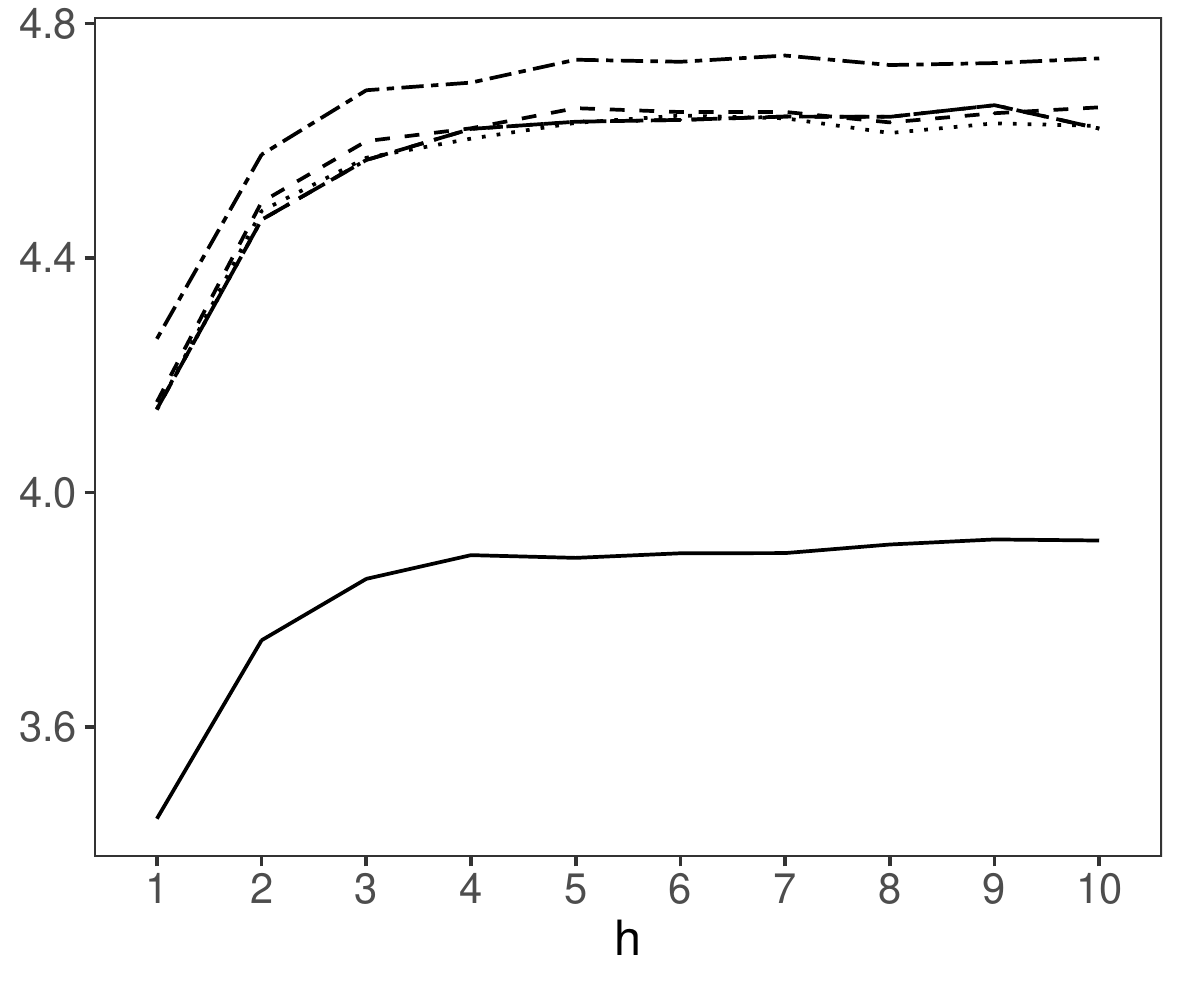}
  \includegraphics[width=5.7cm]{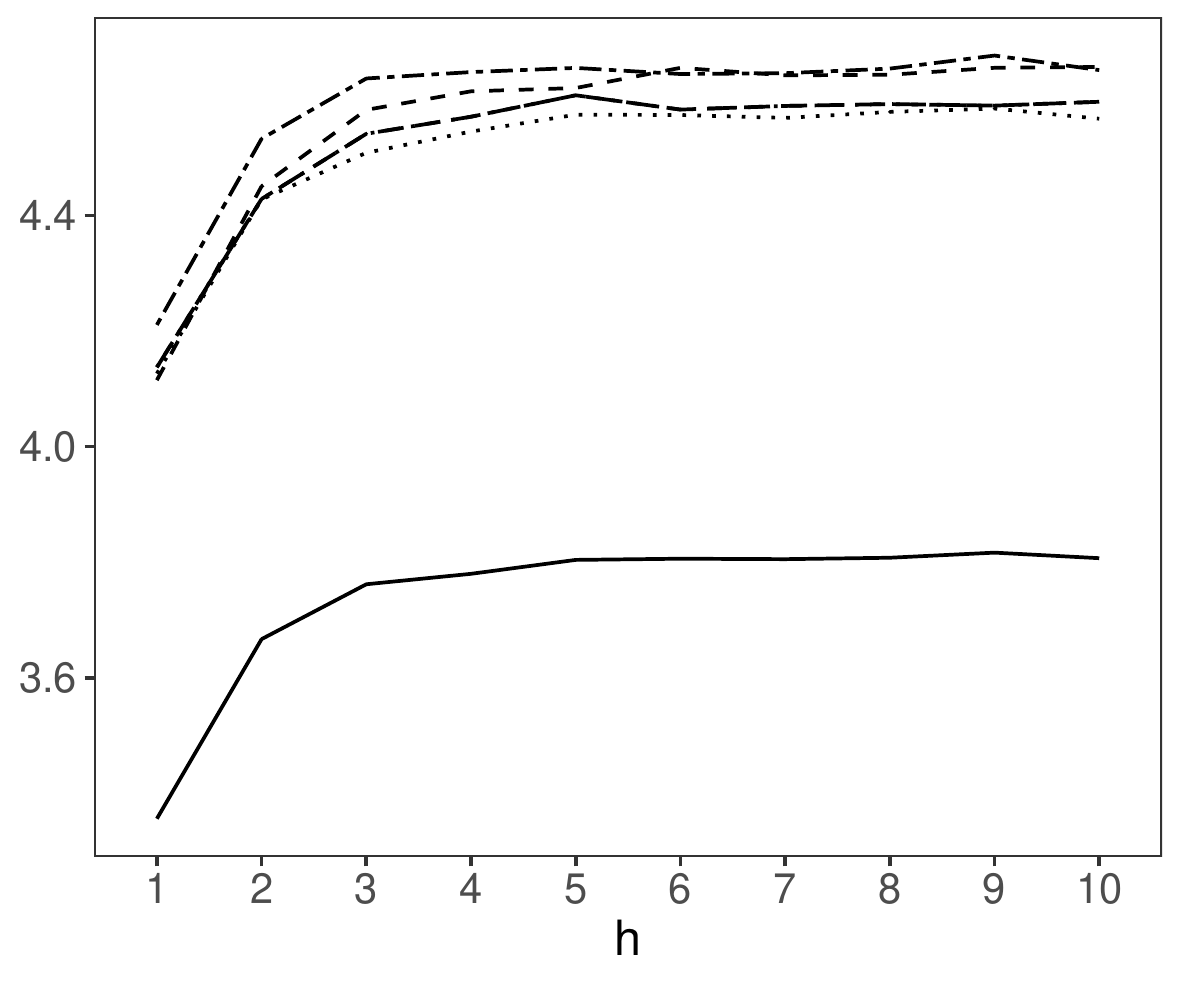}
  \caption{\small{Monte Carlo averages of the coverages (first row) and lengths (second row) of prediction intervals based on the: OLS (dotted line), weighted likelihood (solid line), Rob-YW (dashed line), Rob-Reg (dot-dashed line), Rob-Flt (long dashed line), and Rob-GM (two dashed line) methods for the AR(1) model when $C = \left[ 1\%~\text{(first column)}, 5\%~\text{(second column)}, 10\%~\text{(third column)} \right]$ of the generated training data are contaminated by deliberately inserted IOs.}}
  \label{fig:4}
\end{figure}

\begin{figure}[!htbp]
  \centering
  \includegraphics[width=5.7cm]{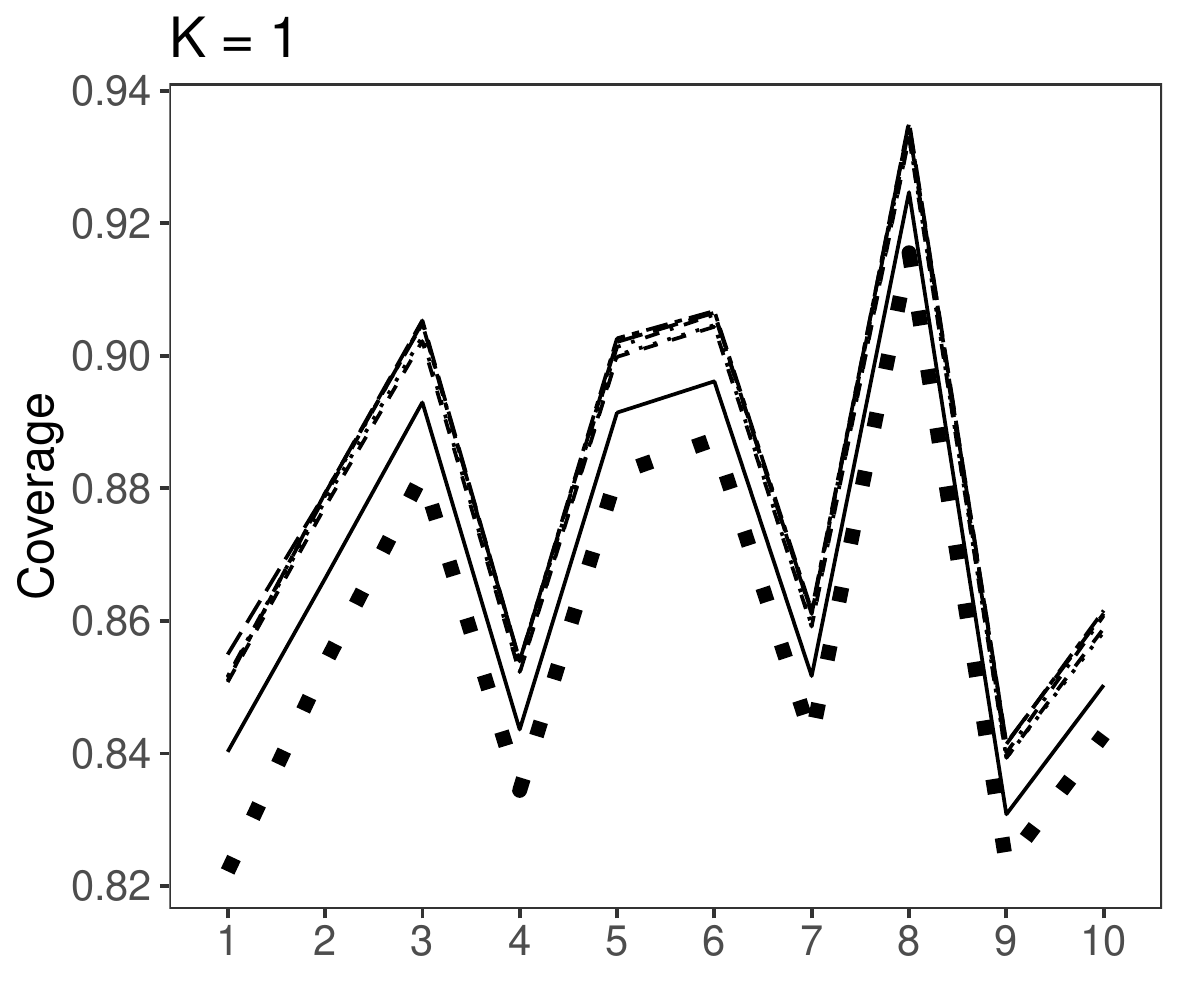}
  \includegraphics[width=5.7cm]{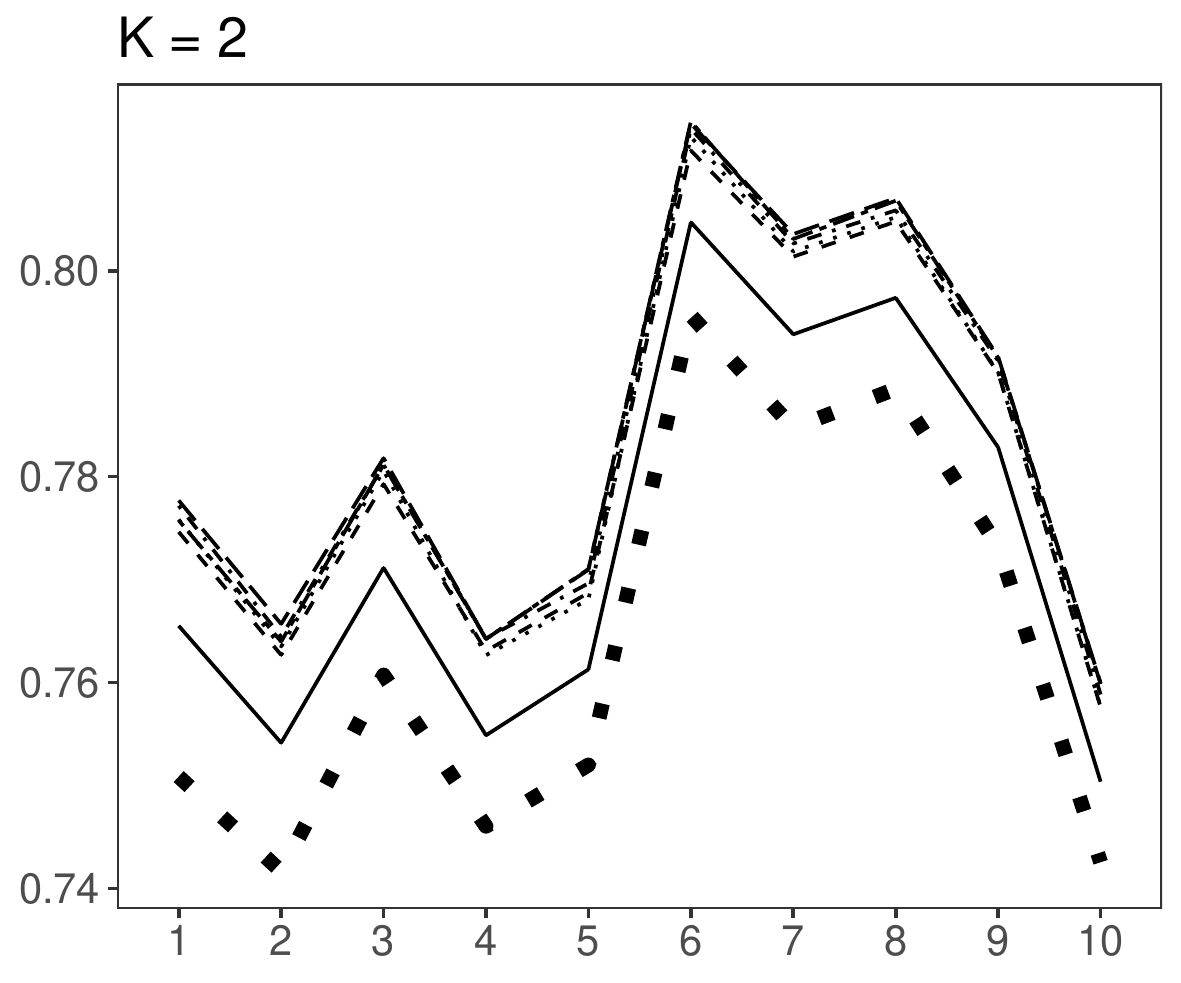}
  \includegraphics[width=5.7cm]{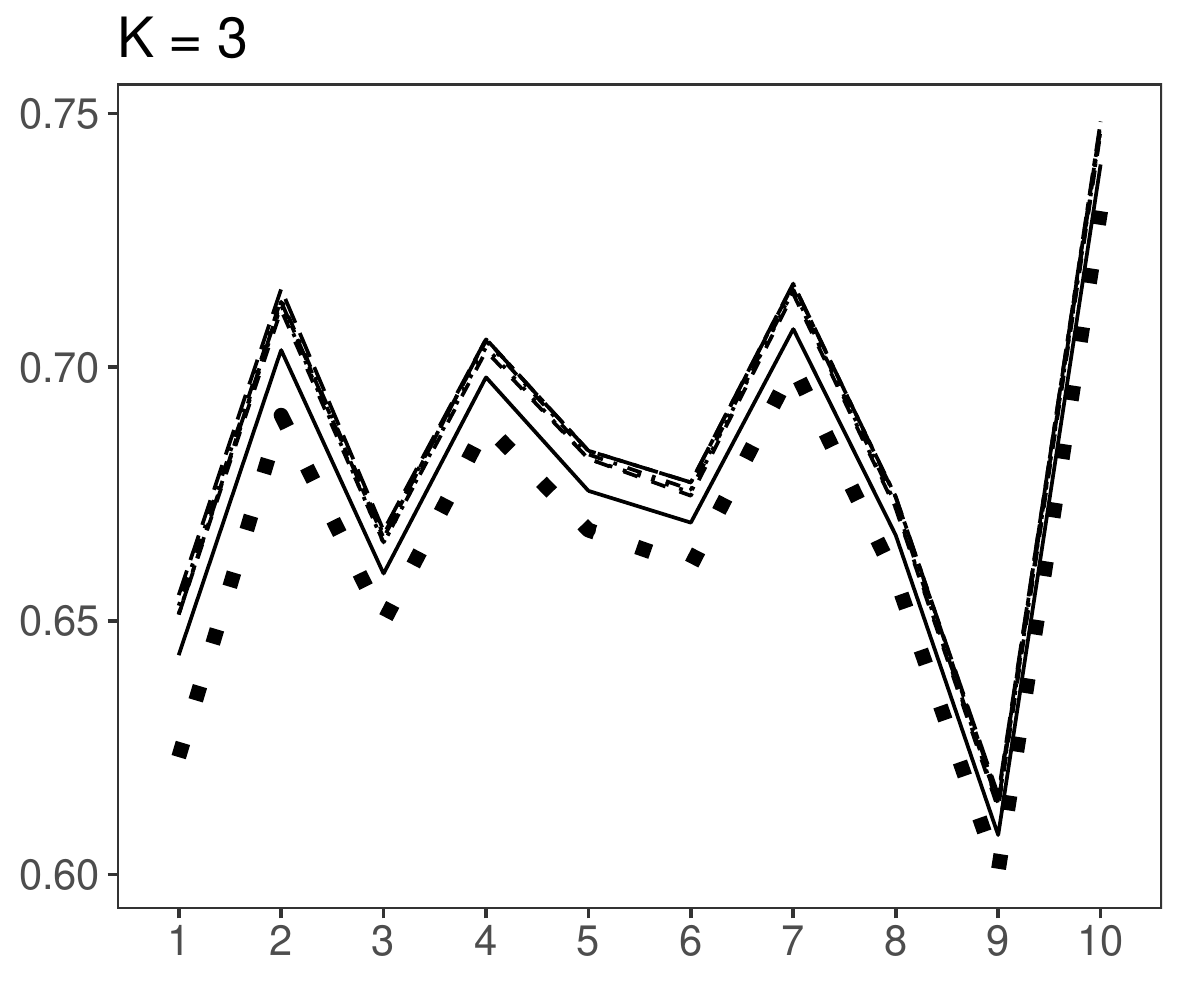}
\\
  \includegraphics[width=5.7cm]{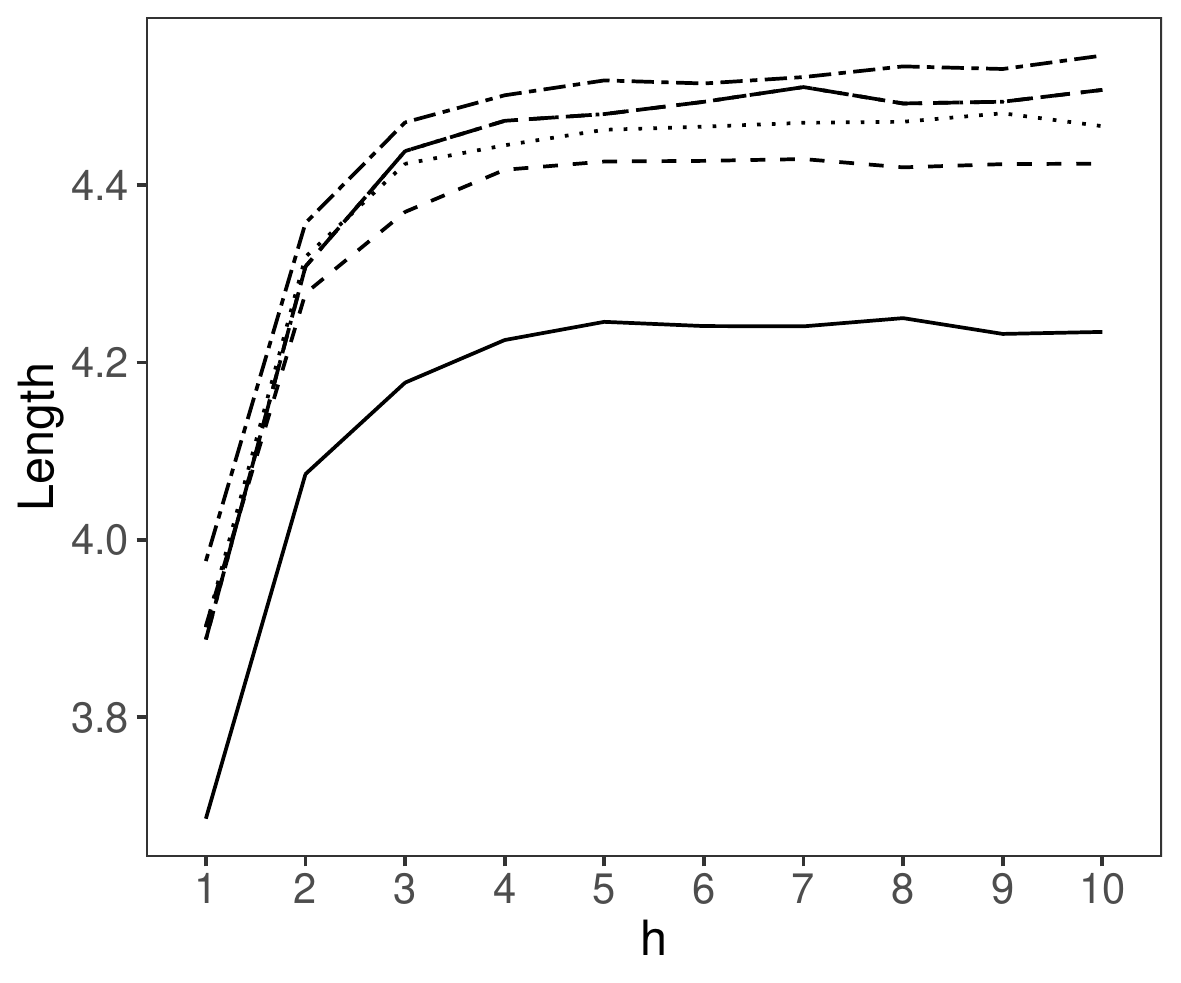}
  \includegraphics[width=5.7cm]{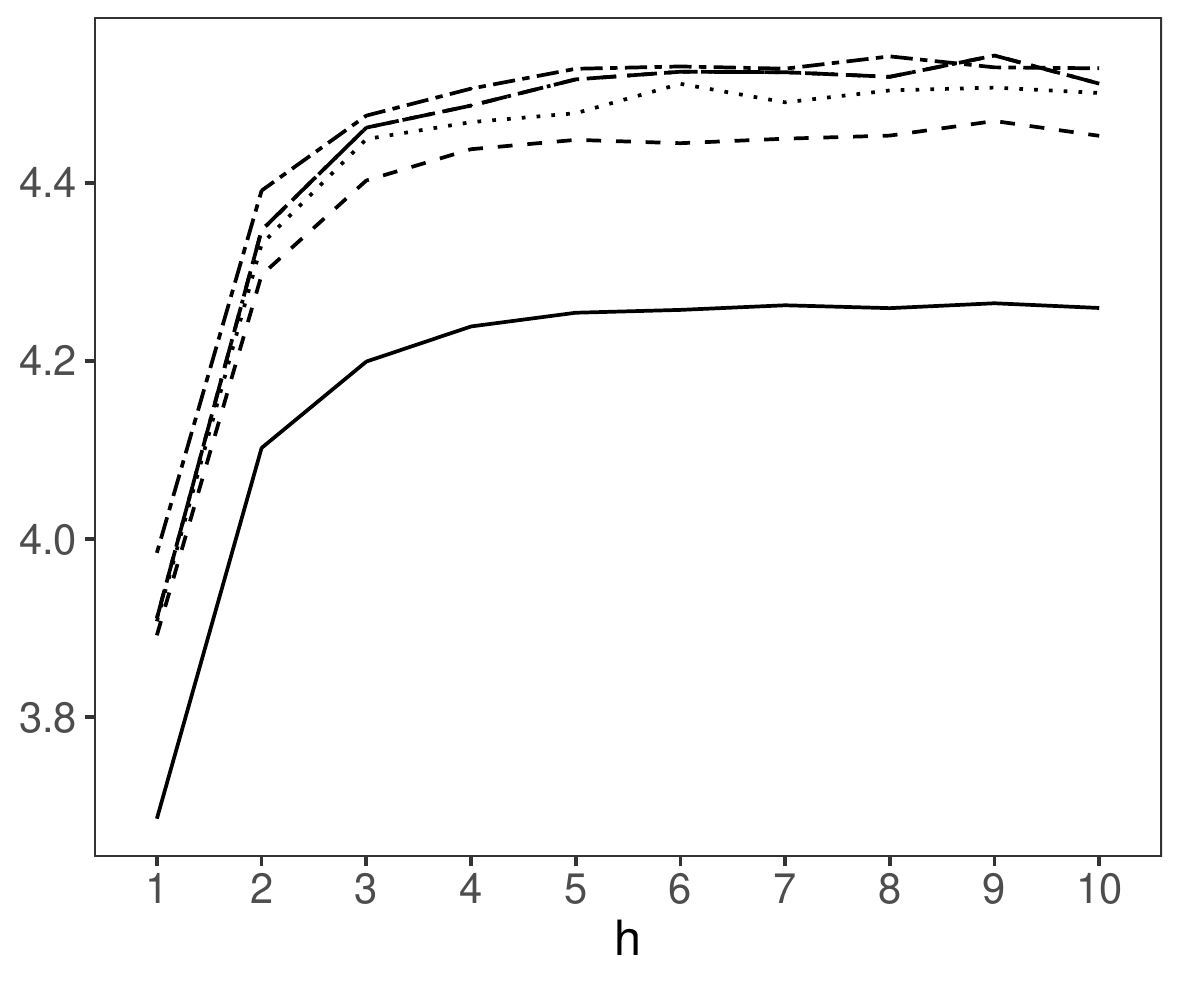}
  \includegraphics[width=5.7cm]{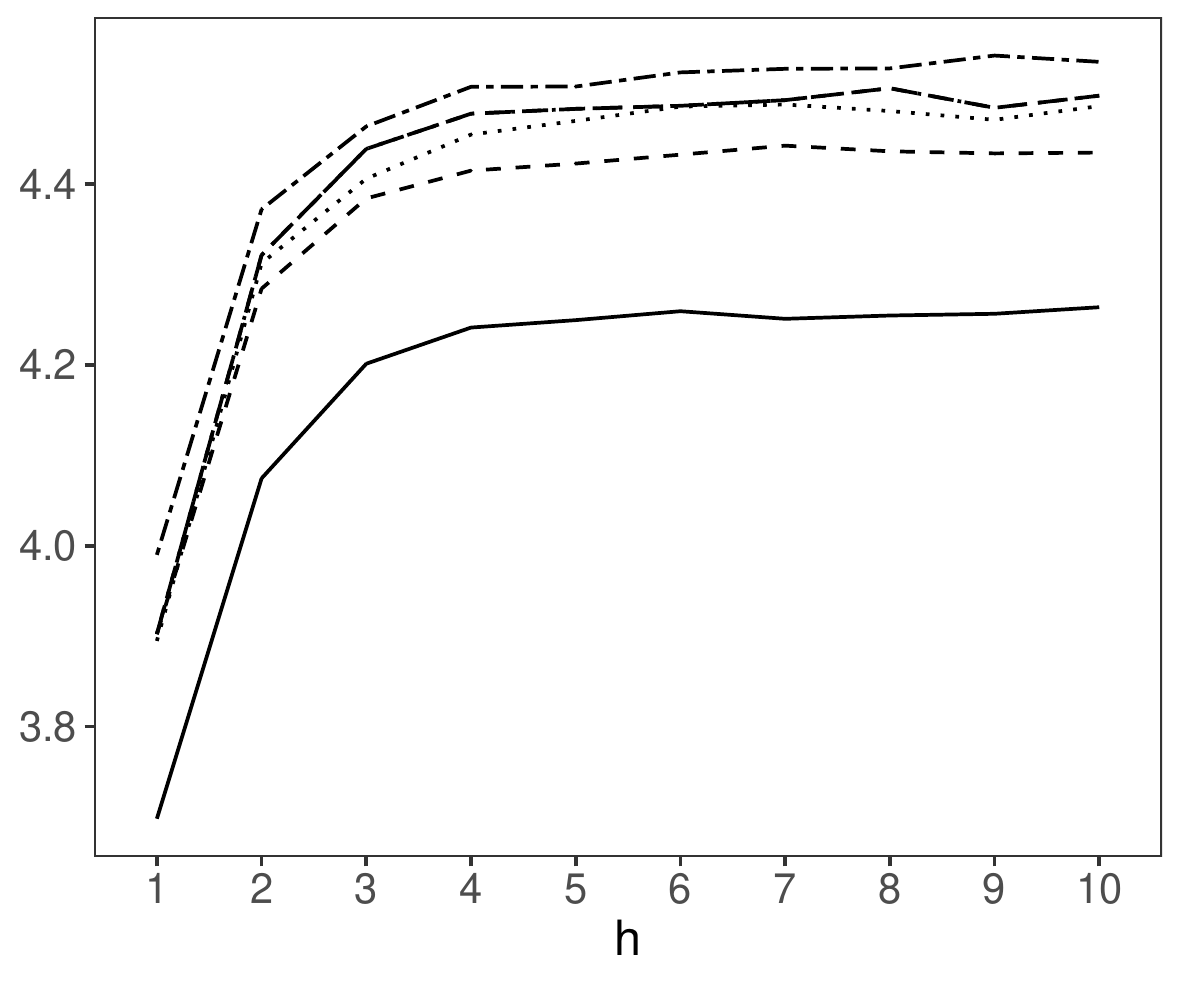}
  \caption{\small{Monte Carlo averages of the coverages (first row) and lengths (second row) of prediction intervals based on the: OLS (dotted line), weighted likelihood (solid line), Rob-YW (dashed line), Rob-Reg (dot-dashed line), Rob-Flt (long dashed line), and Rob-GM (two dashed line) methods for the AR(1) model when $K = \left[ 1~\text{(first column)}, 2~\text{(second column)}, 3~\text{(third column)} \right]$ of the generated future values are contaminated by deliberately inserted AOs.}}
  \label{fig:c2_3}
\end{figure}

\begin{figure}[!htbp]
  \centering
  \includegraphics[width=5.7cm]{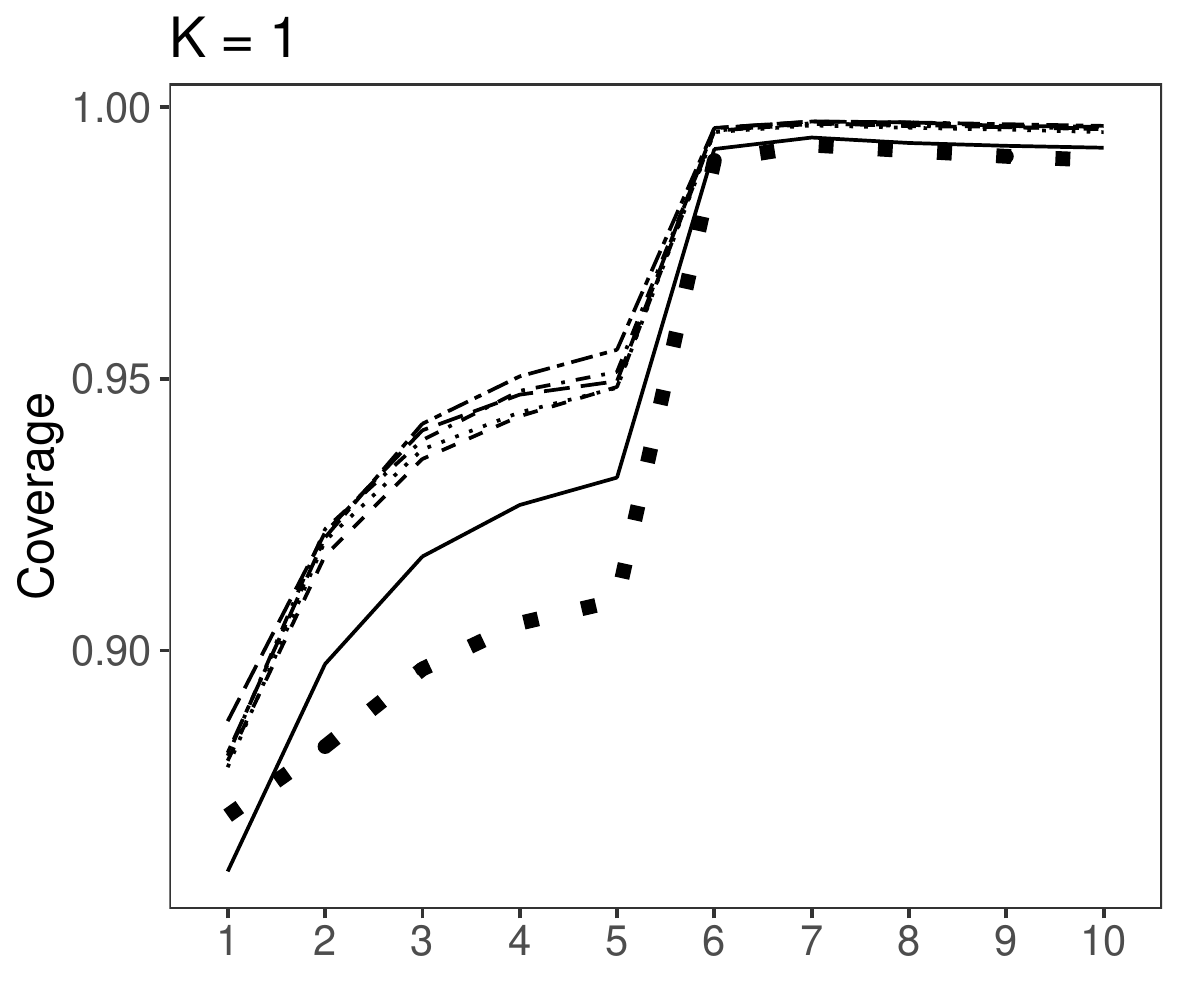}
  \includegraphics[width=5.7cm]{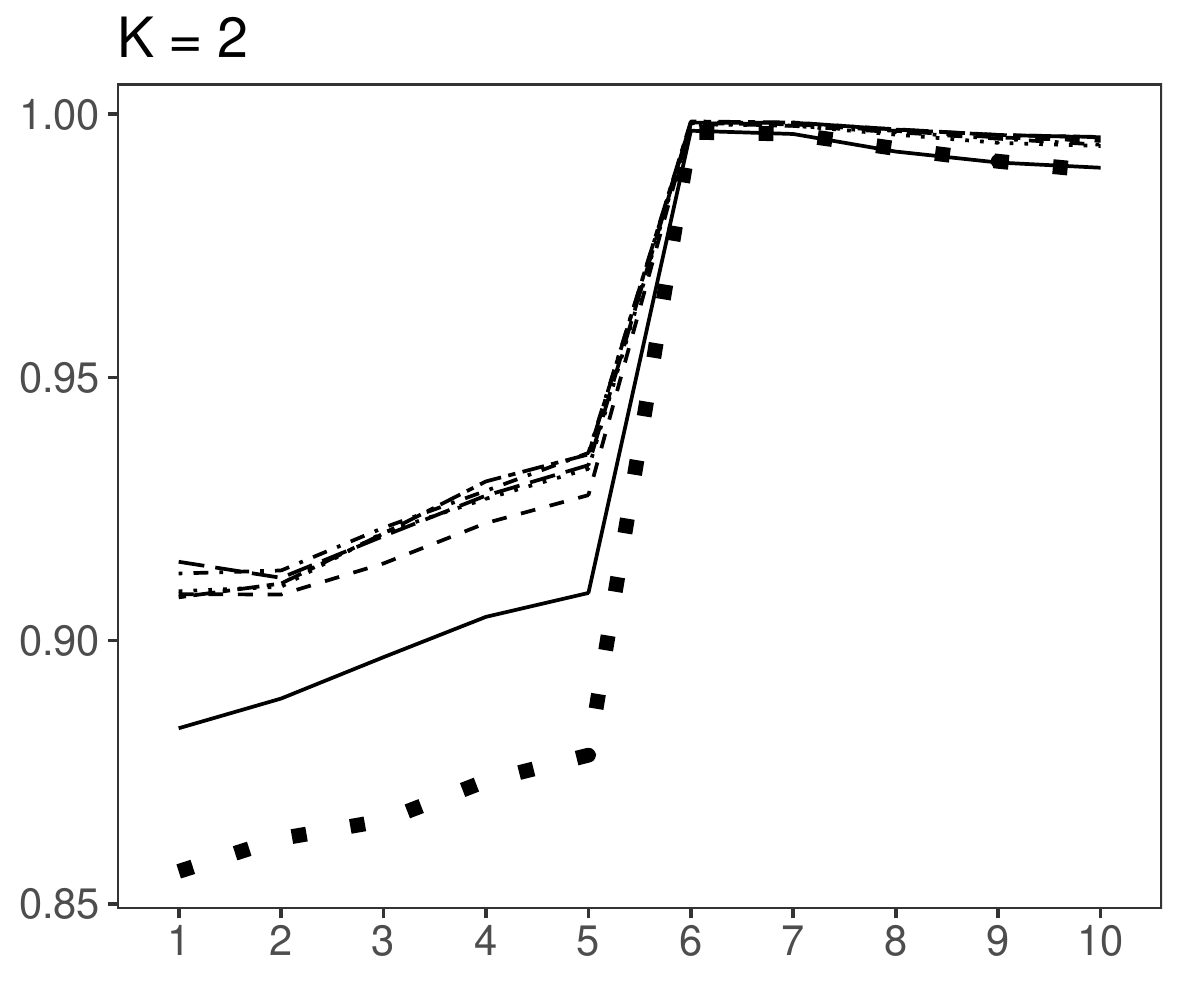}
  \includegraphics[width=5.7cm]{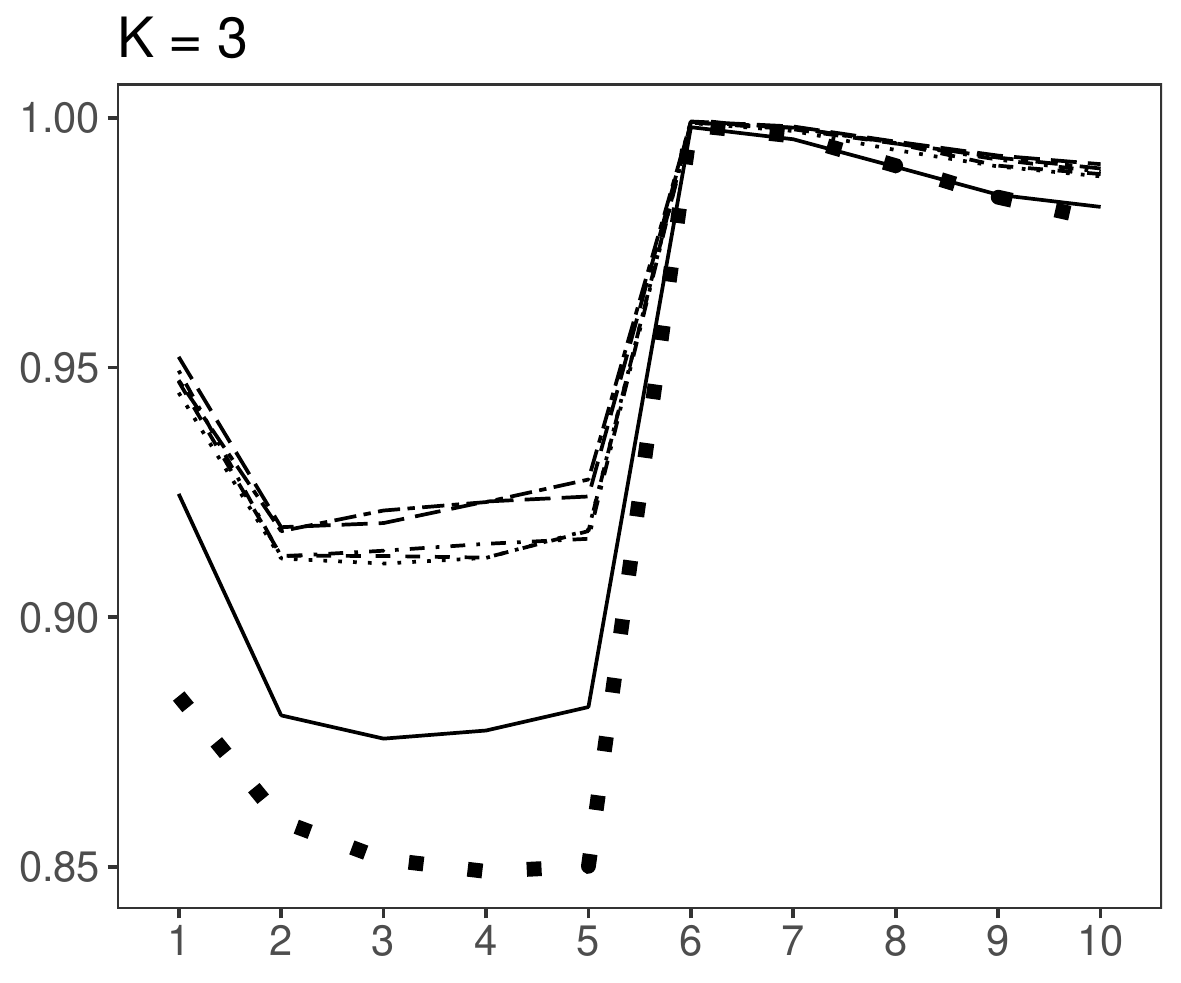}
\\
  \includegraphics[width=5.7cm]{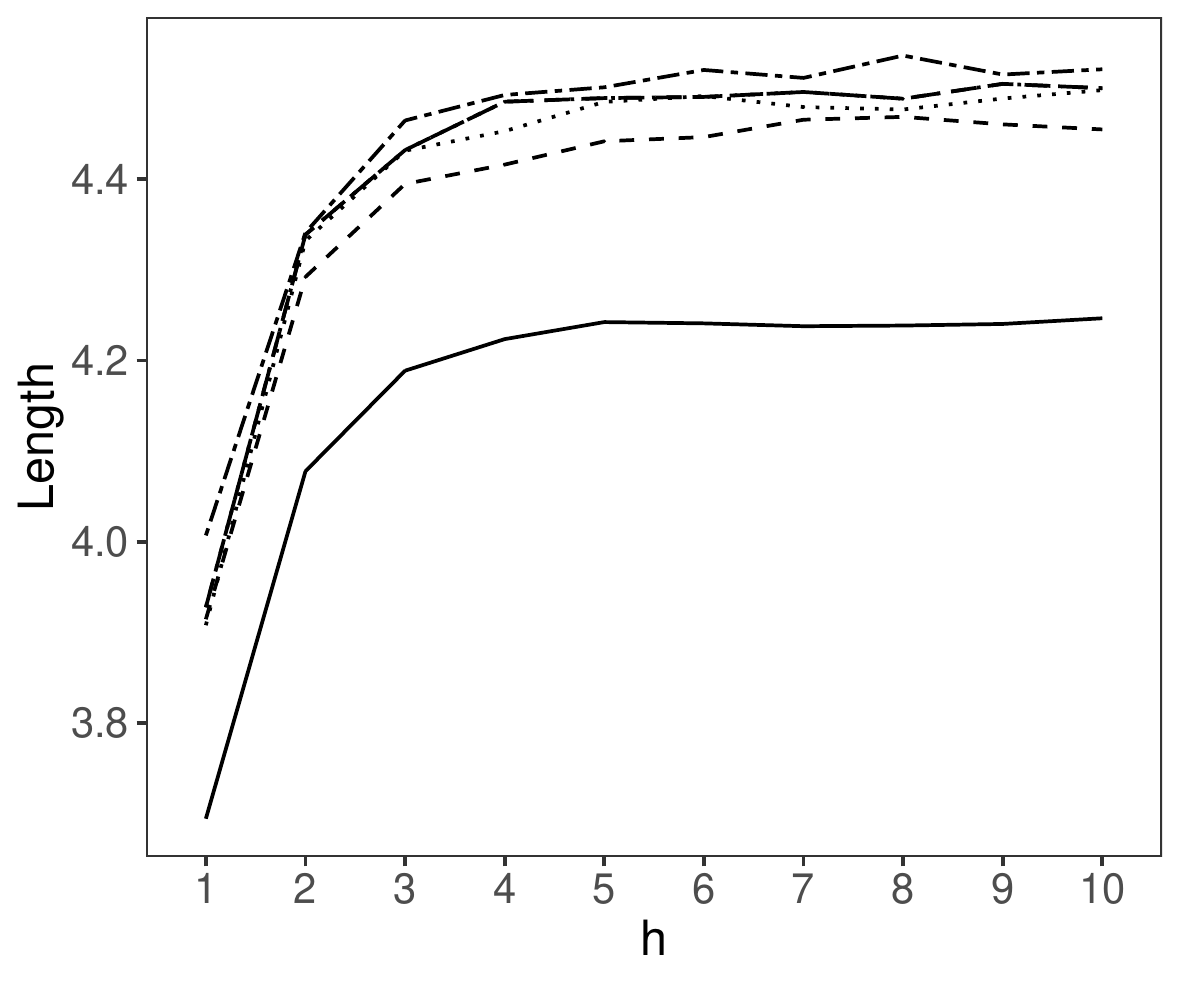}
  \includegraphics[width=5.7cm]{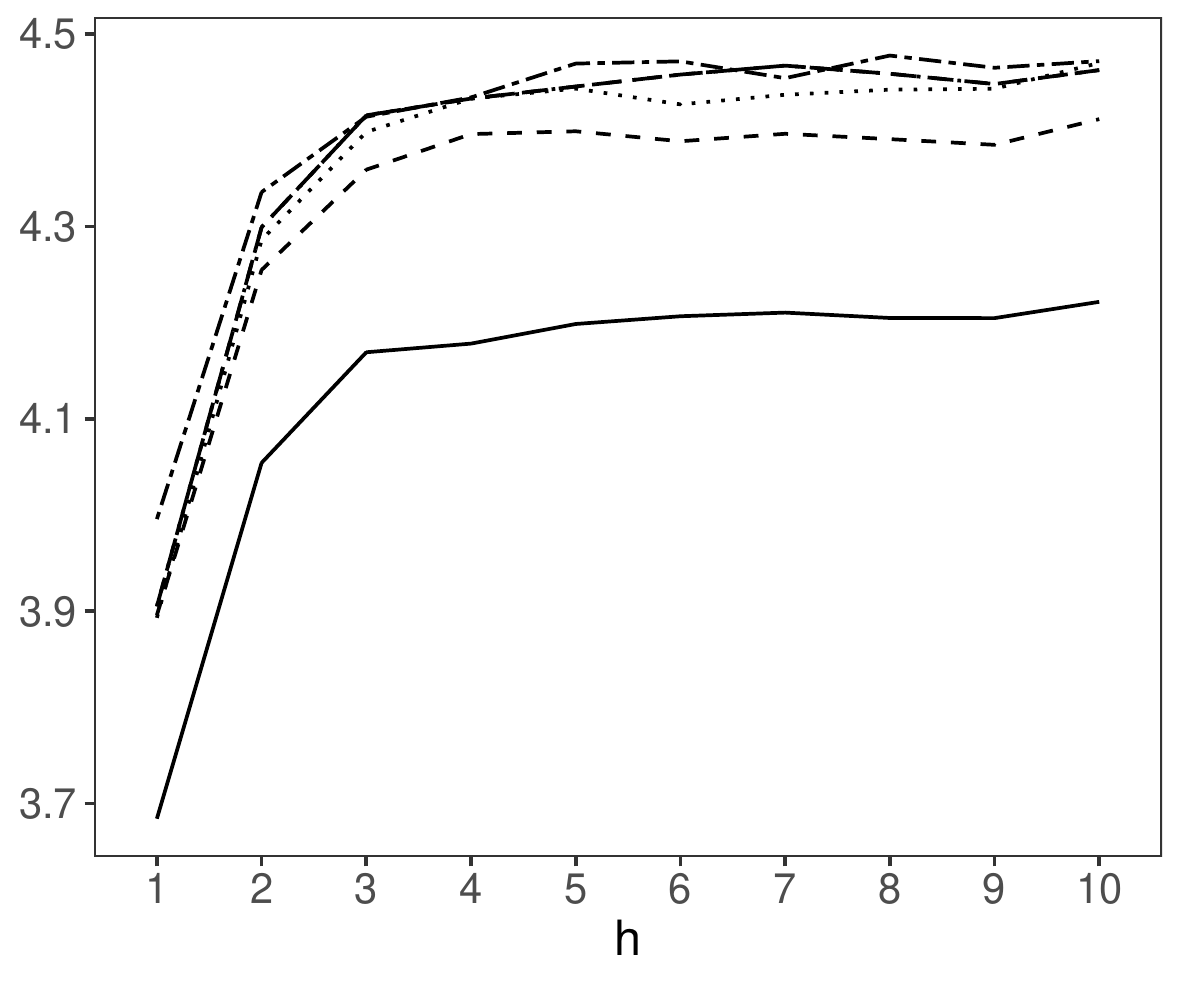}
  \includegraphics[width=5.7cm]{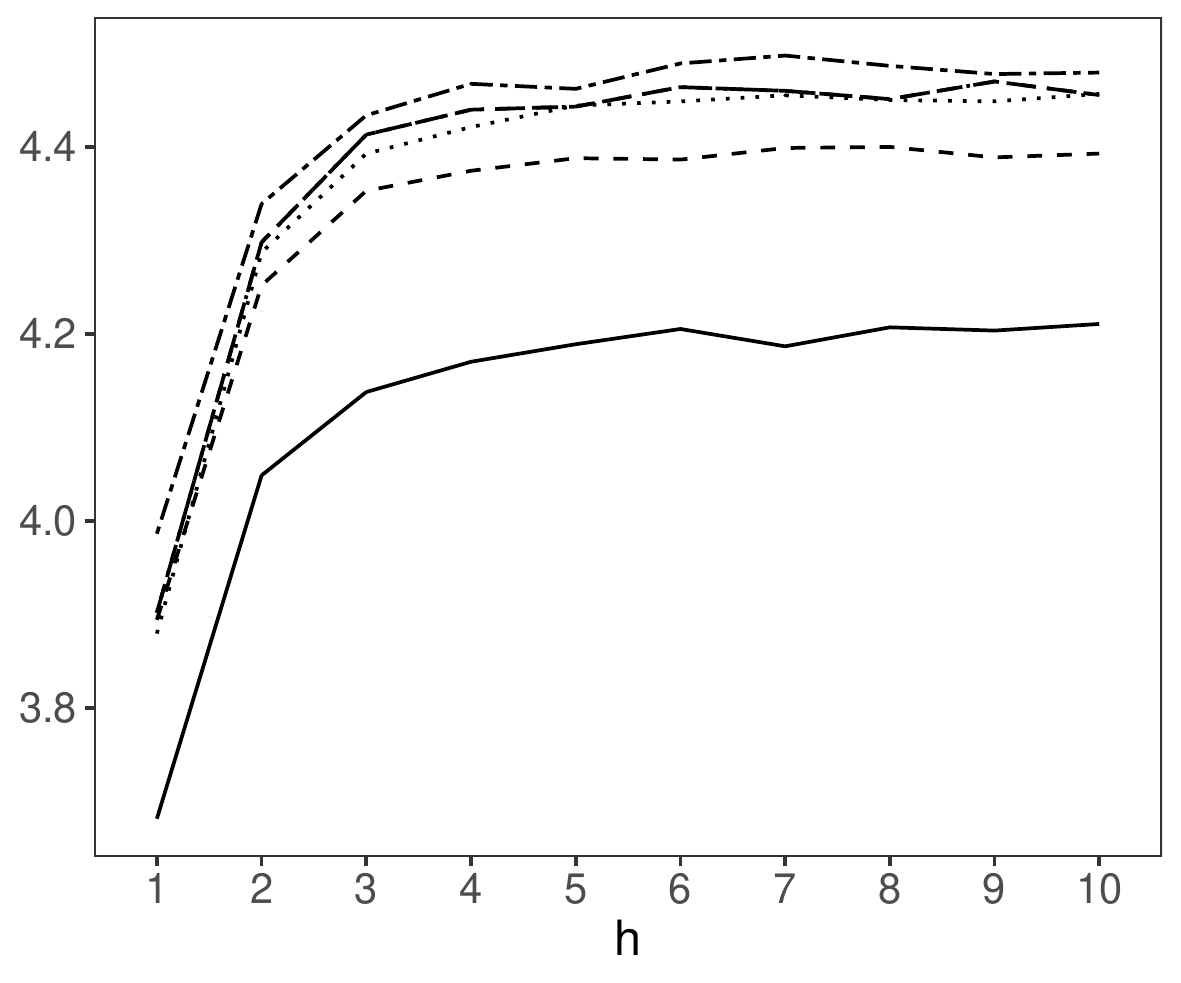}
  \caption{\small{Monte Carlo averages of the coverages (first row) and lengths (second row) of prediction intervals based on the: OLS (dotted line), weighted likelihood (solid line), Rob-YW (dashed line), Rob-Reg (dot-dashed line), Rob-Flt (long dashed line), and Rob-GM (two dashed line) methods for the AR(1) model when $K = \left[ 1~\text{(first column)}, 2~\text{(second column)}, 3~\text{(third column)} \right]$ of the generated future values are contaminated by deliberately inserted IOs.}}
  \label{fig:c2_4}
\end{figure}

For the univariate case, we consider the following AR(1) model:
\begin{equation}
y_t = \phi_0 + \phi_1 y_{t-1} + \epsilon_t, ~~ t = 1, \cdots, T,
\end{equation}
where $\epsilon_t \sim N(0,1)$. The modeling scenarios are adapted such that
\begin{inparaenum}
\item[1)] no clear outlying data points are generated, 
\item[2)] $C = \left[ 1\%, 5\%, 10\% \right]$ of only the generated training data are contaminated by deliberately inserted AOs or IOs, and
\item[3)] $K = \left[ 1, 2, 3 \right]$ of the generated future values are contaminated by deliberately inserted AOs or IOs.
\end{inparaenum}
AOs are generated by contaminating $T C$ randomly selected observations with the value 5, while IOs are generated by contaminating $T C$ randomly selected innovations with the value 3. For the first scenario, where no outlying observations are deliberately inserted into the dataset, we consider the parameter values $\phi_0 = 0$ and $\phi_1 \in \left[ -0.9, -0.5, -0.1, 0.1, 0.5, 0.9 \right]$ to assess the effect of parameter values on the bootstrap prediction intervals. For the second and third scenarios, the parameter values are fixed at $\phi_0 = 0$ and $\phi_1 = 0.5$, and we check the effects of the number of outliers on the prediction intervals. For these cases, the average coverage probabilities when the bootstrap-based quantiles are replaced with the quantiles of the normal distribution are also calculated to explore whether bootstrapping improves the proposed and other statistical methods. The results are given in Figures~\ref{fig:1}-\ref{fig:c2_4}.


Our results show that when no fixed outlying data point is inserted into the dataset:
\begin{inparaenum}
\item[1)] all the bootstrap prediction intervals tend to produce similar average coverage probabilities when the time series data are persistent ($\phi_1 = \pm 0.9$), with only the Rob-YW slightly differing from other methods;
\item[2)] the bootstrap procedure based on the weighted likelihood methodology generally produces improved coverage probabilities than other methods when the generated data are not persistent;
\item[3)] when the generated data are not persistent, the proposed method produces coverage probabilities close to the nominal level and generally results in slight under-coverage (over-coverage is observed only for the case when $\phi_1 = 0.5$ and $h > 4$, but is observed for all bootstrap prediction intervals based on different methods); and 
\item[4)] the proposed bootstrap procedure has a smaller prediction interval length than other methods when the data are not persistent, while all bootstrap methods tend to produce similar interval lengths when $\phi_1 = \pm 0.9$. One exception is that Rob-Reg has smaller interval lengths than other methods when $\phi_1 = \pm 0.9$ and $h > 4$.
\end{inparaenum}
Note that, when the generated data have no deliberately inserted outliers, the proposed bootstrap method is expected to perform similarly to other methods. However, it performs better than other bootstrap prediction intervals in some cases, because some outlying observations with small magnitudes may be produced by the data generating process. The weighted likelihood method reduces the effects of such observations, giving the proposed method a slightly better performance than other methods in some cases.

When the training data have AOs, over-coverage ($\geq 0.98$, in general) is observed for all bootstrap prediction intervals, excluding the proposed method because of the high forecasting errors produced in line with the outlying points. In contrast, the proposed procedure is less affected by these points and provides coverage probabilities close to the nominal level. In addition, the proposed idea has significantly narrower prediction intervals than other methods when AOs are present in the data (see Figure~\ref{fig:3}).

When the generated training data have IOs, compared with other methods, the proposed bootstrap method generally performs poorly for small $h$ in terms of empirical coverage probability. Under-coverage is observed for the proposed method while the other methods result in over-coverage. As in the AO case, the proposed bootstrap method produces significantly narrower prediction interval lengths compared with other methods. When comparing the performance of the bootstrap methods for the AO and IO cases (Figures~\ref{fig:3} and~\ref{fig:4}), the results demonstrate that the bootstrap methods produce slightly lower coverage values when AOs are present in the generated datasets than those of IOs. This is due to fact that the bootstrap methods tend to produce wider prediction intervals when the data are contaminated by AOs compared to IO case. As is mentioned in Section~\ref{sec:outlier}, an AO at time $t$ influences corresponding residual as well as the future residuals (at time $t+1, \cdots$) while an IO at time $t$ influences only the corresponding residual. Therefore, compared with the IO case, more high-valued forecast errors are present when the data have AOs. As a result, the bootstrap methods produces wider prediction intervals when AOs are present in the data compared with IOs, and thus, the prediction intervals cover more future values.

Moreover, our findings for the coverage probabilities indicate that bootstrap-based quantiles generally perform better than the quantiles of the normal distribution when the time series is generated without outliers. When the training data are contaminated by AOs, the proposed bootstrap method outperforms normal quantiles for small forecast horizons, while the quantiles obtained from the proposed method and the normal distribution produce similar coverage performances for mid-term and long-term forecast horizons. In contrast, when the training data are contaminated by IOs, the quantiles of the normal distribution perform better than all bootstrap-based quantiles for moderate and large values of $h$, while the bootstrap-based quantiles perform better when $h$ is small. Our conclusions about the comparison between bootstrap-based and normal quantiles are restricted to data structures considered in this study, and may not be readily generalizable.

When the generated future values include AOs (see Figure~\ref{fig:c2_3}), under-coverage is observed for all the bootstrap methods and normal quantiles. In this case, all the methods perform considerably less in terms of coverage probability compared with the second scenario, and their coverage performances get worse as the number of outliers increase. On the other hand, when the generated future observations have IOs (see Figure~\ref{fig:c2_4}), for all bootstrap methods, under-coverage is observed for short-term forecast horizons while over-coverage is observed for long-term forecast horizons. For this scenario, the proposed bootstrap method produces smaller prediction interval length, among others.

\subsection{Numerical results for the VAR model}

The experimental design of our simulation study for the VAR case is somewhat similar to that of \cite{FresoliRuisPascual2015}. We consider the following VAR(2) model:
\begin{equation*}
Y_t = \Phi_0 + \Phi_1 Y_{t-1} + \Phi_2 Y_{t-2} + \epsilon_t,
\end{equation*}
where
\[ 
\Phi_0 =
0,
~~
\Phi_1 =
\begin{bmatrix}
0.9 & 0.0 \\
-0.5 & -0.7
\end{bmatrix},
~~
\Phi_2 =
\begin{bmatrix}
-0.2 & 0.0 \\
0.8 & -0.1
\end{bmatrix},
~~
\Sigma_{\epsilon} =
\begin{bmatrix}
1.0 & 0.5 \\
0.5 & 1.0
\end{bmatrix}.
\]
The distribution of the error process depends on the modeling scenarios; three error distributions-Gaussian $N(0,1)$, $t_5$ (heavy-tailed) and $\chi^2(4)$ (skewed)-are considered when the data have no clear outliers while the errors are distributed as $N(0,1)$ and $t_5$ when AOs or IOs are present in the data. To generate AOs, $C = \left[ 1\%, 5\%, 10\% \right]$ of the generated data are randomly selected, and both components contaminated by the value 5 while $T C$ innovation terms are randomly selected and only the first component of the innovations are contaminated by the value 5 to generate the IOs.

\begin{figure}[!htbp]
  \centering
  \includegraphics[width=5.9cm]{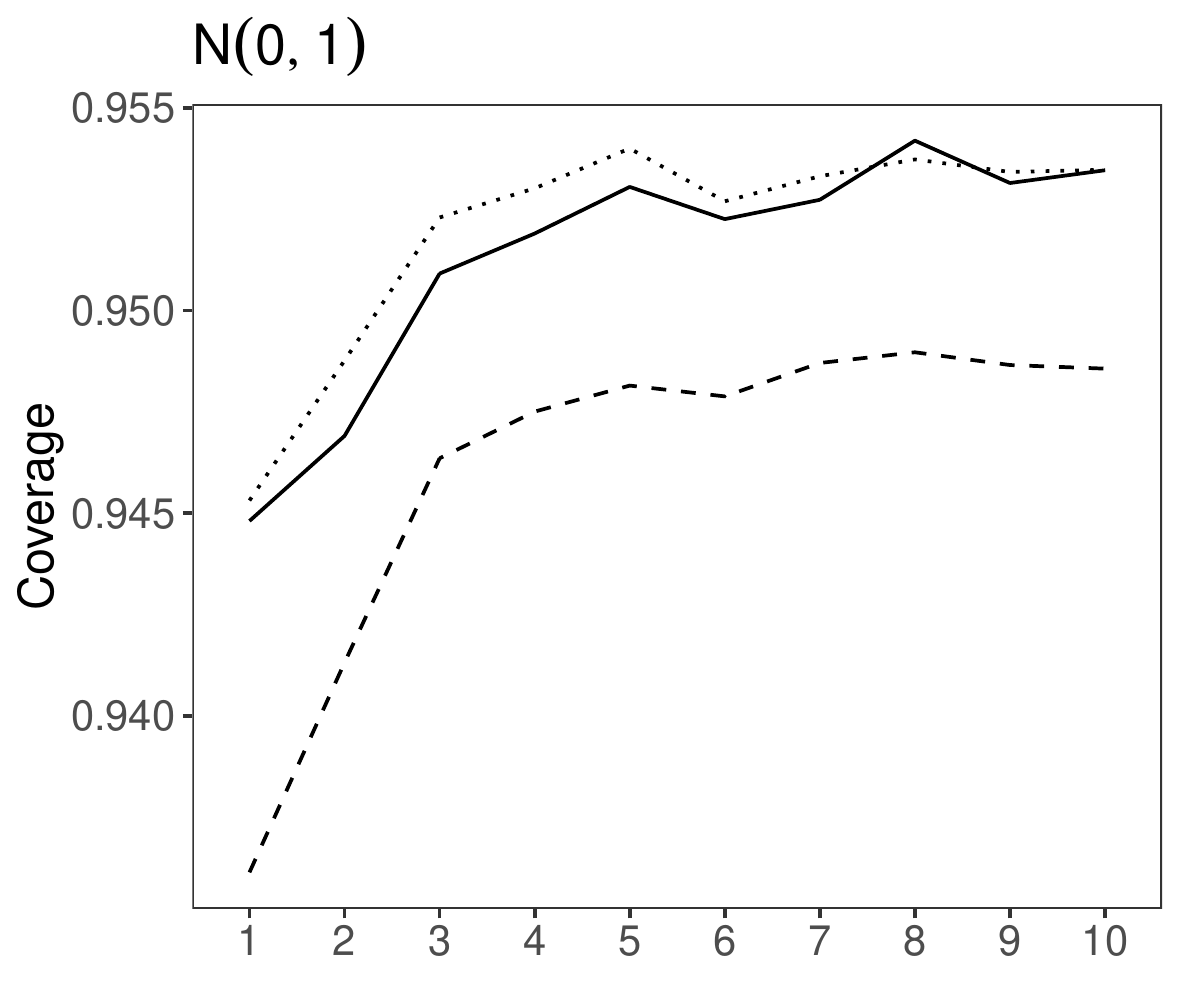}
  \includegraphics[width=5.9cm]{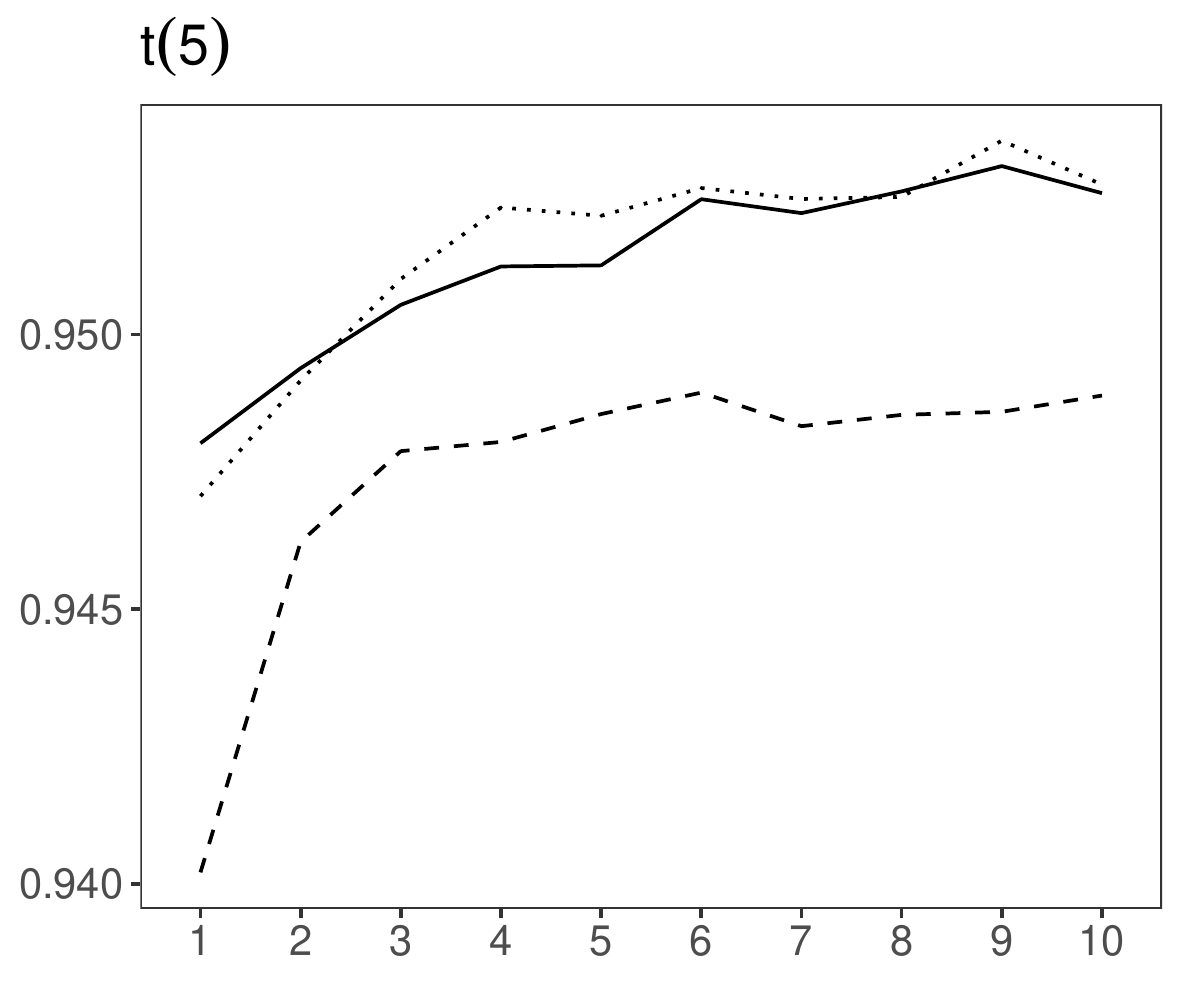}
  \includegraphics[width=5.9cm]{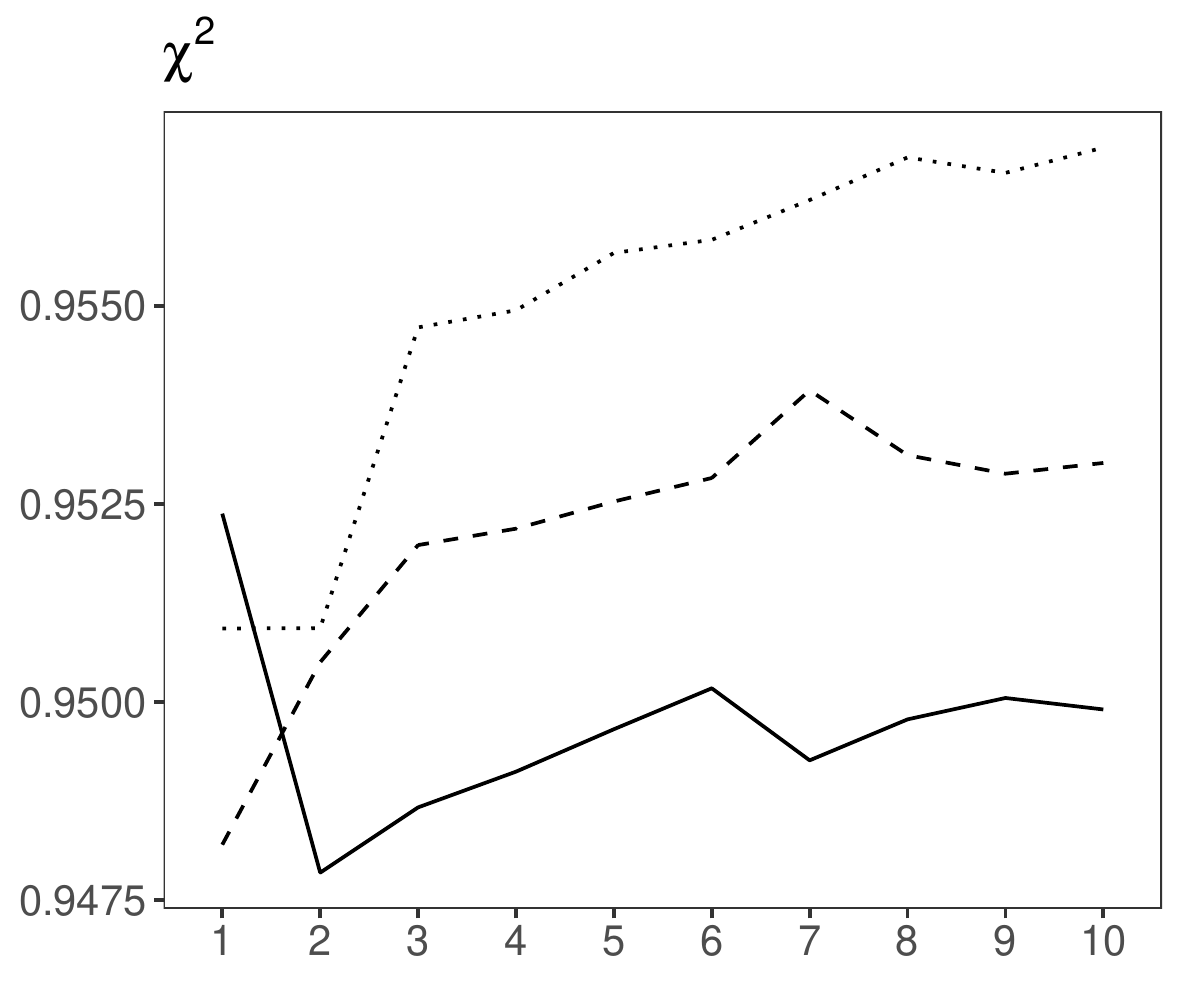}
\\
  \includegraphics[width=5.9cm]{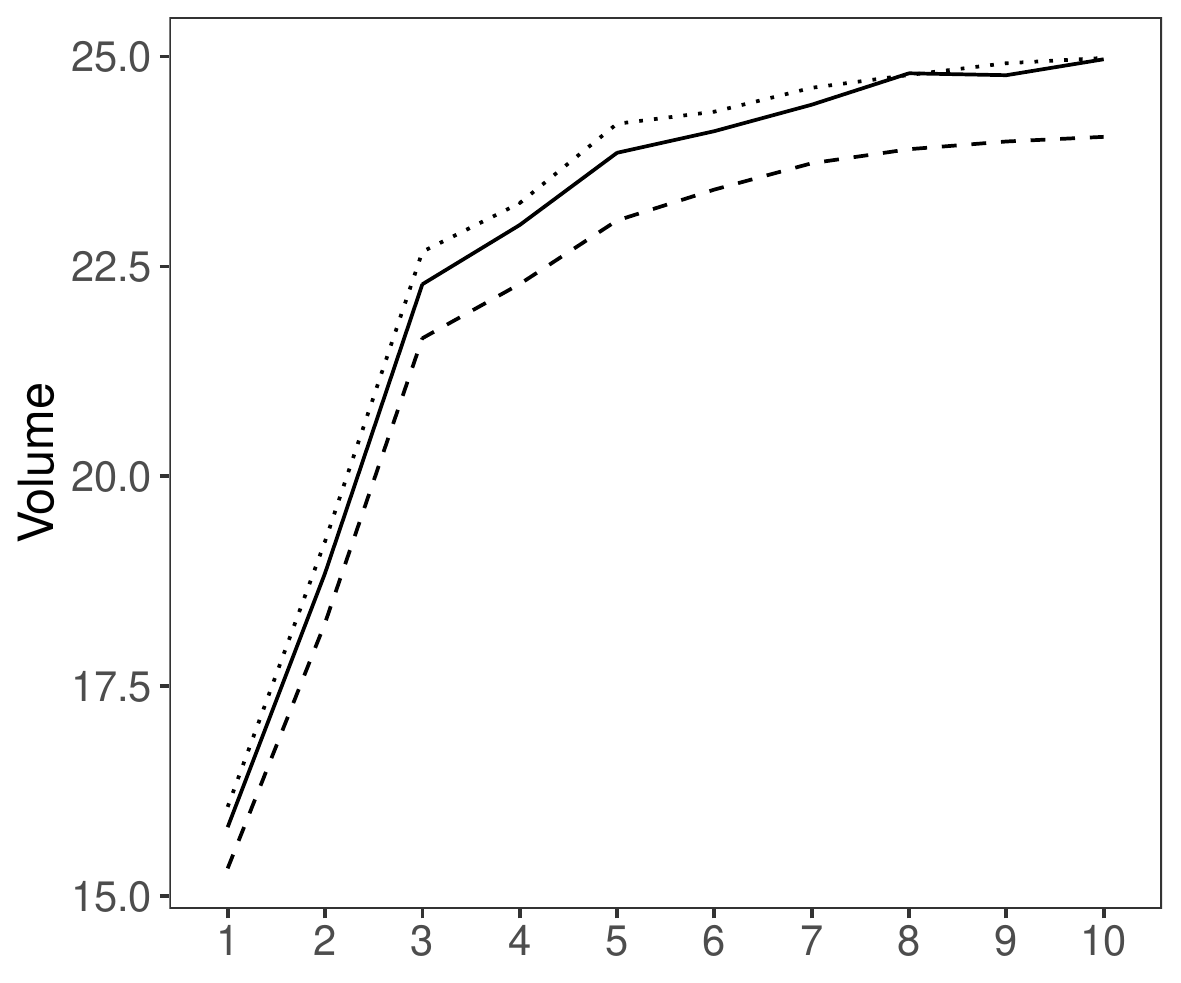}
  \includegraphics[width=5.9cm]{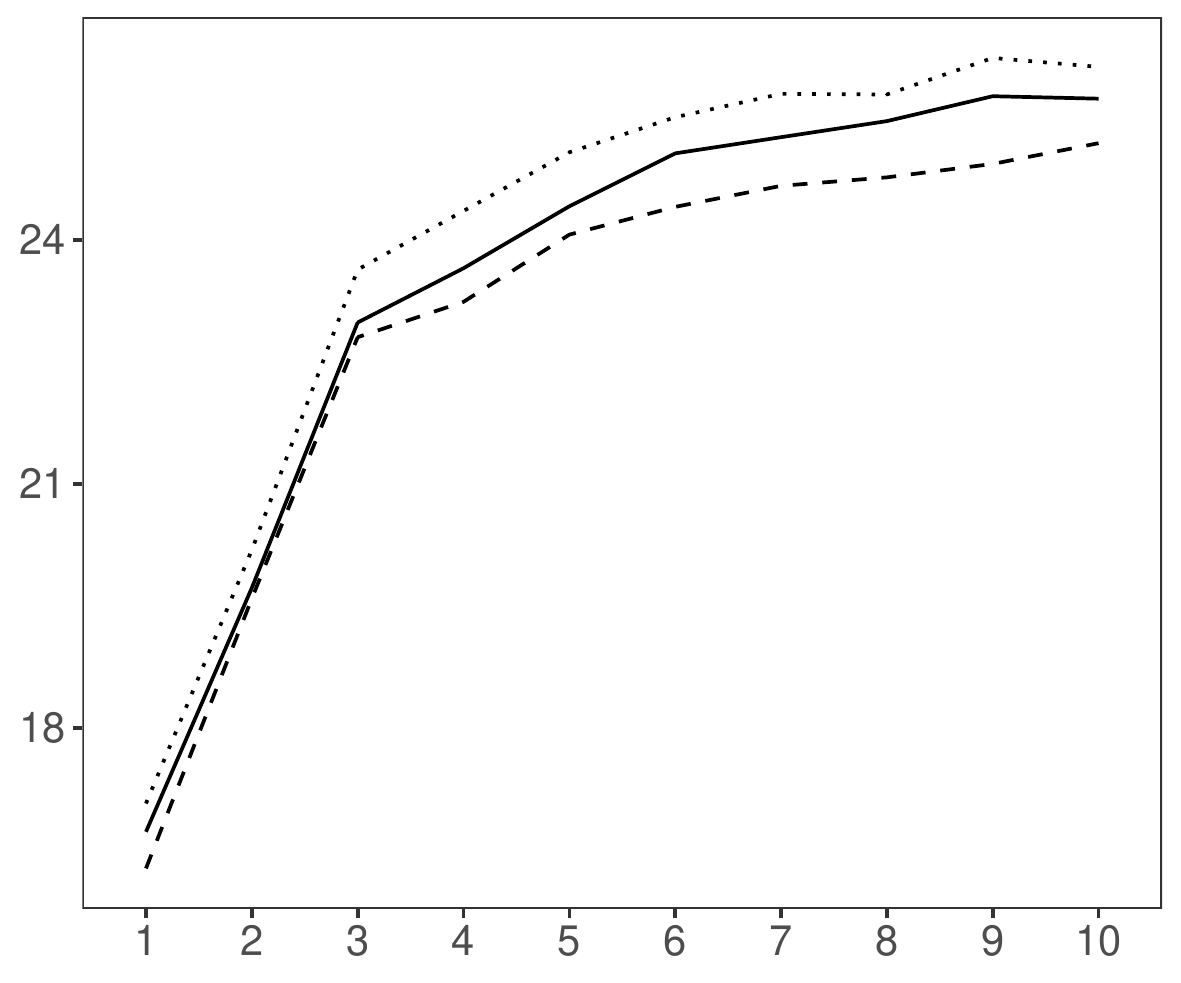}
  \includegraphics[width=5.9cm]{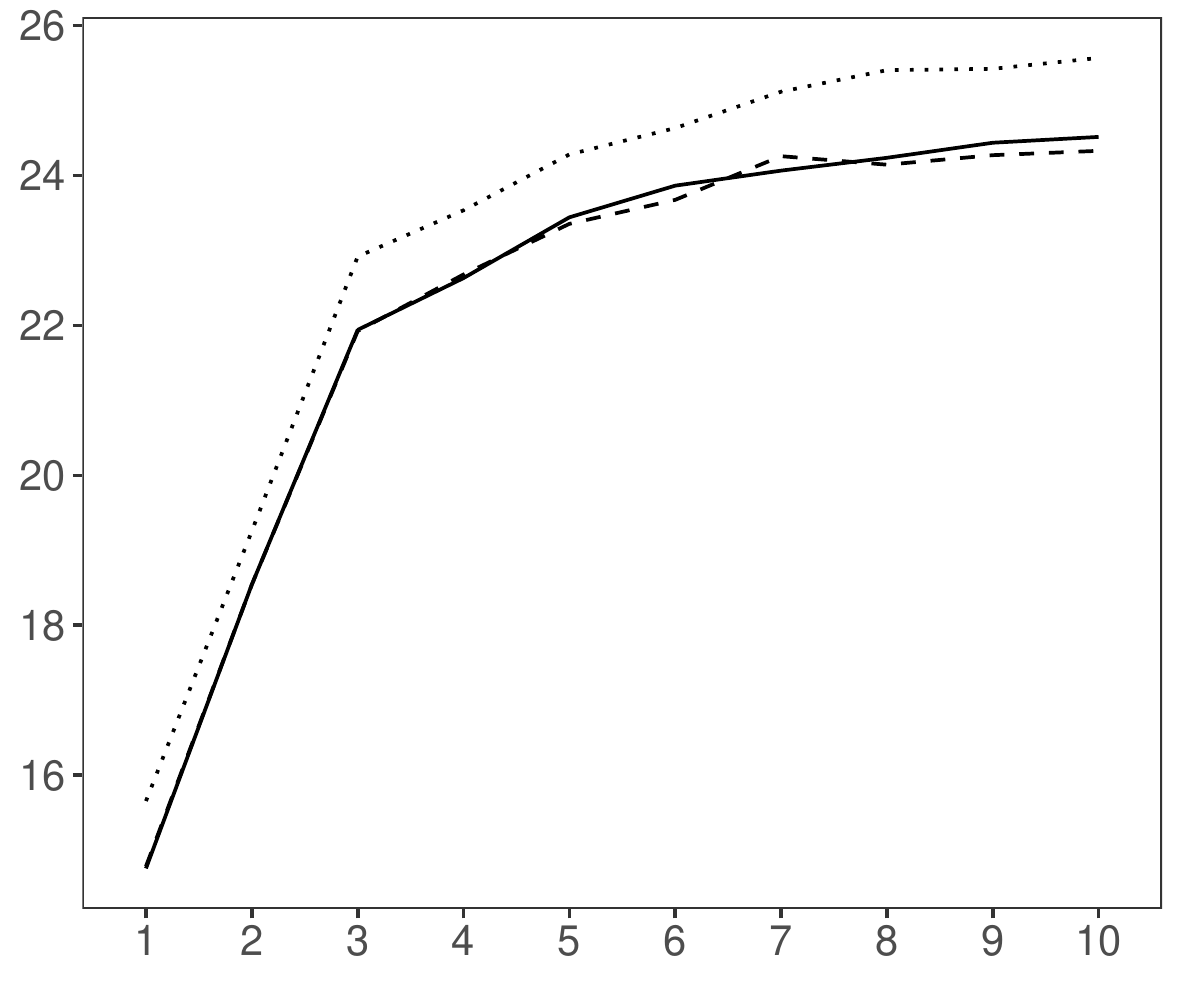}
\\
  \includegraphics[width=5.9cm]{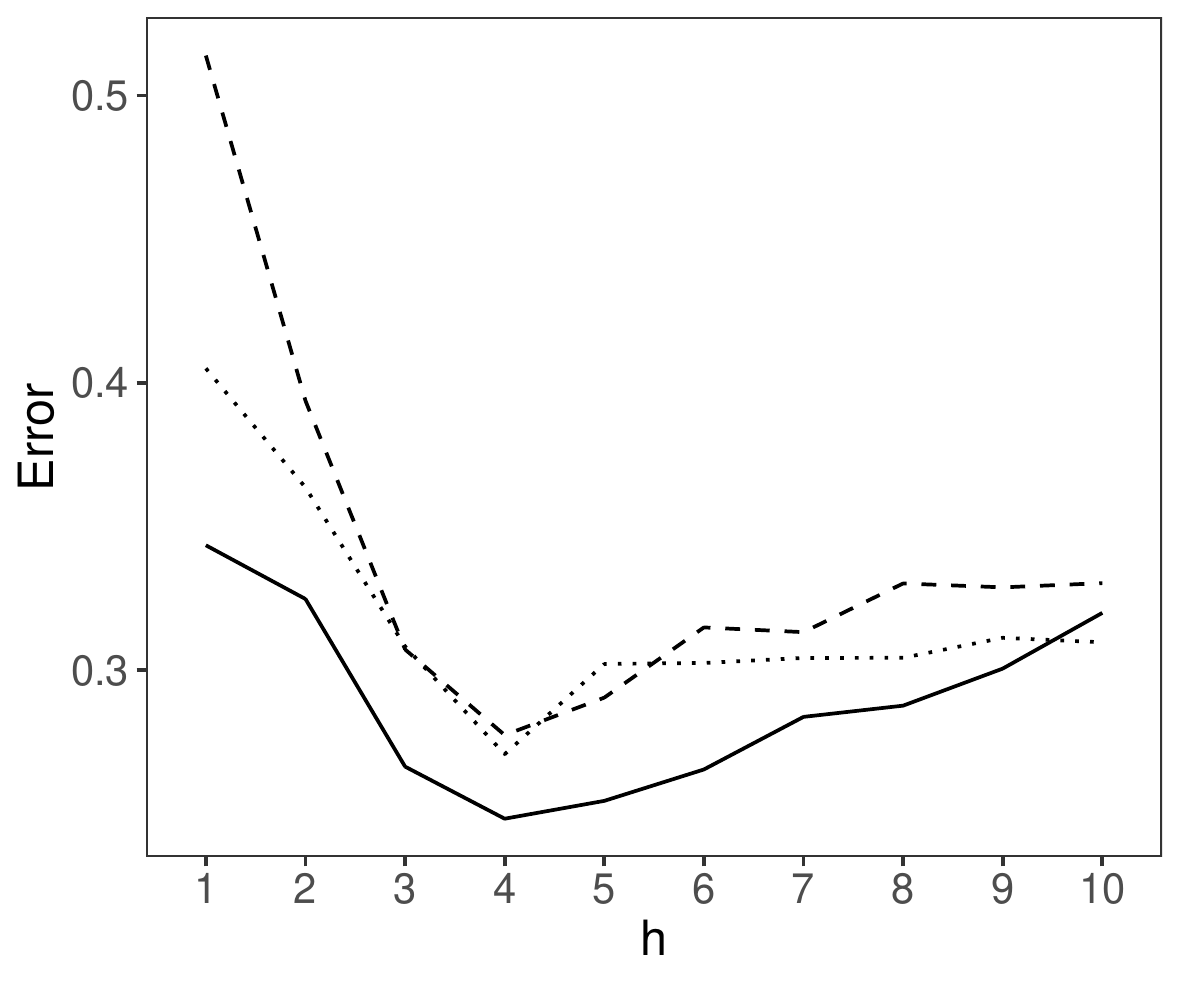}
  \includegraphics[width=5.9cm]{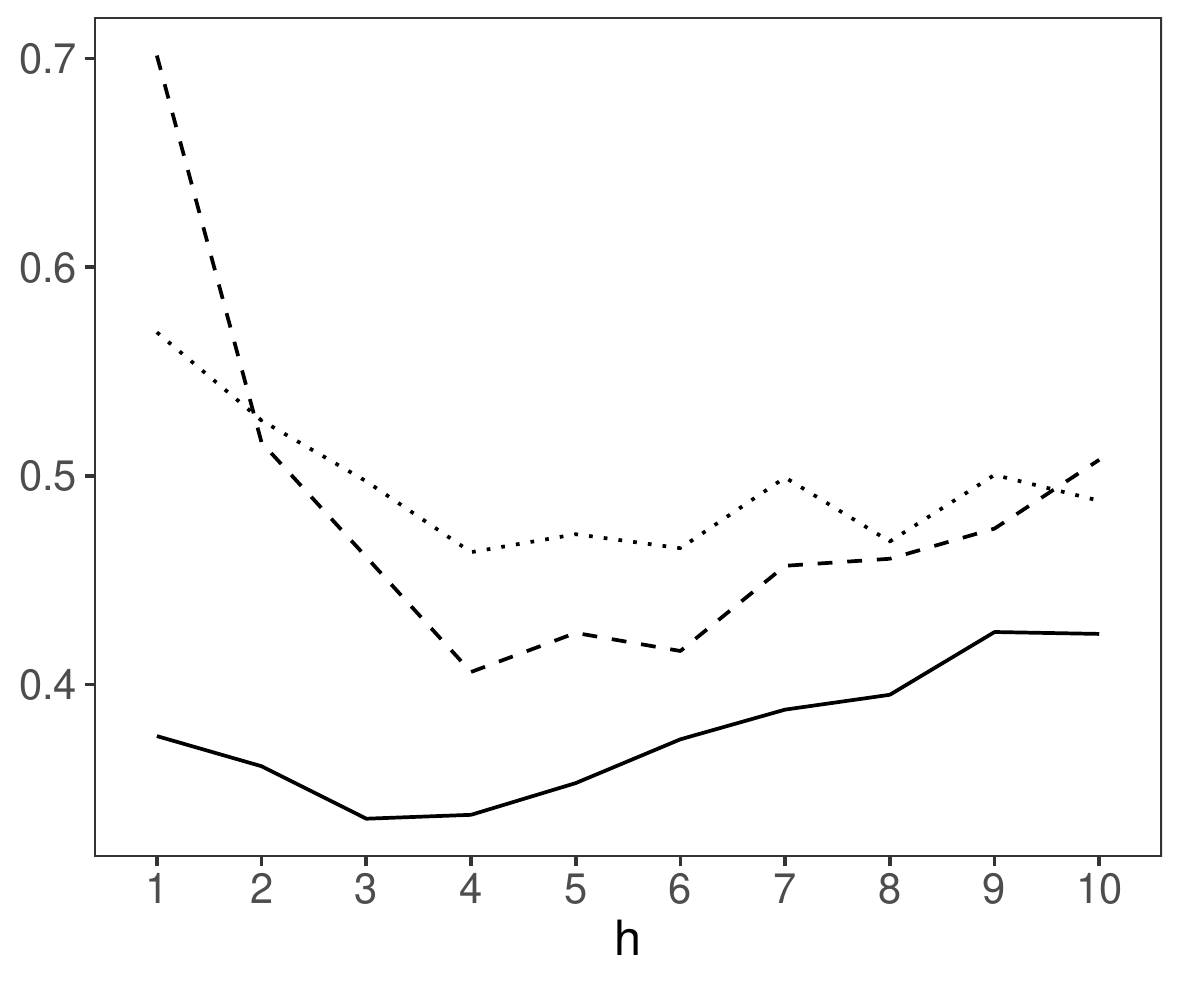}
  \includegraphics[width=5.9cm]{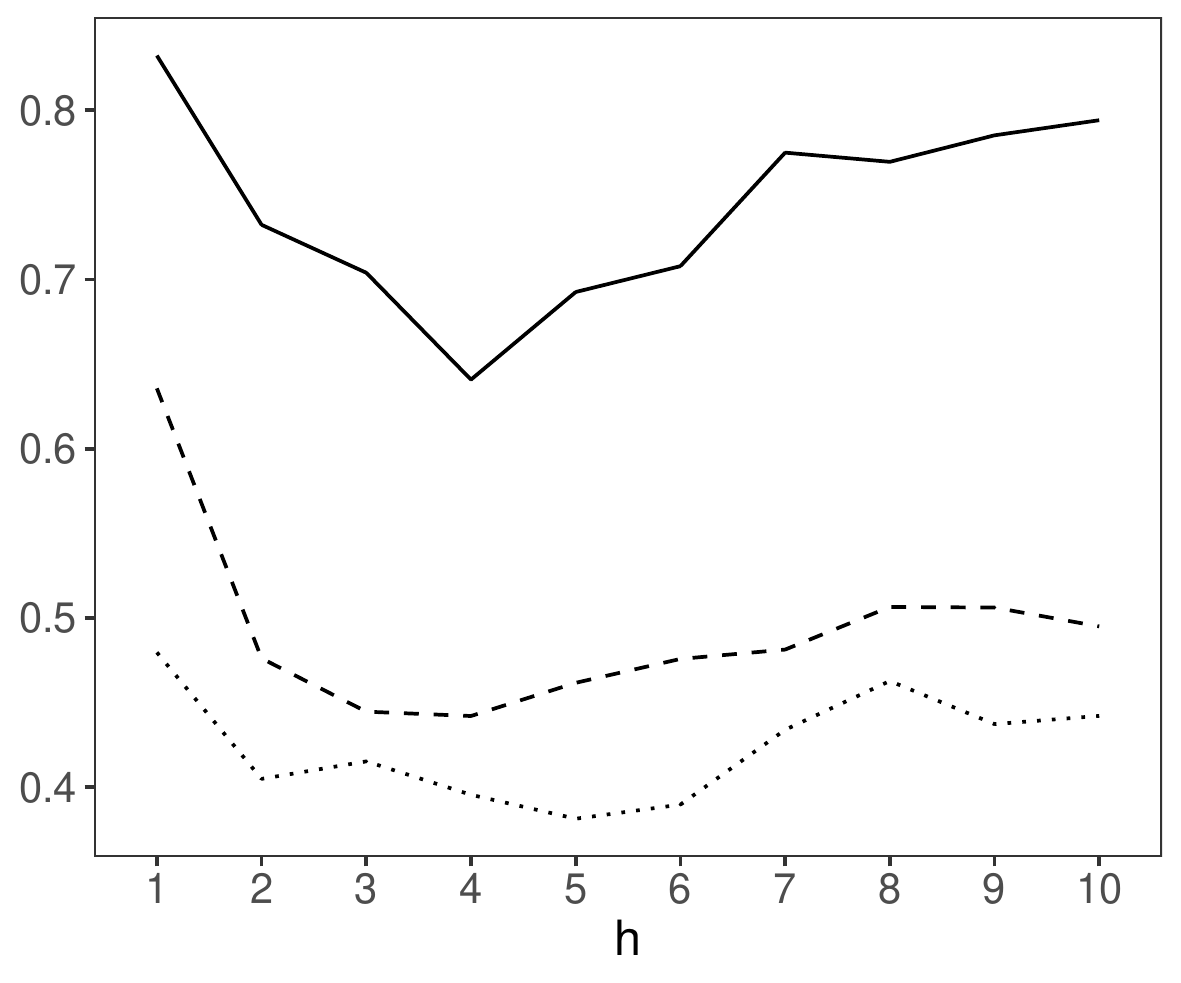}
  \caption{\small{Monte Carlo averages of the performance metrics for the VAR(2) model when no fixed outlying data point is inserted into the dataset: coverage probability (first row), the volume of the Bonferroni cube (second row), and squared error between the empirical and bootstrap Bonferroni cubes (third row). Methods: OLS (dotted line), weighted likelihood (solid line), and Rob-VAR (dashed line).}}
  \label{fig:5}
\end{figure}

\begin{figure}[!htbp]
  \centering
  \includegraphics[width=5.9cm]{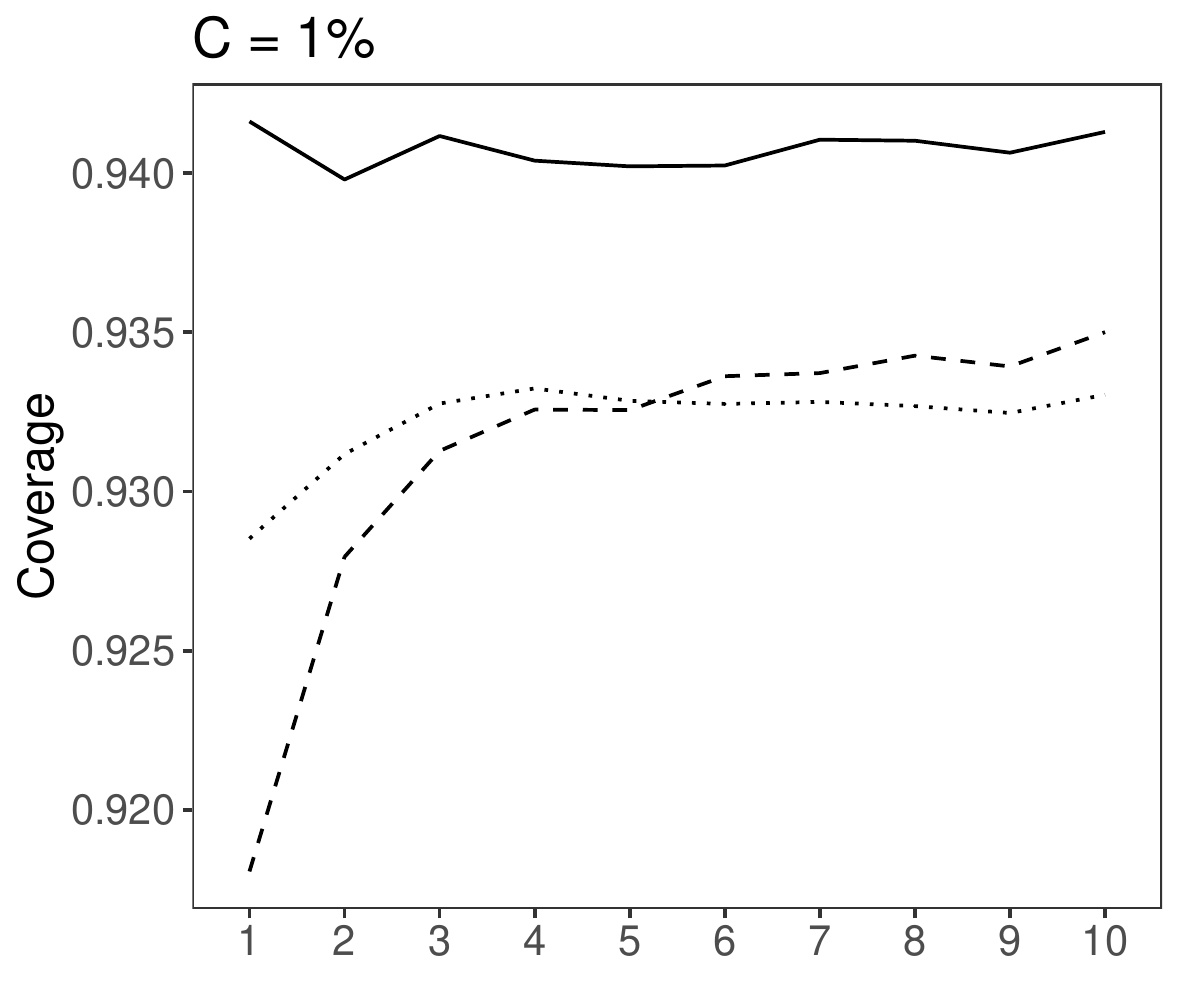}
  \includegraphics[width=5.9cm]{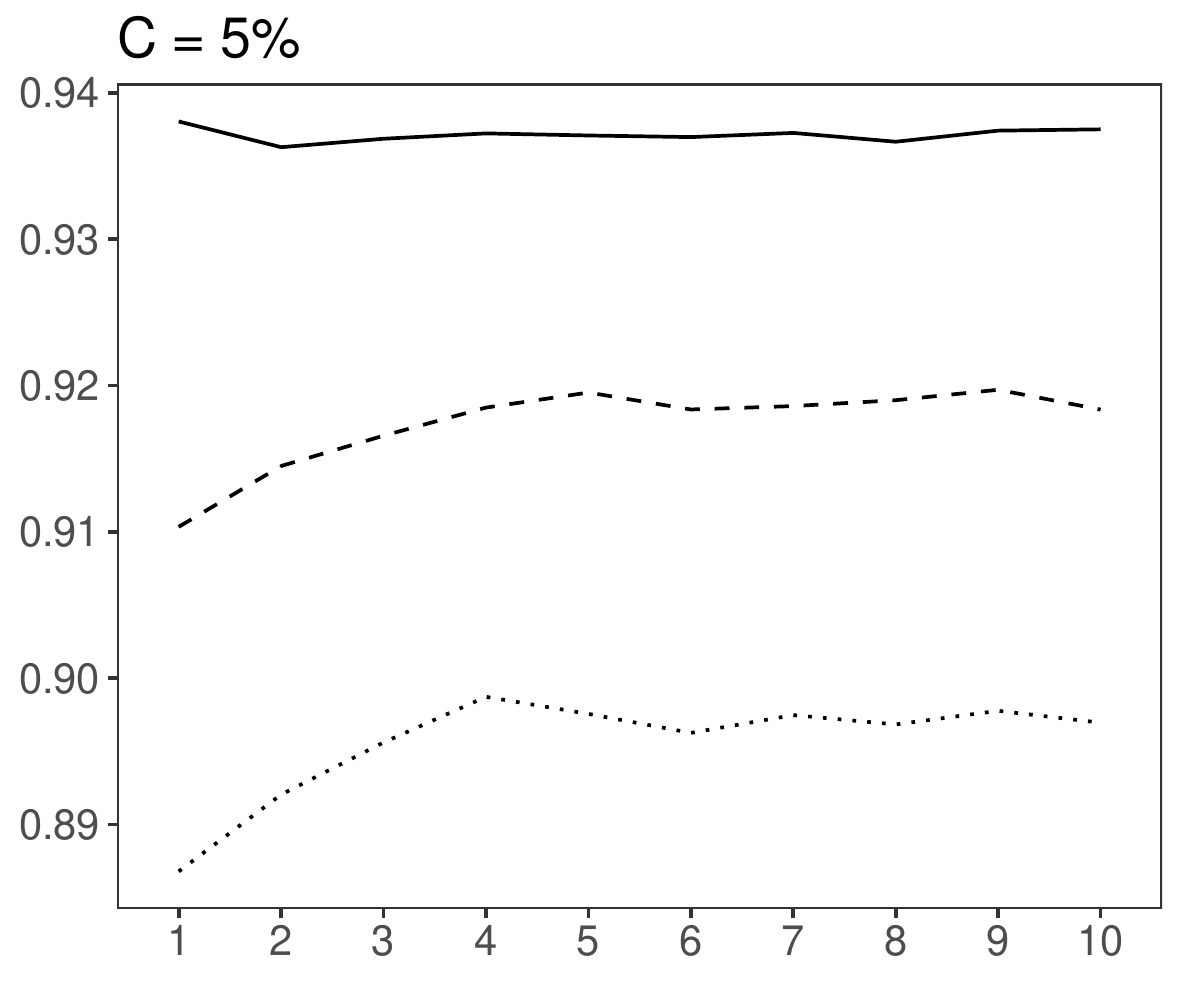}
  \includegraphics[width=5.9cm]{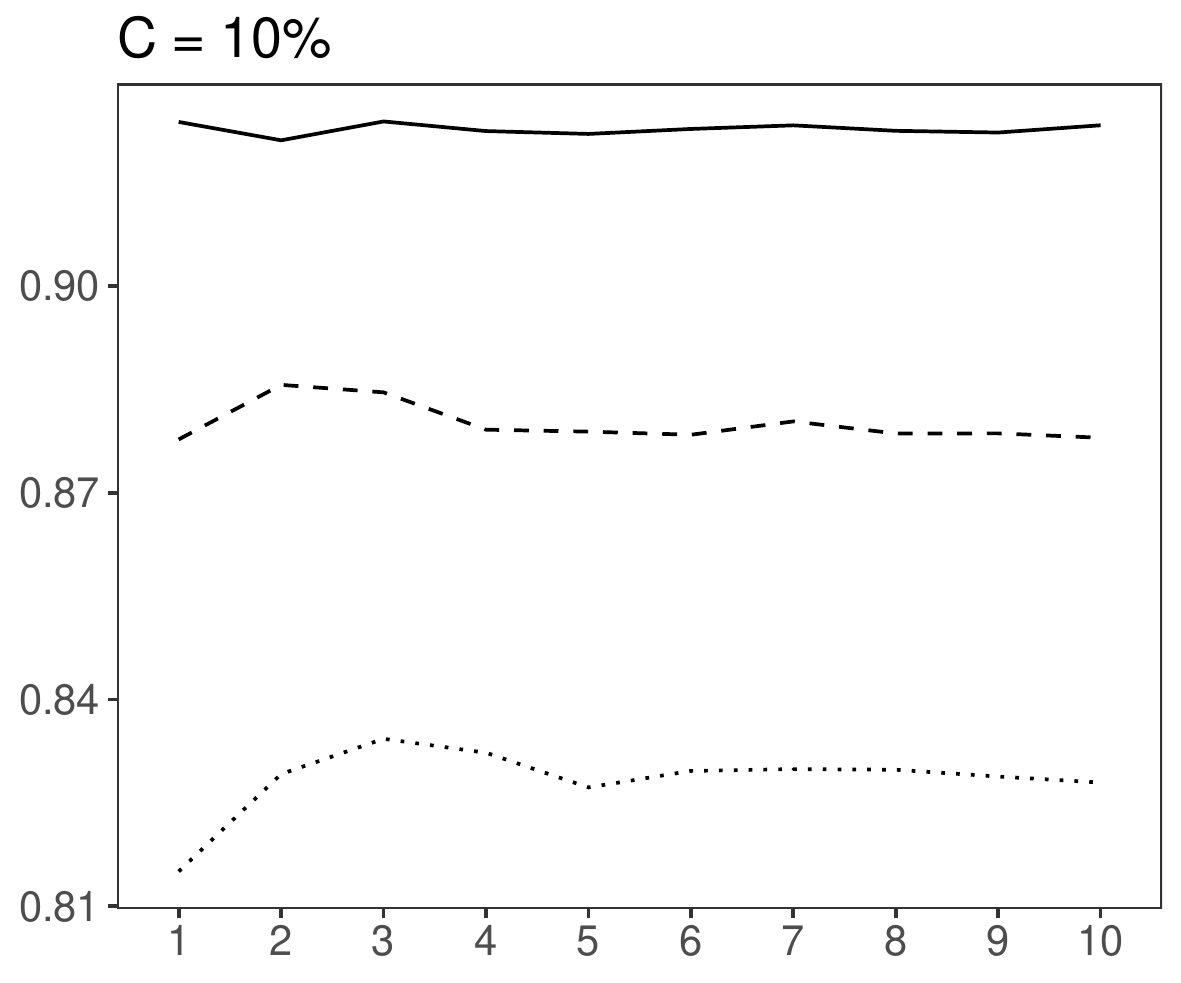}
\\
  \includegraphics[width=5.9cm]{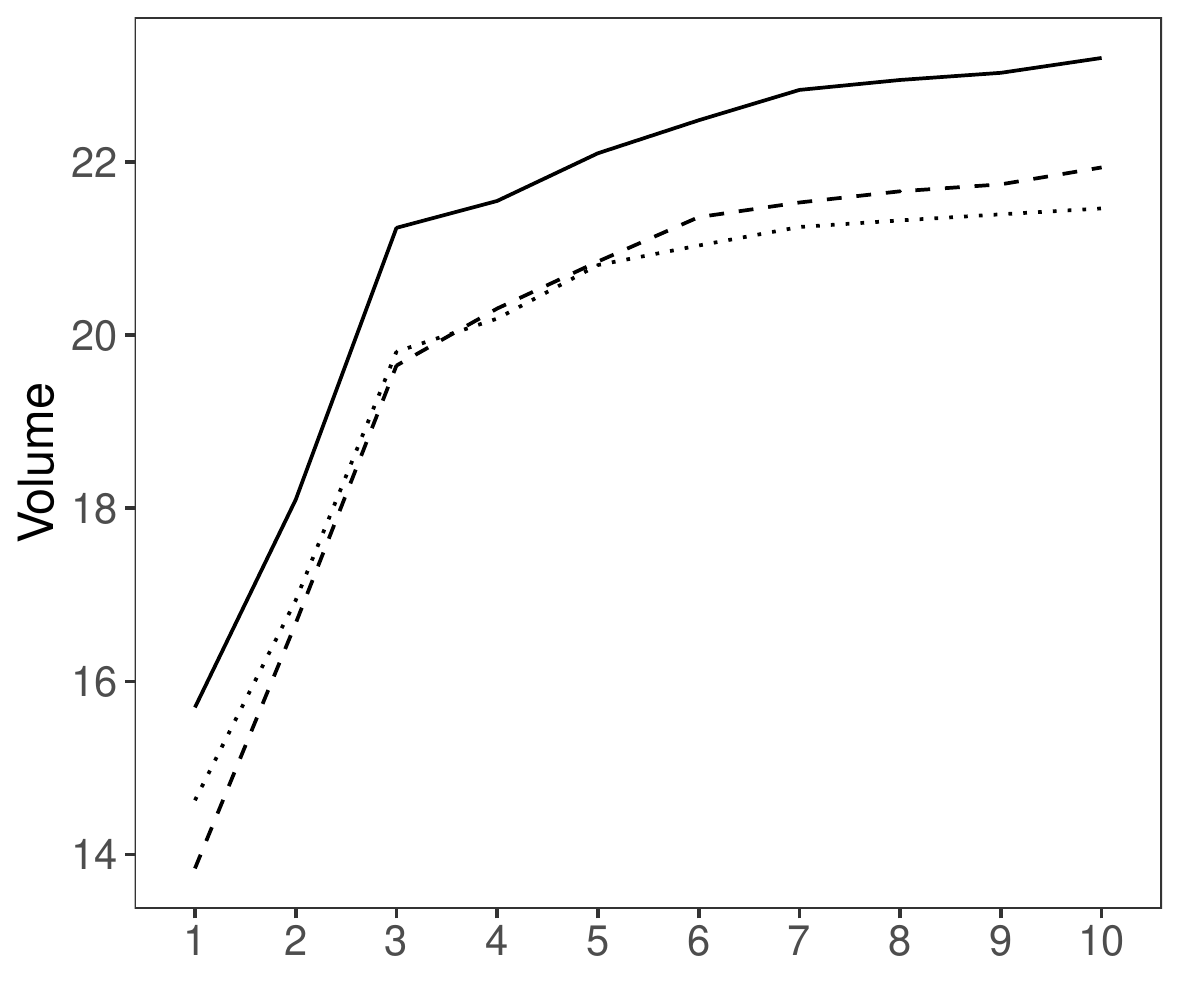}
  \includegraphics[width=5.9cm]{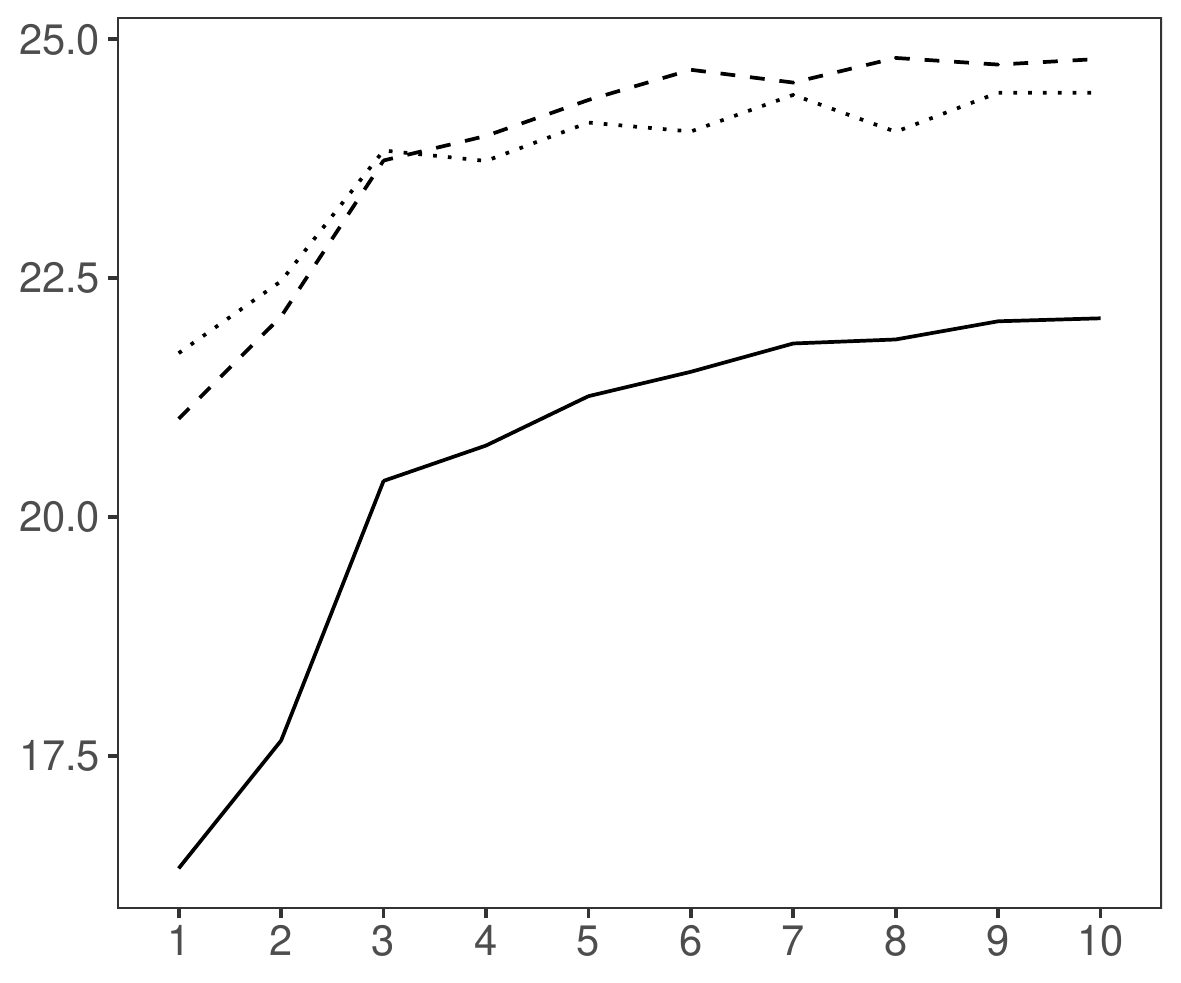}
  \includegraphics[width=5.9cm]{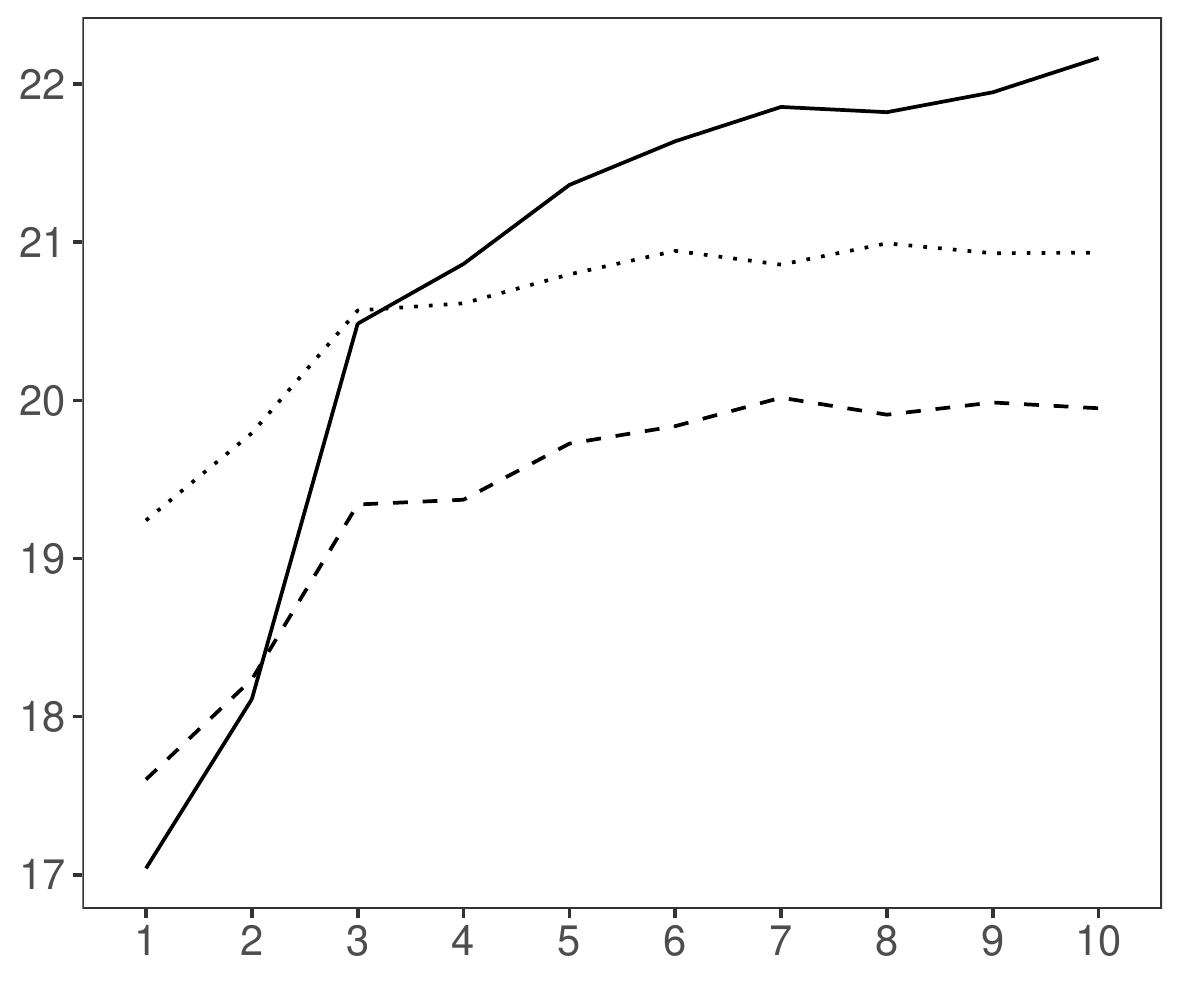}
\\
  \includegraphics[width=5.9cm]{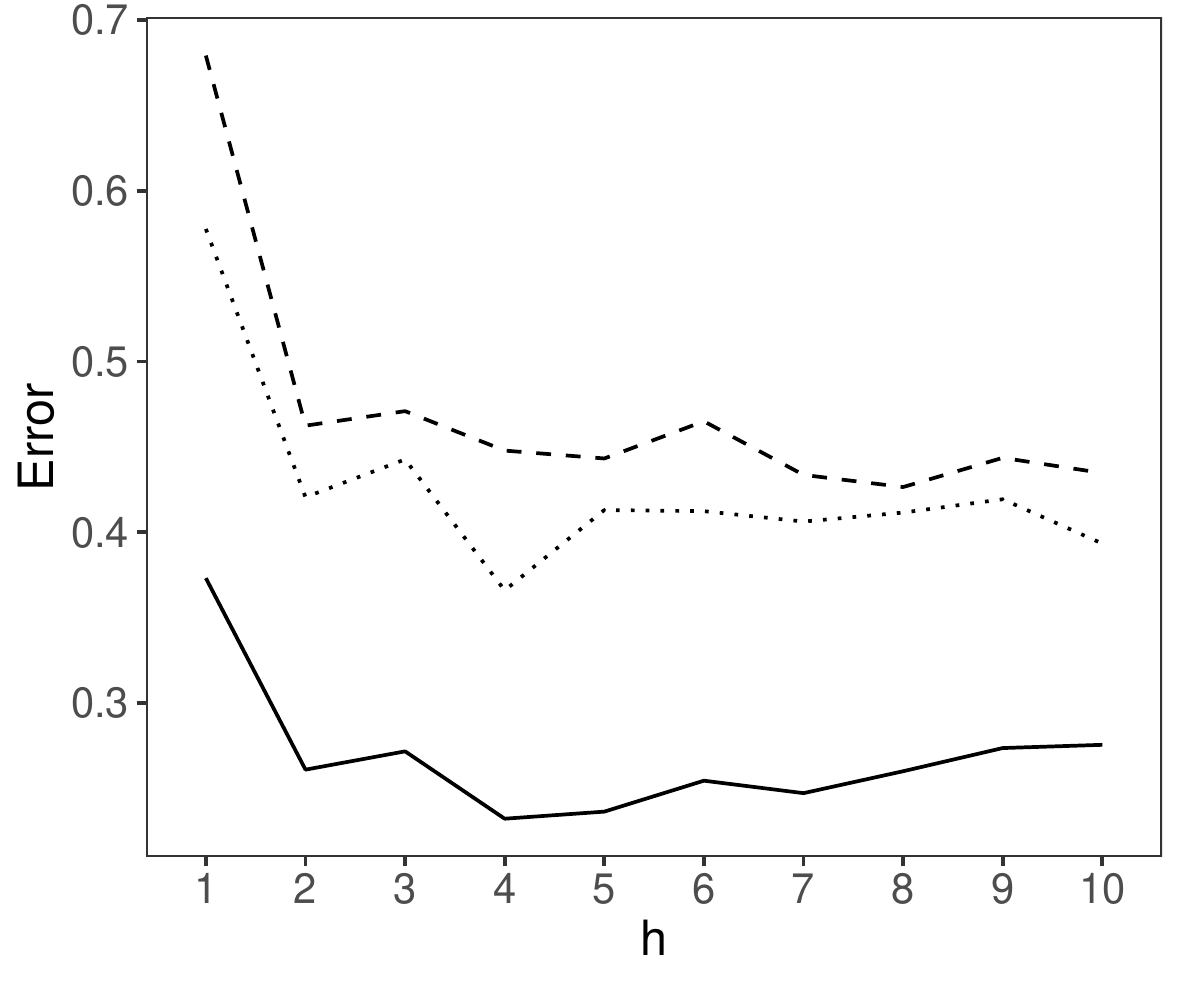}
  \includegraphics[width=5.9cm]{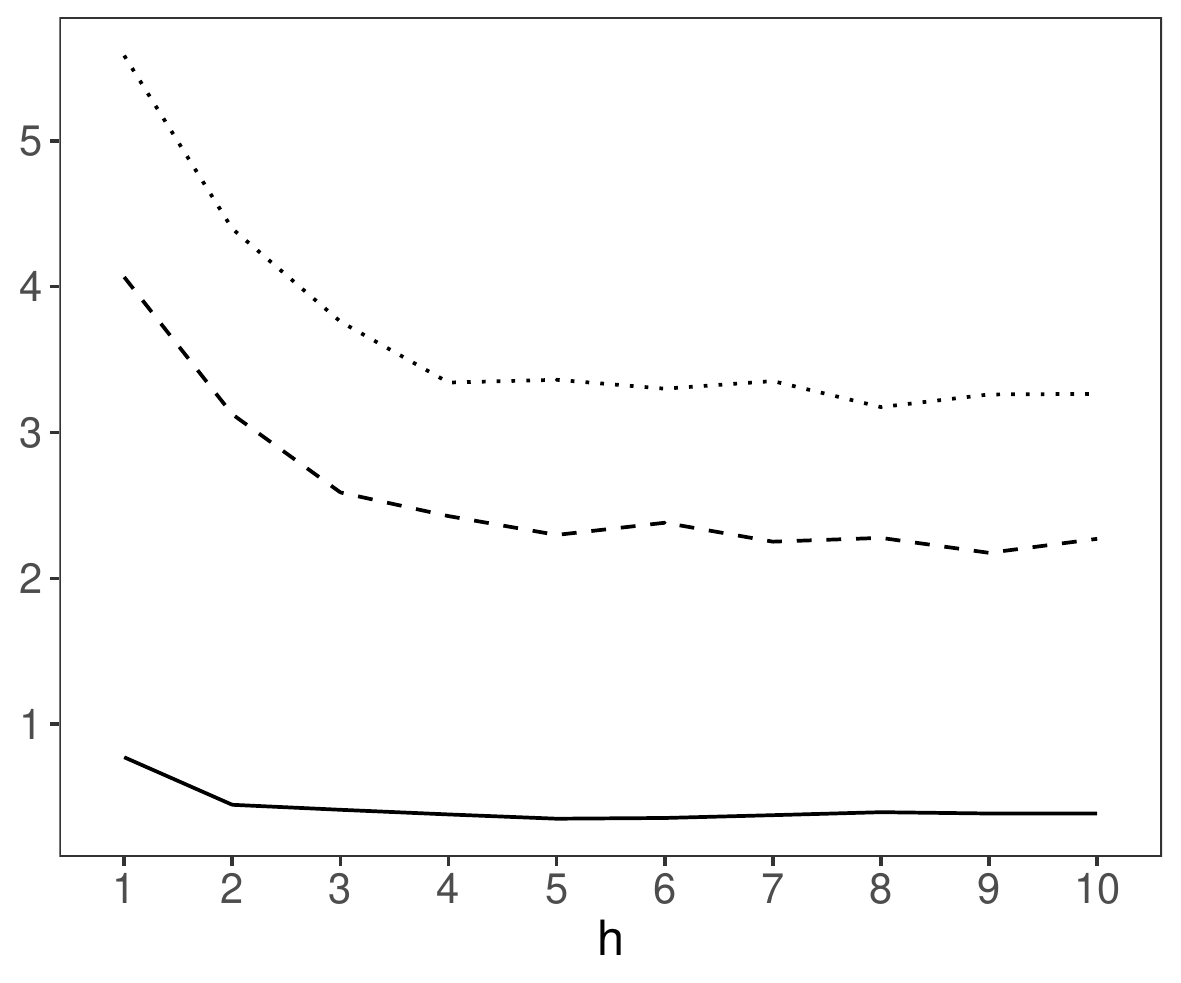}
  \includegraphics[width=5.9cm]{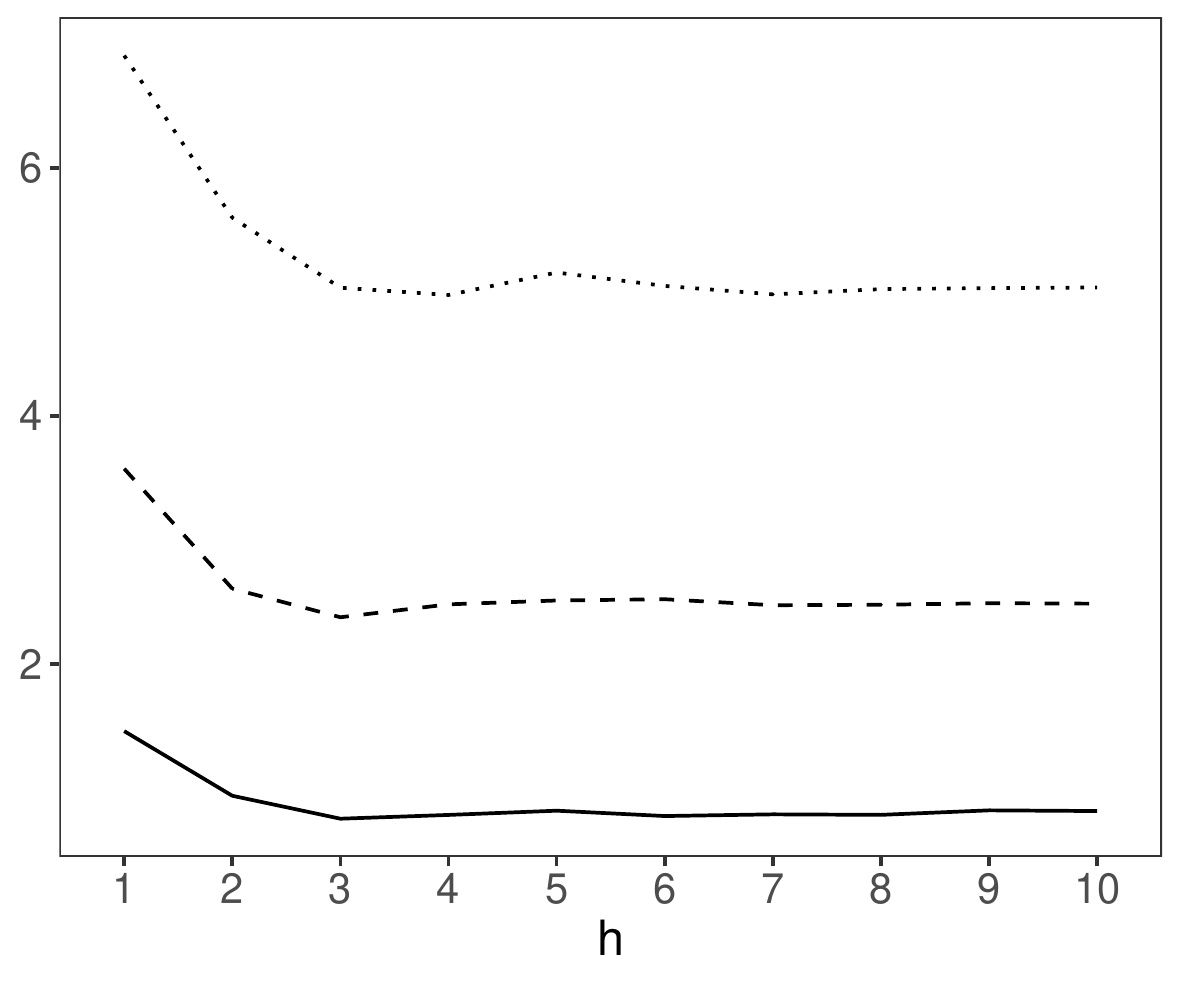}
  \caption{\small{Monte Carlo averages of the performance metrics for the VAR(2) model when $C = \left[ 1\%~\text{(first column)}, 5\%~\text{(second column)}, 10\%~\text{(third column)} \right]$ of the generated data are contaminated by deliberately inserted AOs and $\epsilon_t \sim N(0, 1)$: coverage probability (first row), volume of the Bonferroni cube (second row), and squared error between the empirical and bootstrap Bonferroni cubes (third row). Methods: OLS (dotted line), weighted likelihood (solid line), and Rob-VAR (dashed line).}}
  \label{fig:6}
\end{figure}

\begin{figure}[!htbp]
  \centering
  \includegraphics[width=5.9cm]{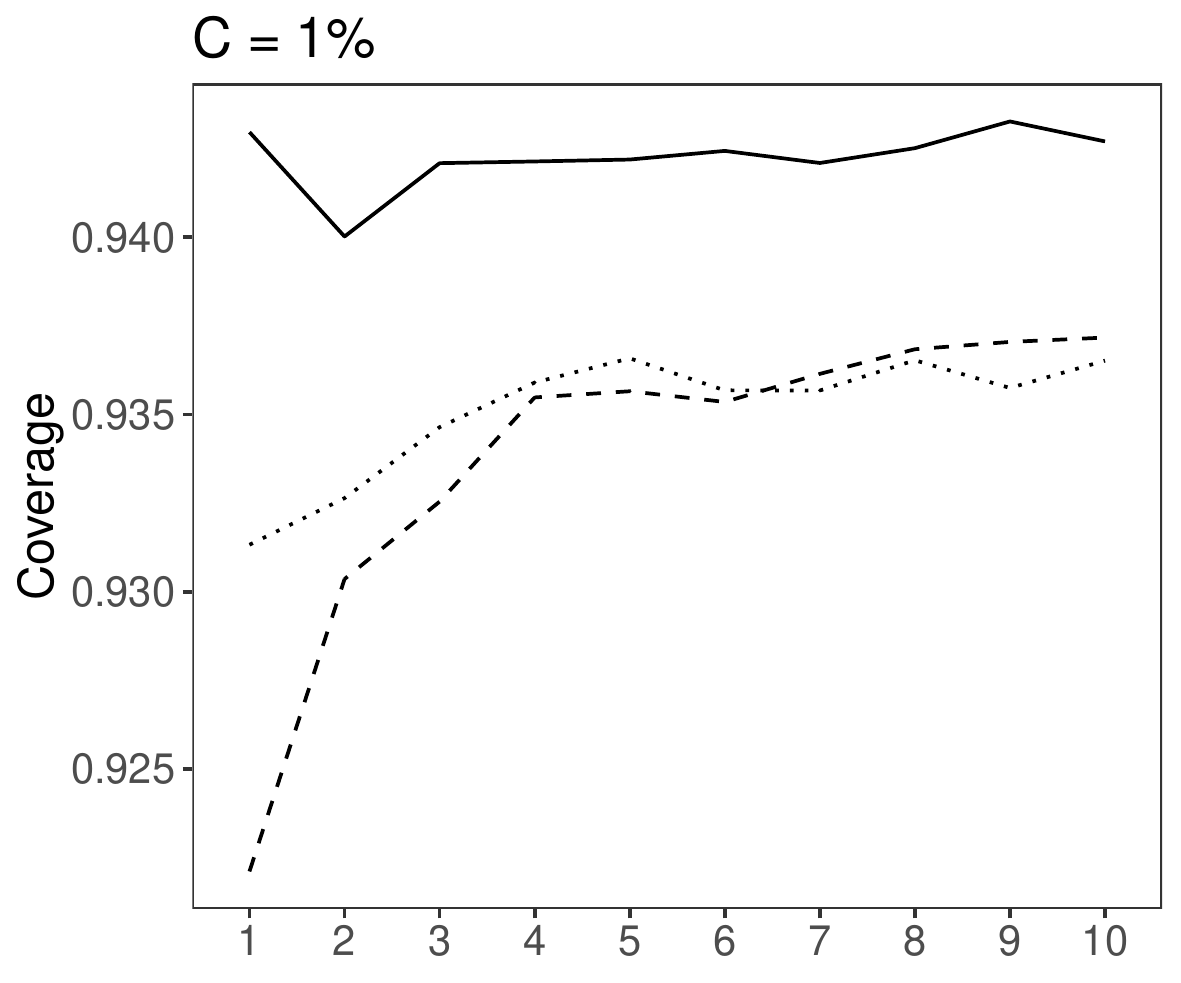}
  \includegraphics[width=5.9cm]{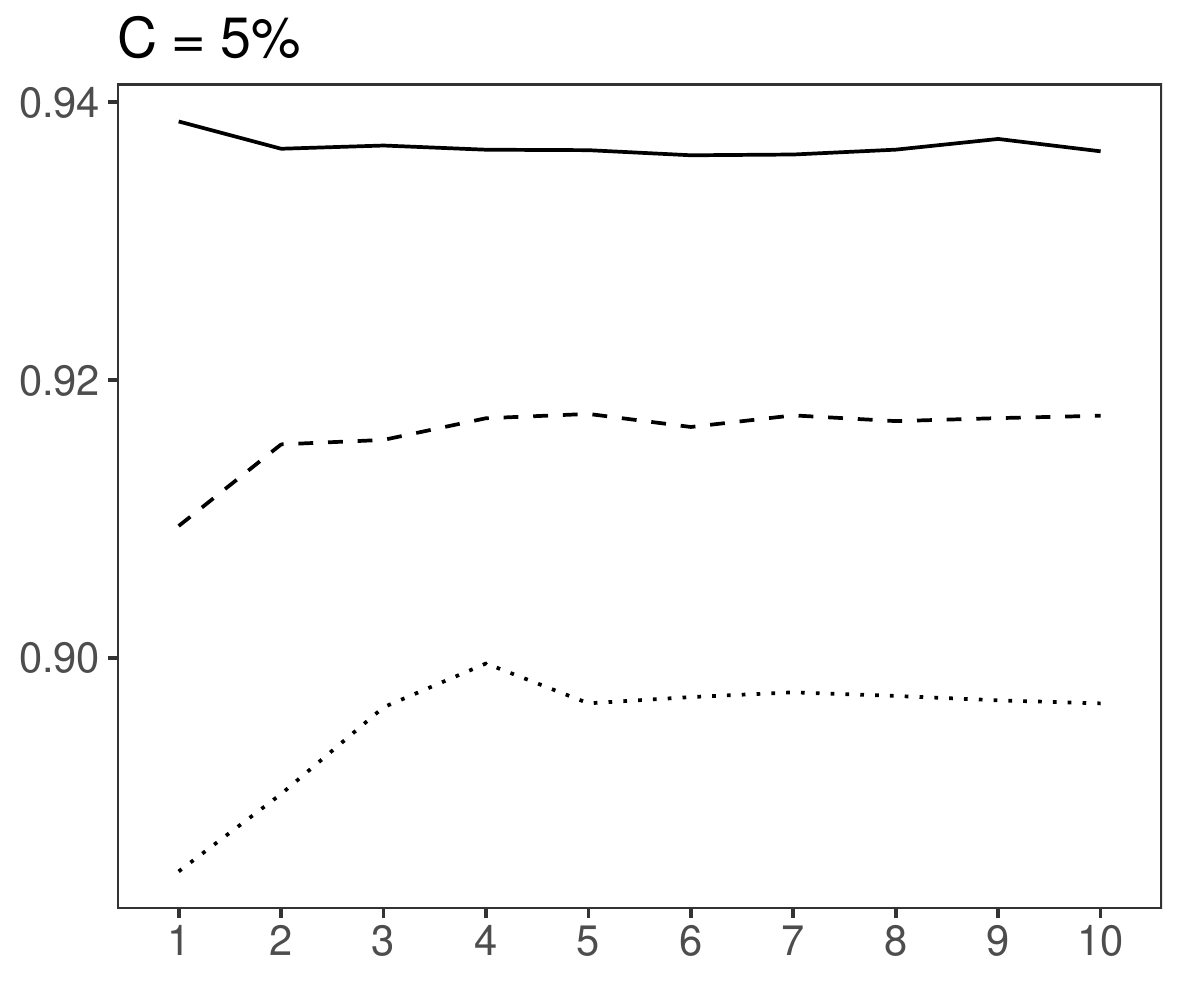}
  \includegraphics[width=5.9cm]{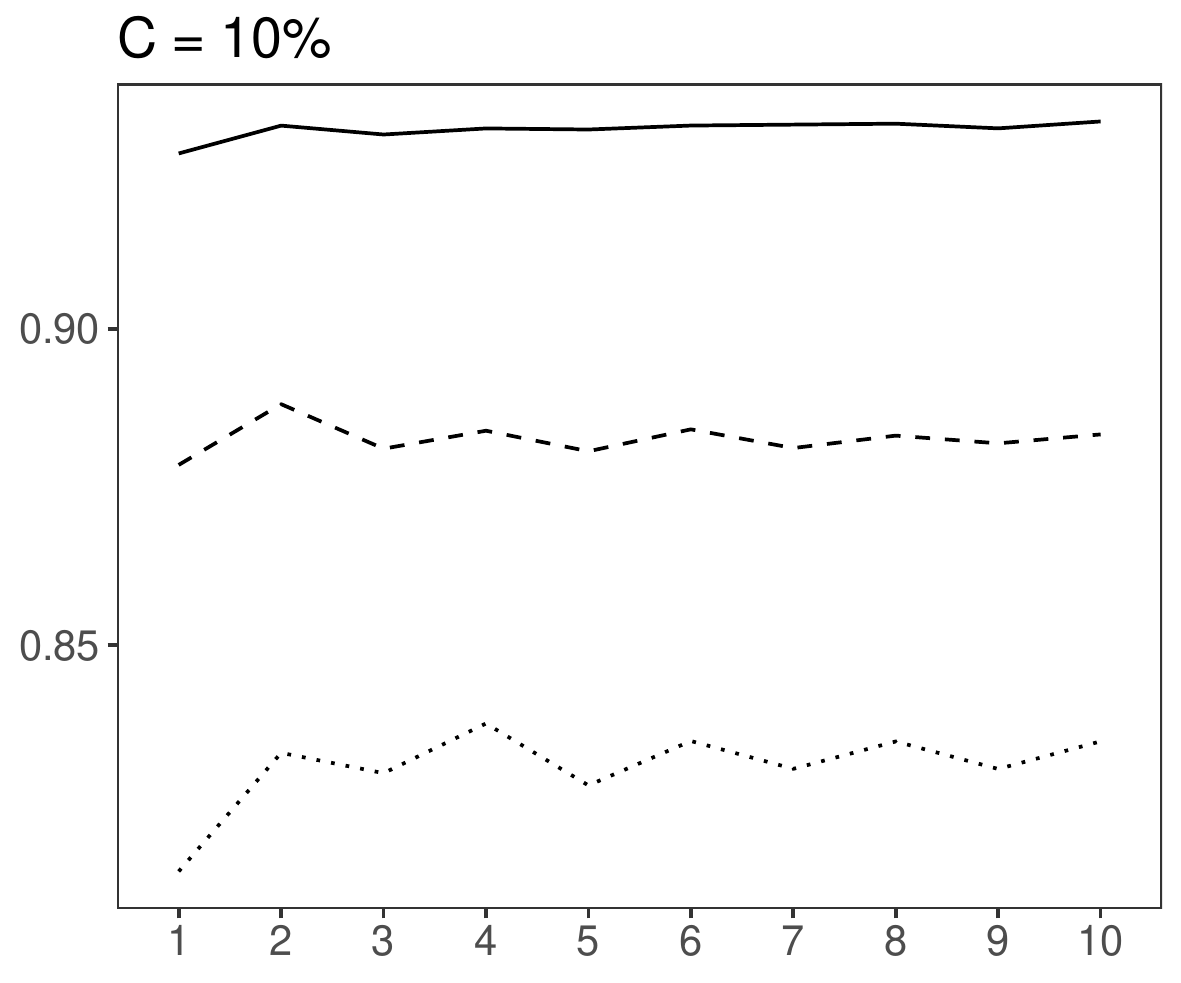}
\\
  \includegraphics[width=5.9cm]{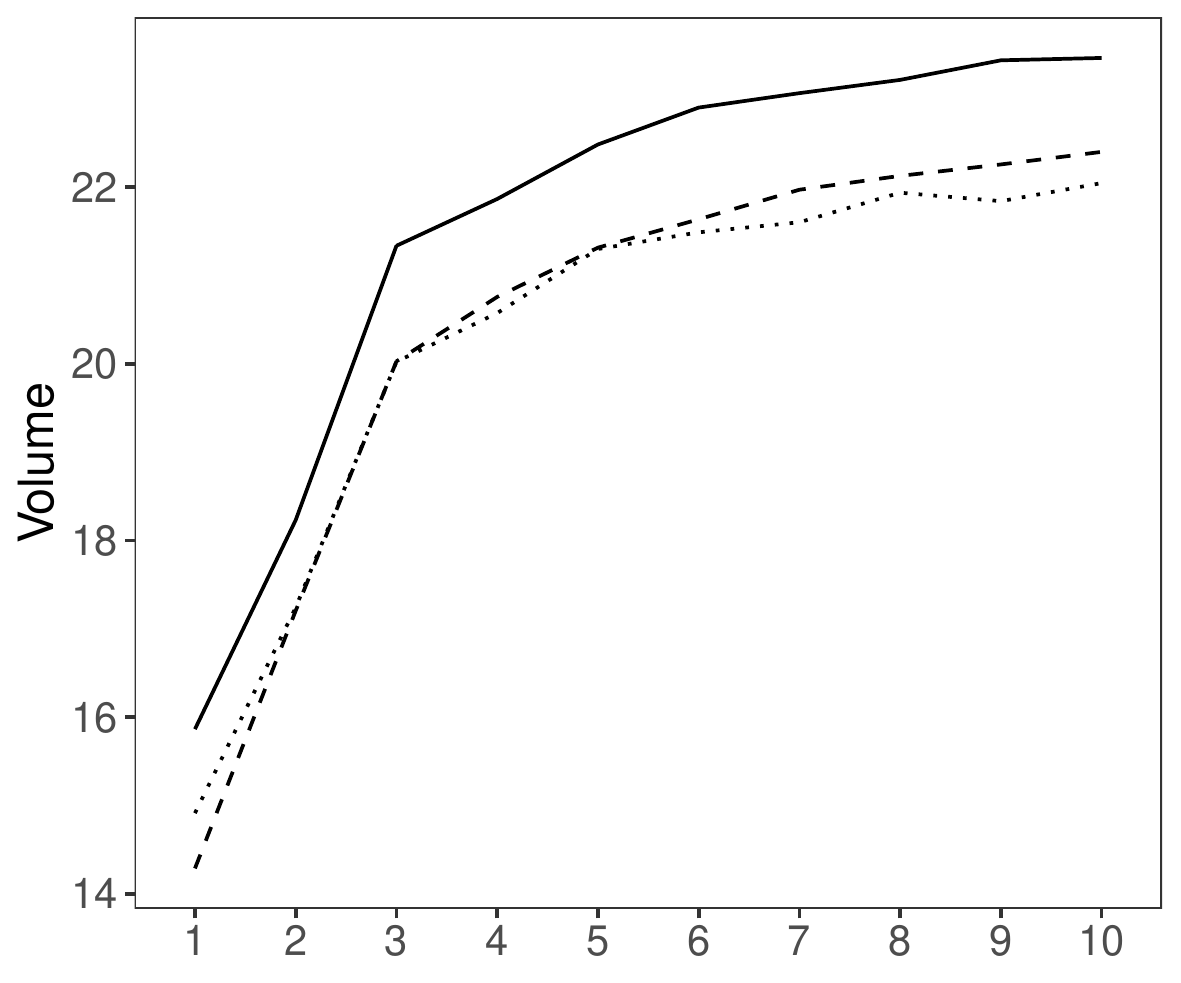}
  \includegraphics[width=5.9cm]{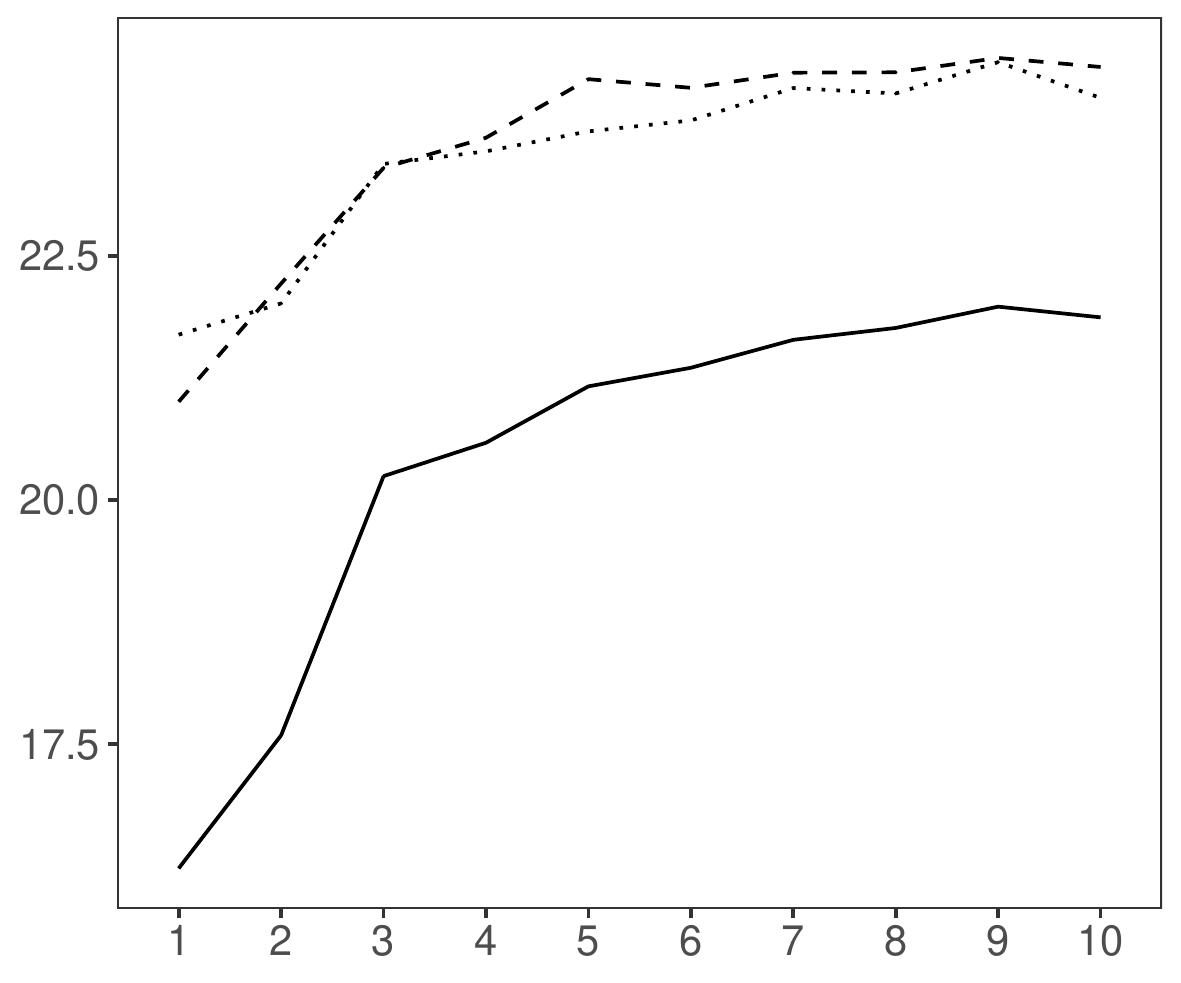}
  \includegraphics[width=5.9cm]{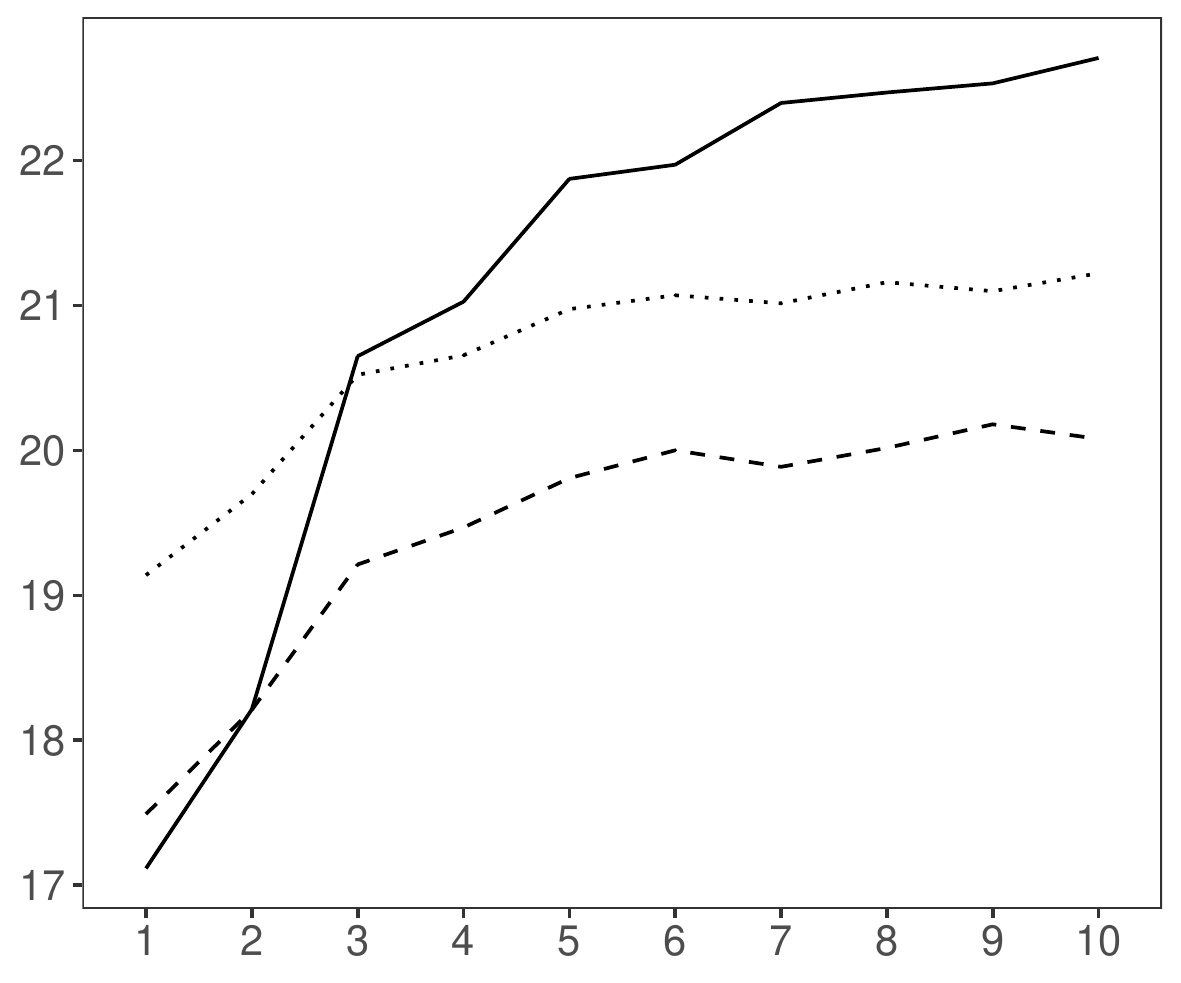}
\\
  \includegraphics[width=5.9cm]{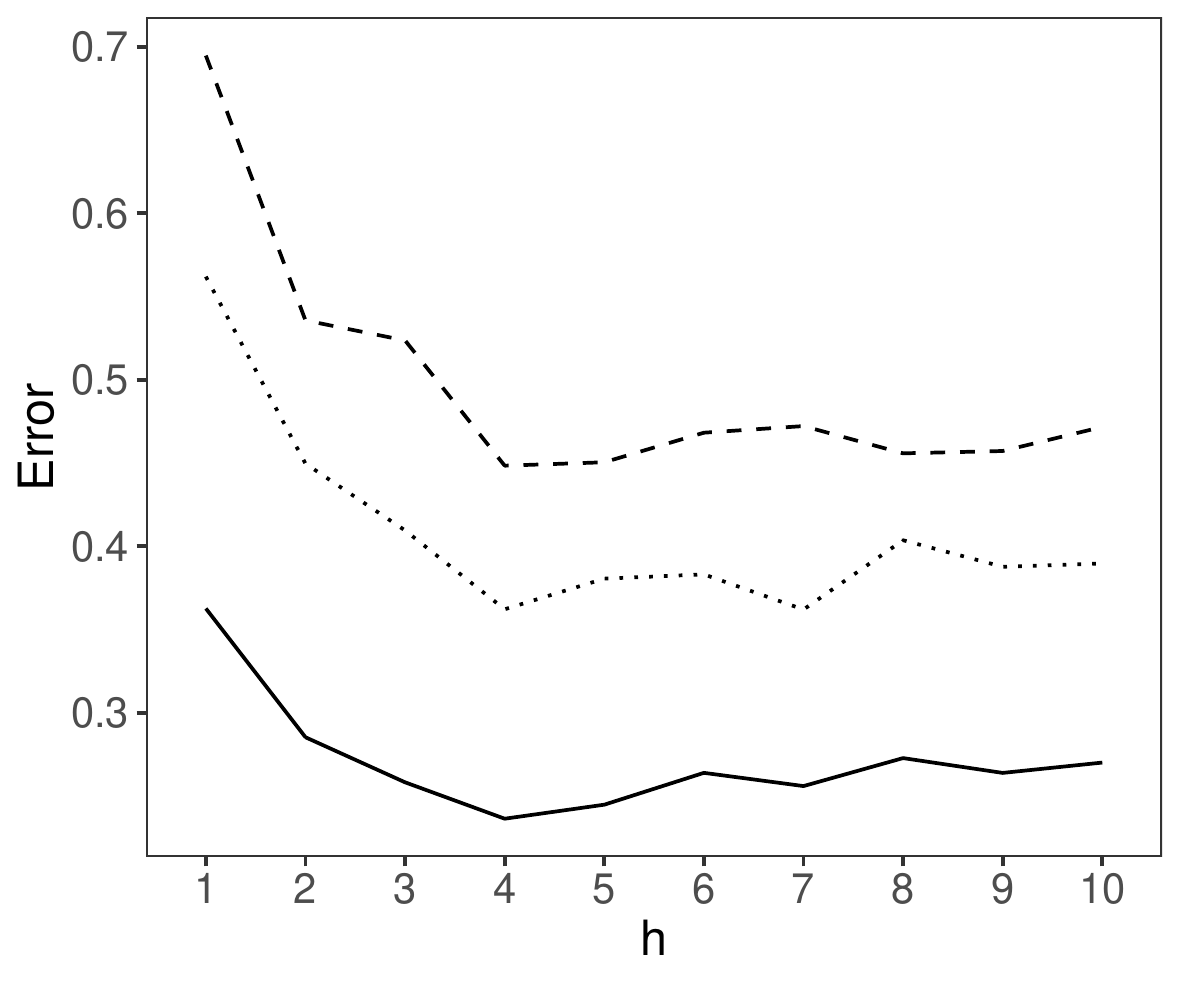}
  \includegraphics[width=5.9cm]{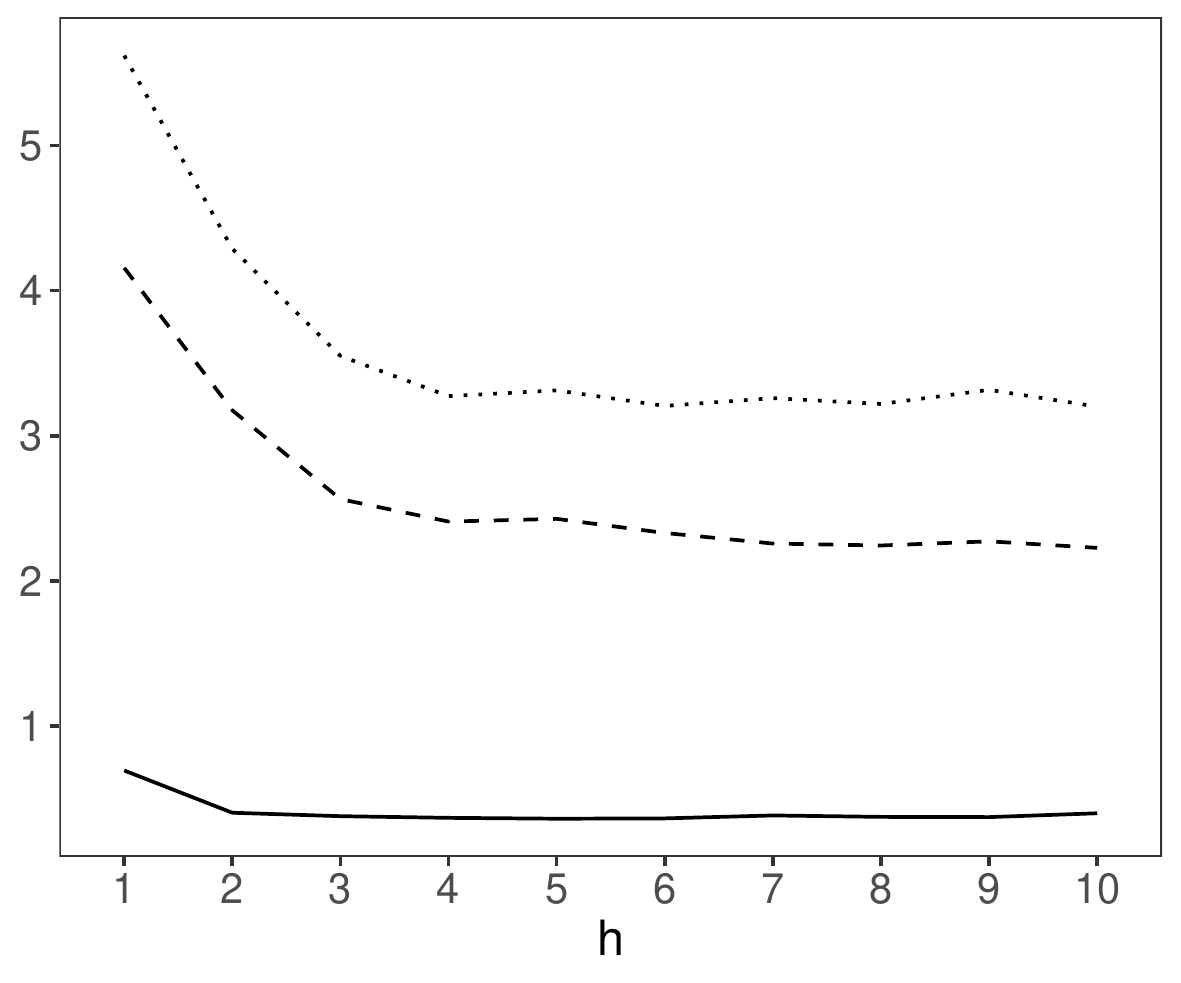}
  \includegraphics[width=5.9cm]{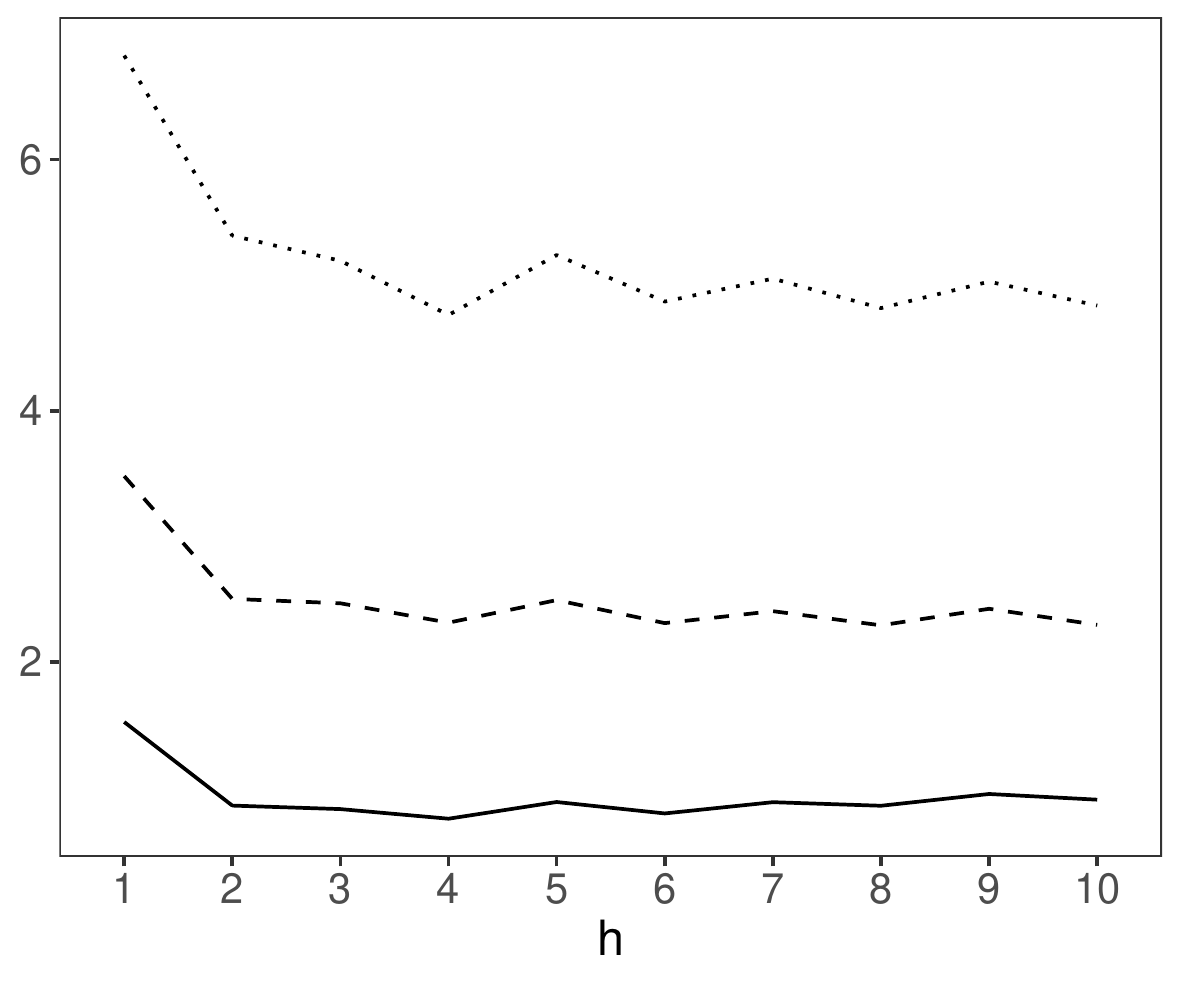}
  \caption{\small{Monte Carlo averages of the performance metrics for the VAR(2) model when $C = \left[ 1\%~\text{(first column)}, 5\%~\text{(second column)}, 10\%~\text{(third column)} \right]$ of the generated data are contaminated by deliberately inserted IOs and $\epsilon_t \sim N(0, 1)$: coverage probability (first row), volume of the Bonferroni cube (second row), and squared error between the empirical and bootstrap Bonferroni cubes (third row). Methods: OLS (dotted line), weighted likelihood (solid line), and Rob-VAR (dashed line).}}
  \label{fig:7}
\end{figure}

\begin{figure}[!htbp]
  \centering
  \includegraphics[width=5.9cm]{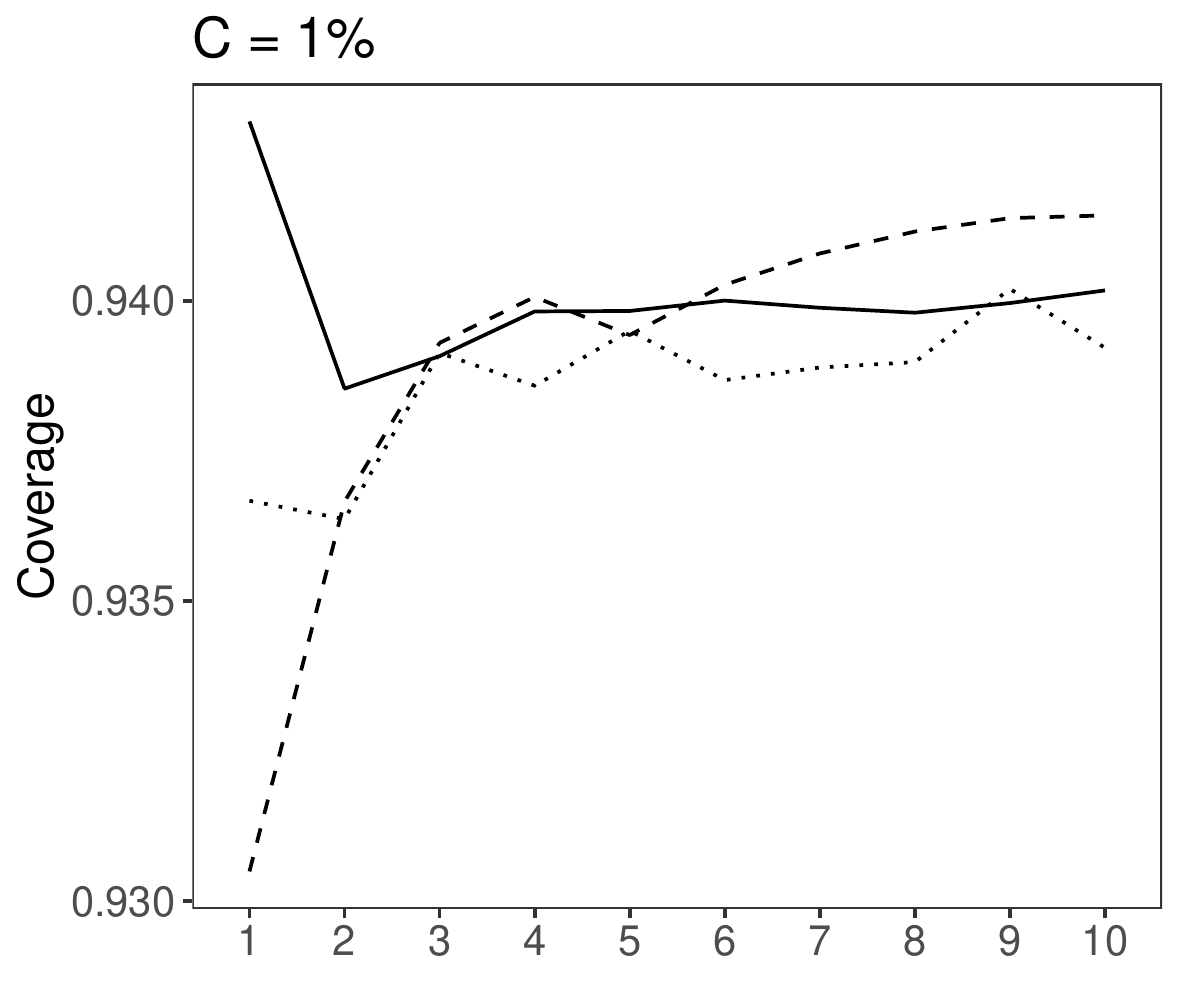}
  \includegraphics[width=5.9cm]{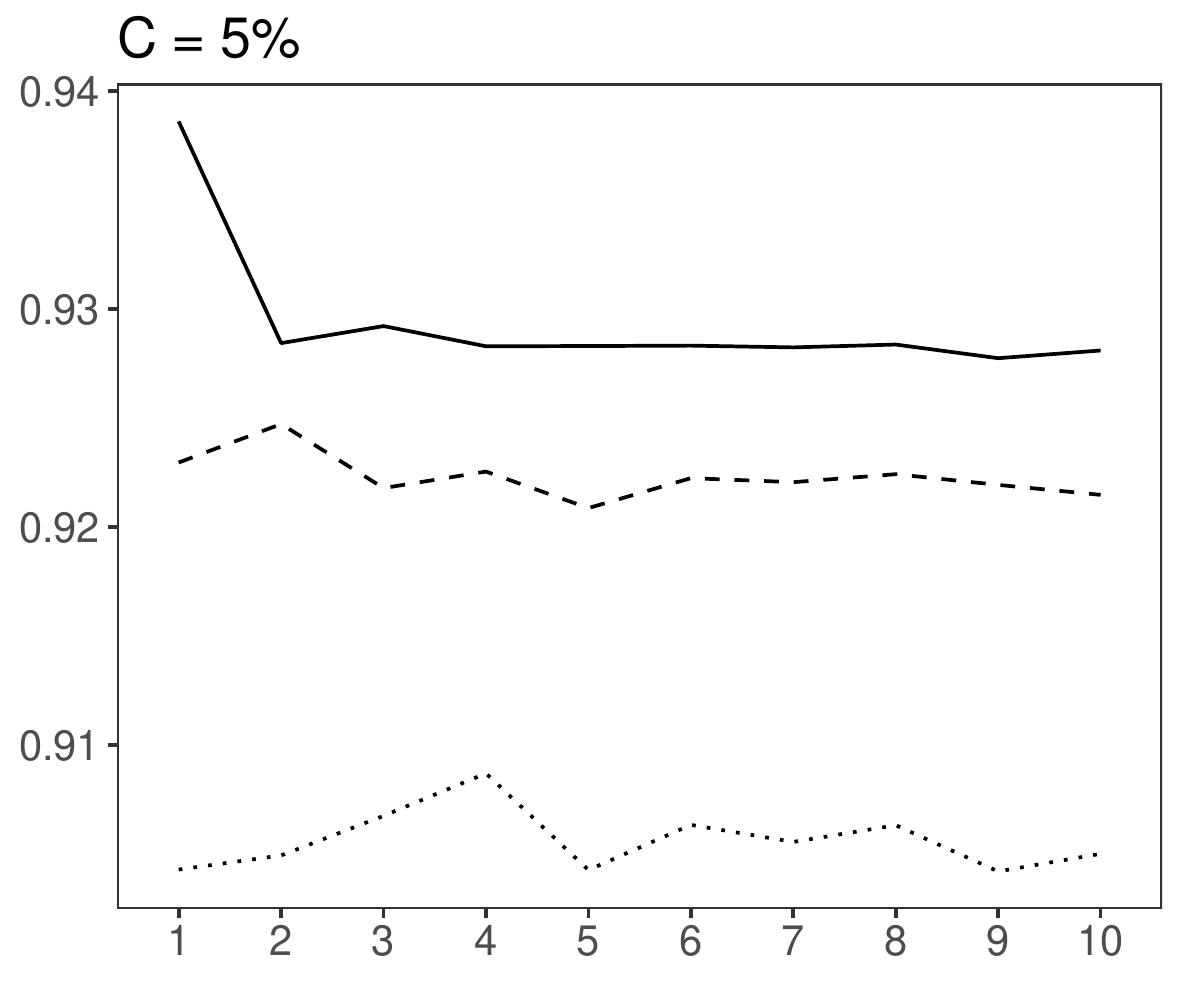}
  \includegraphics[width=5.9cm]{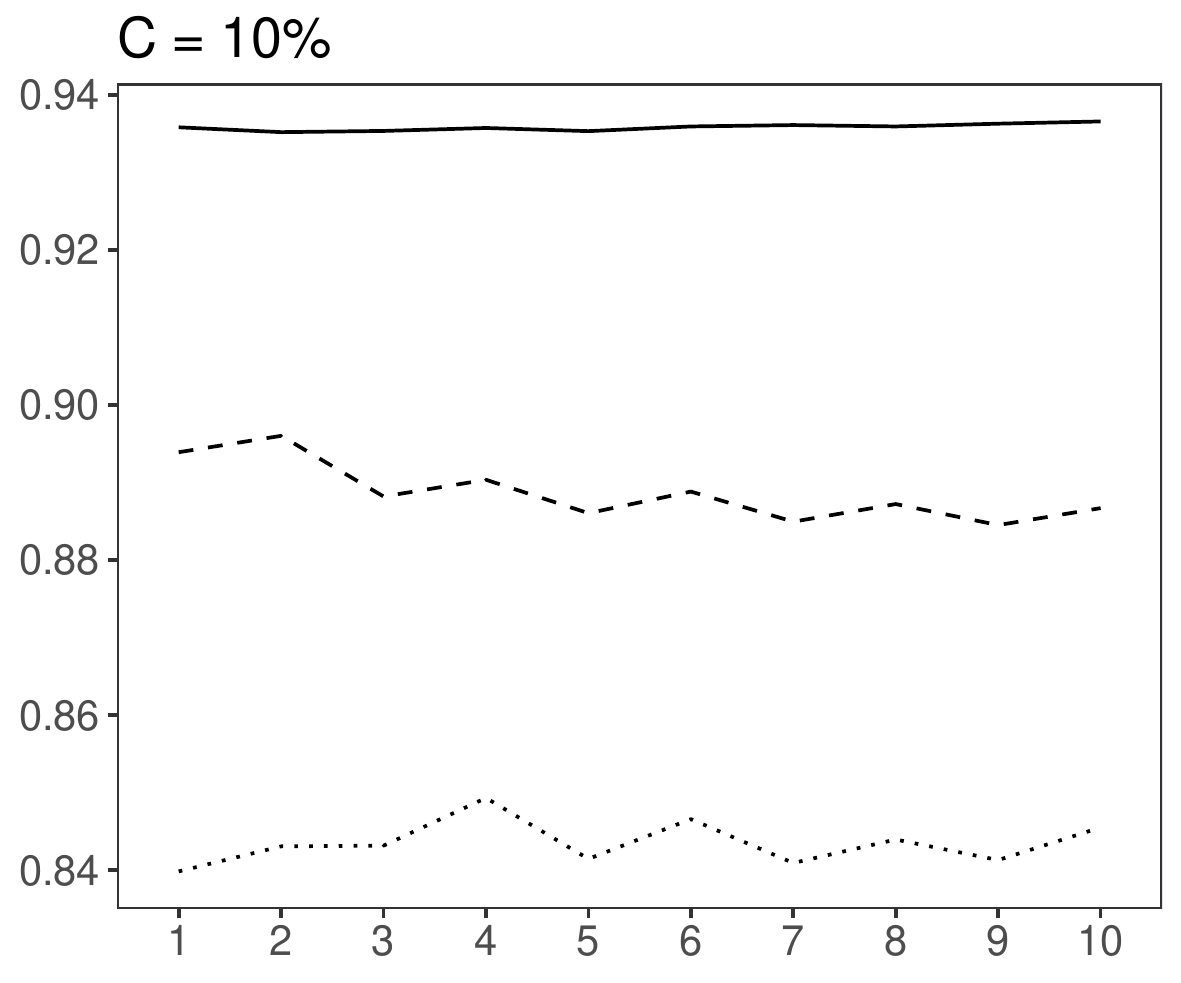}
\\
  \includegraphics[width=5.9cm]{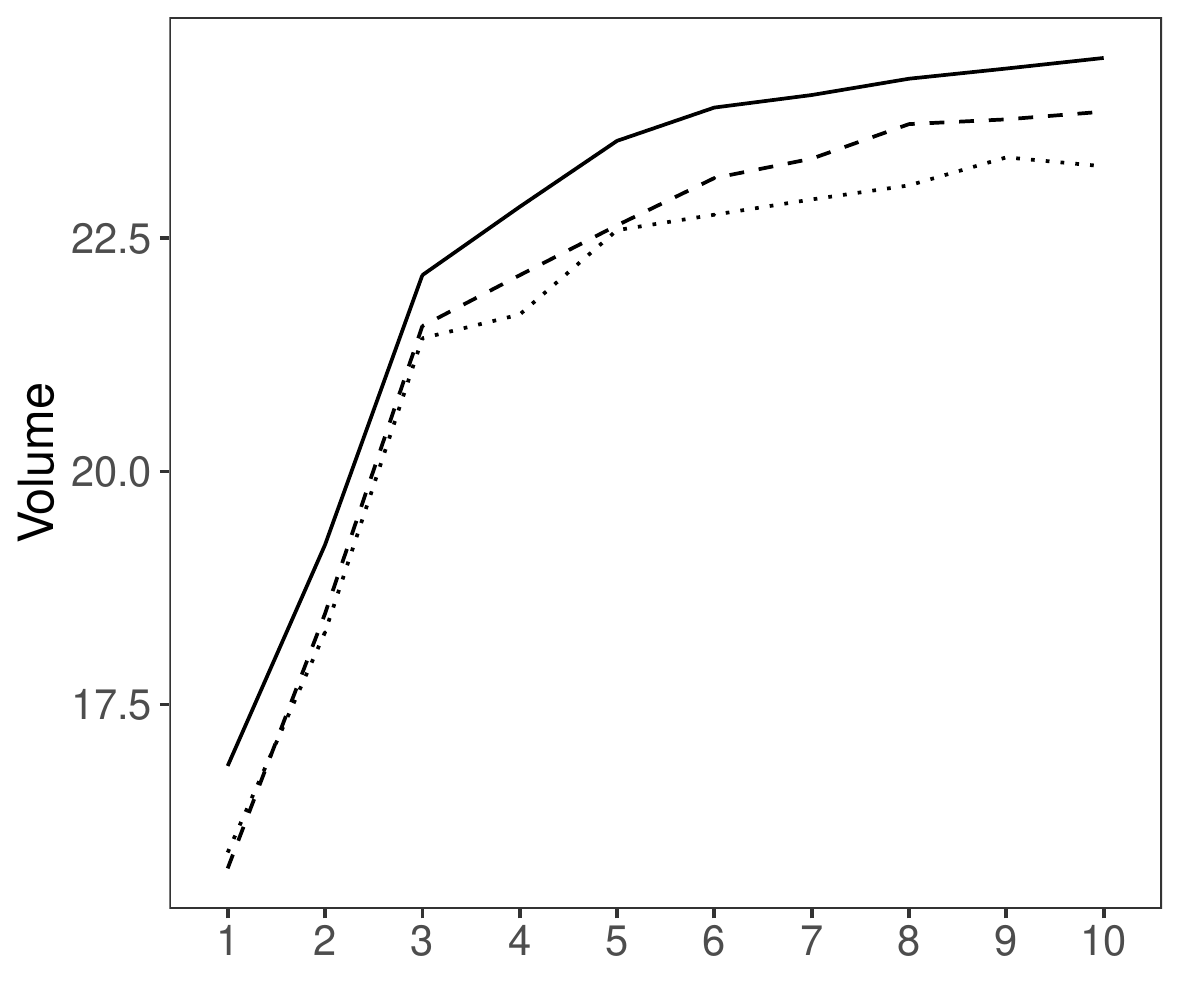}
  \includegraphics[width=5.9cm]{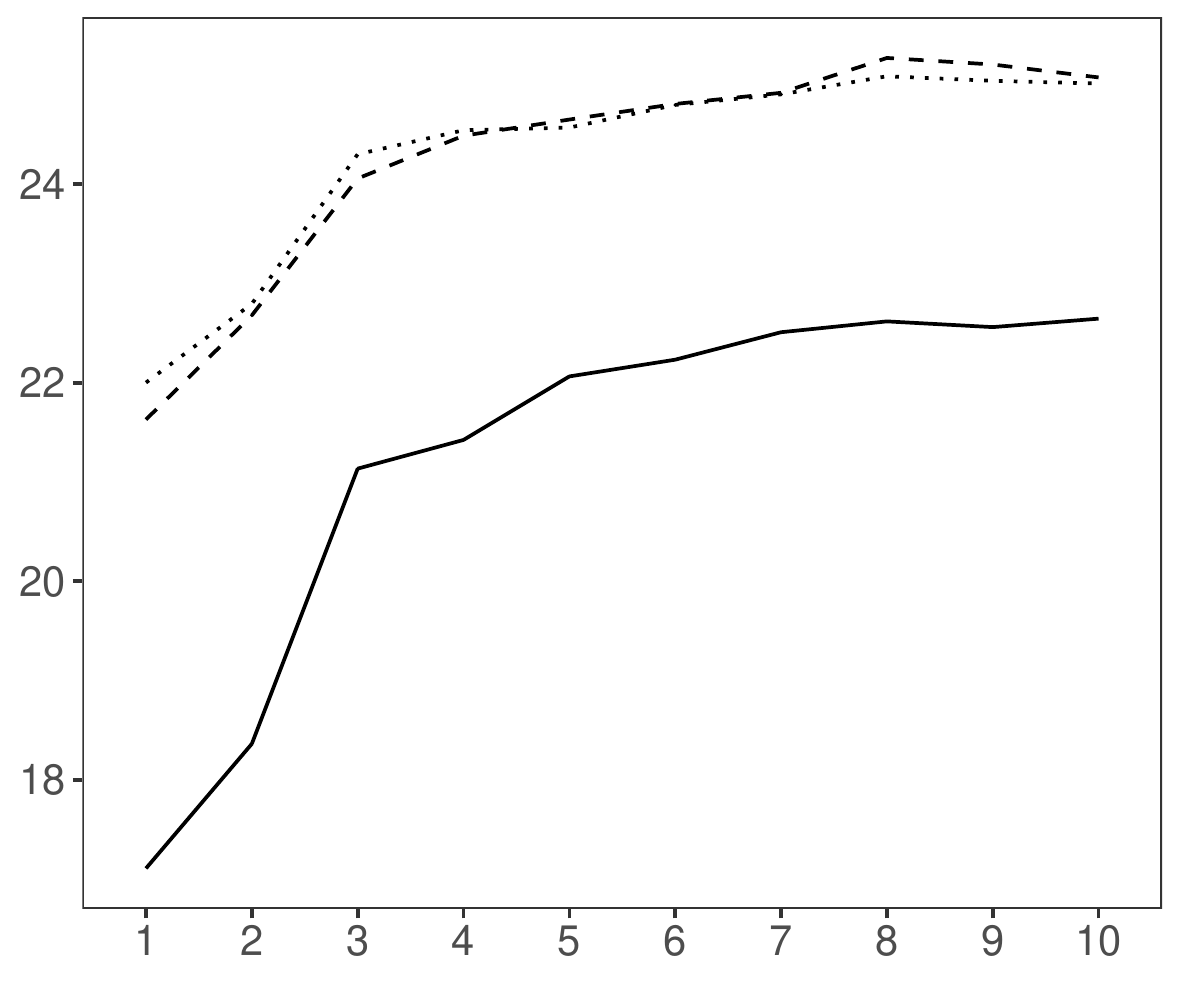}
  \includegraphics[width=5.9cm]{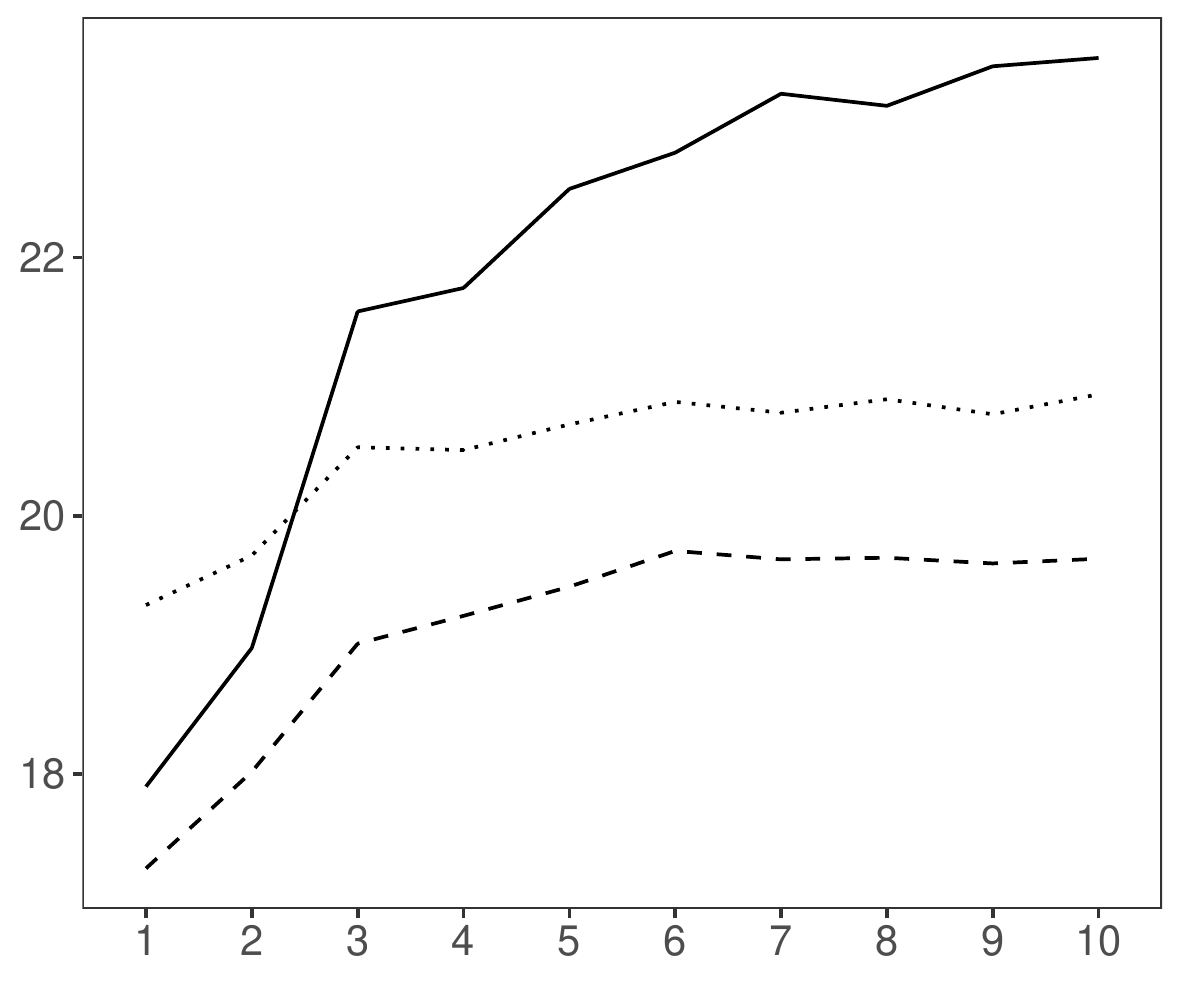}
\\
  \includegraphics[width=5.9cm]{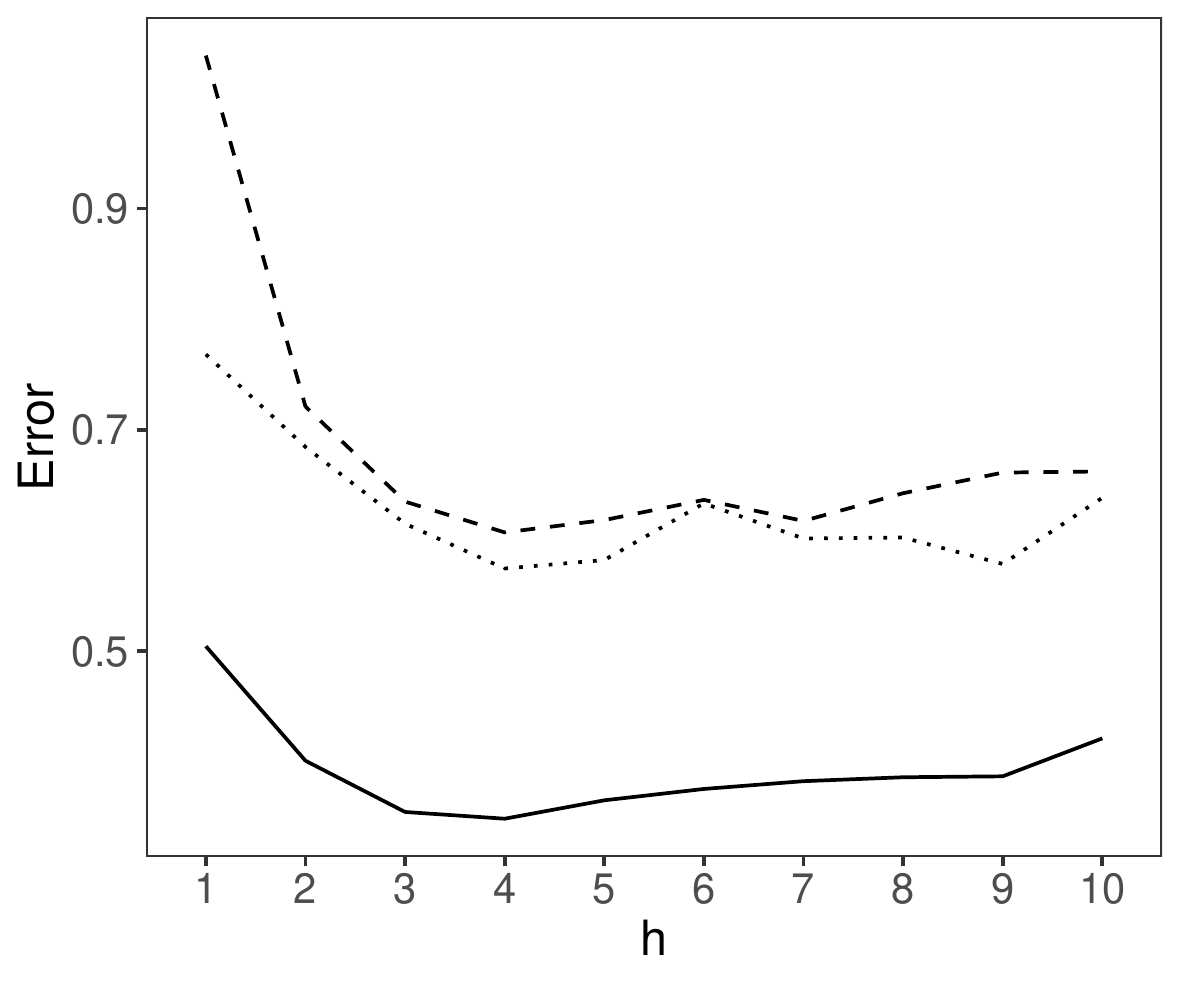}
  \includegraphics[width=5.9cm]{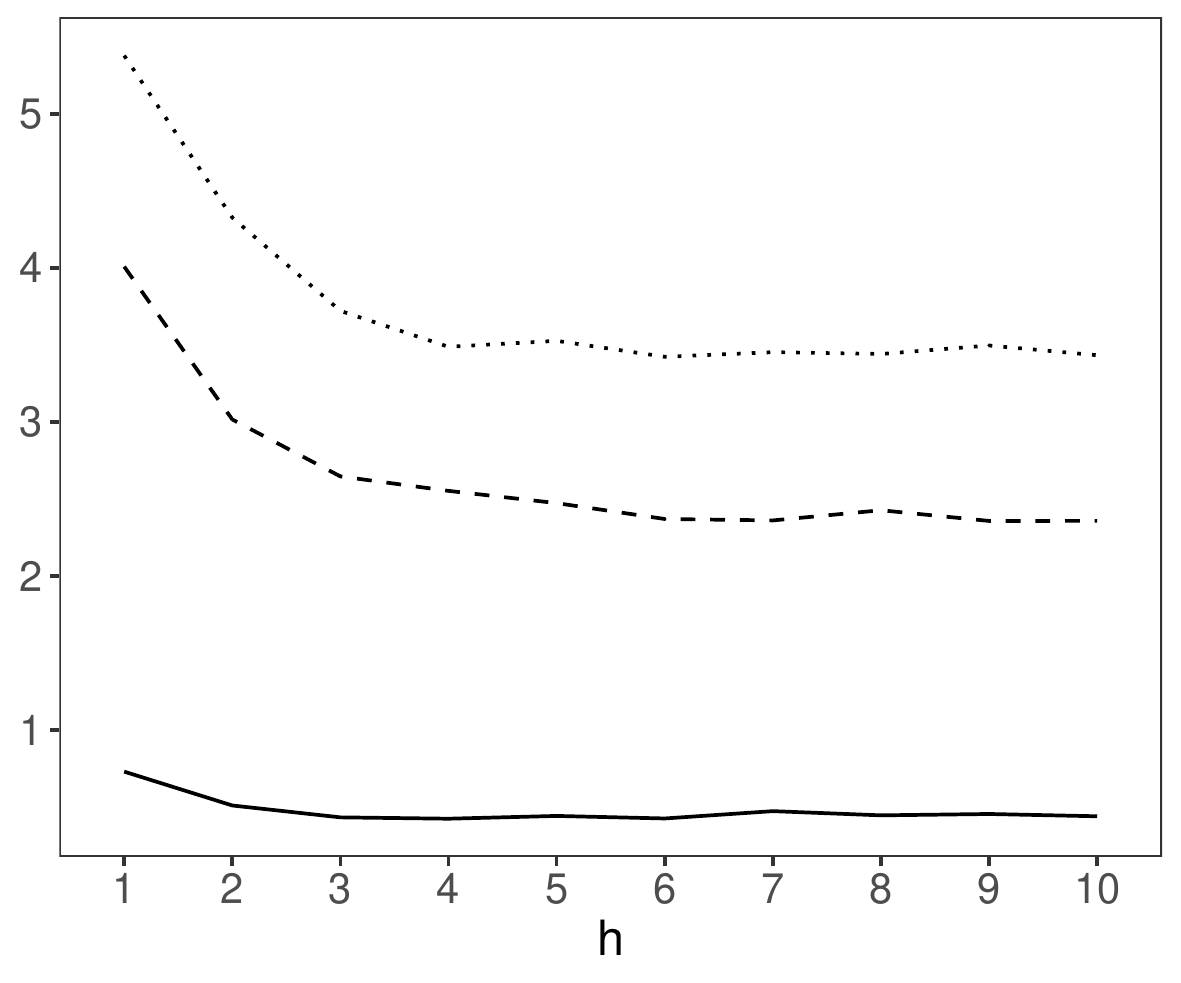}
  \includegraphics[width=5.9cm]{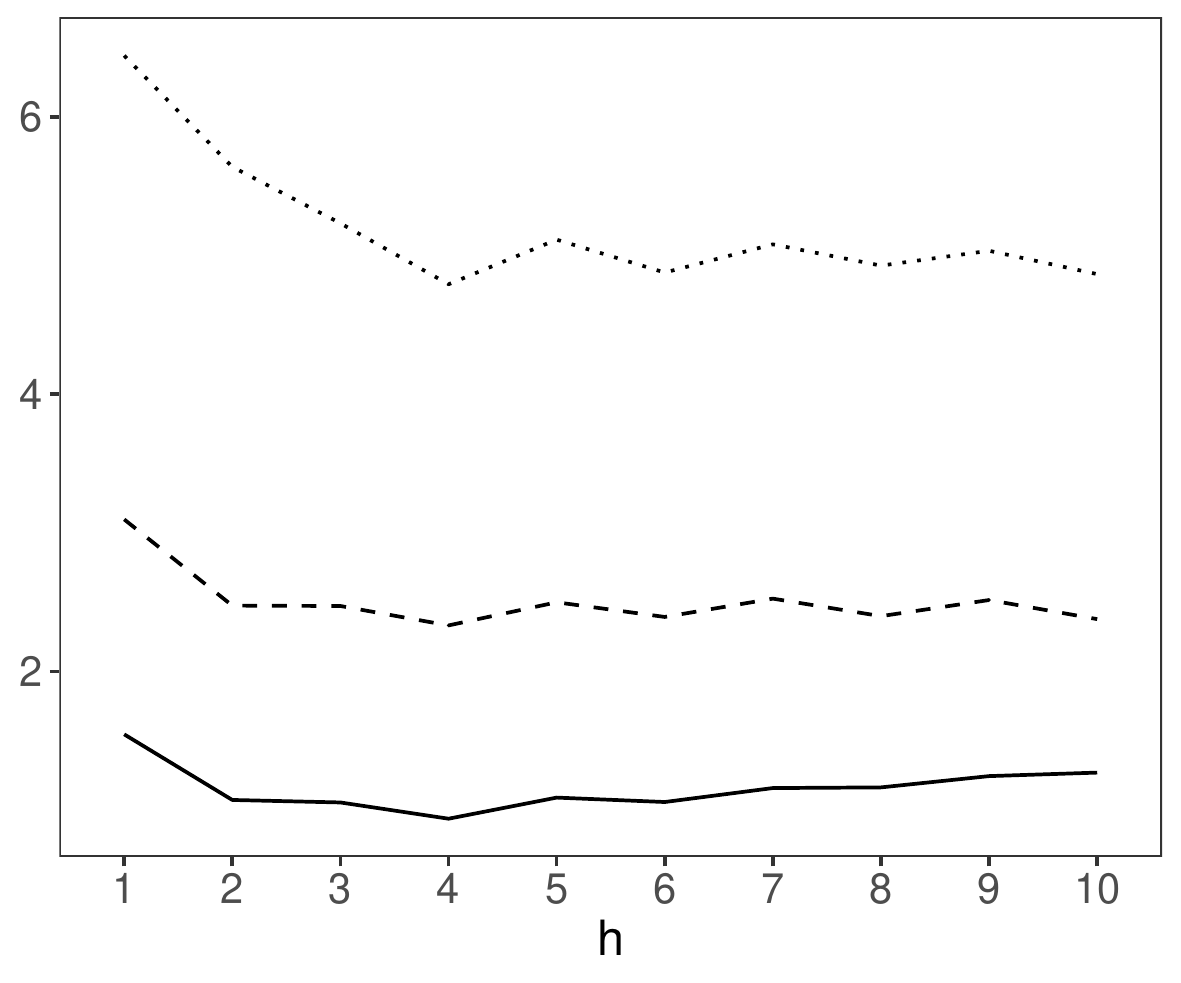}
  \caption{\small{Monte Carlo averages of the performance metrics for the VAR(2) model when $C = \left[ 1\%~\text{(first column)}, 5\%~\text{(second column)}, 10\%~\text{(third column)} \right]$ of the generated data are contaminated by deliberately inserted AOs and $\epsilon_t \sim t_5$: coverage probability (first row), volume of the Bonferroni cube (second row), and squared error between the empirical and bootstrap Bonferroni cubes (third row). Methods: OLS (dotted line), weighted likelihood (solid line), and Rob-VAR (dashed line).}}
  \label{fig:8}
\end{figure}

\begin{figure}[!htbp]
  \centering
  \includegraphics[width=5.9cm]{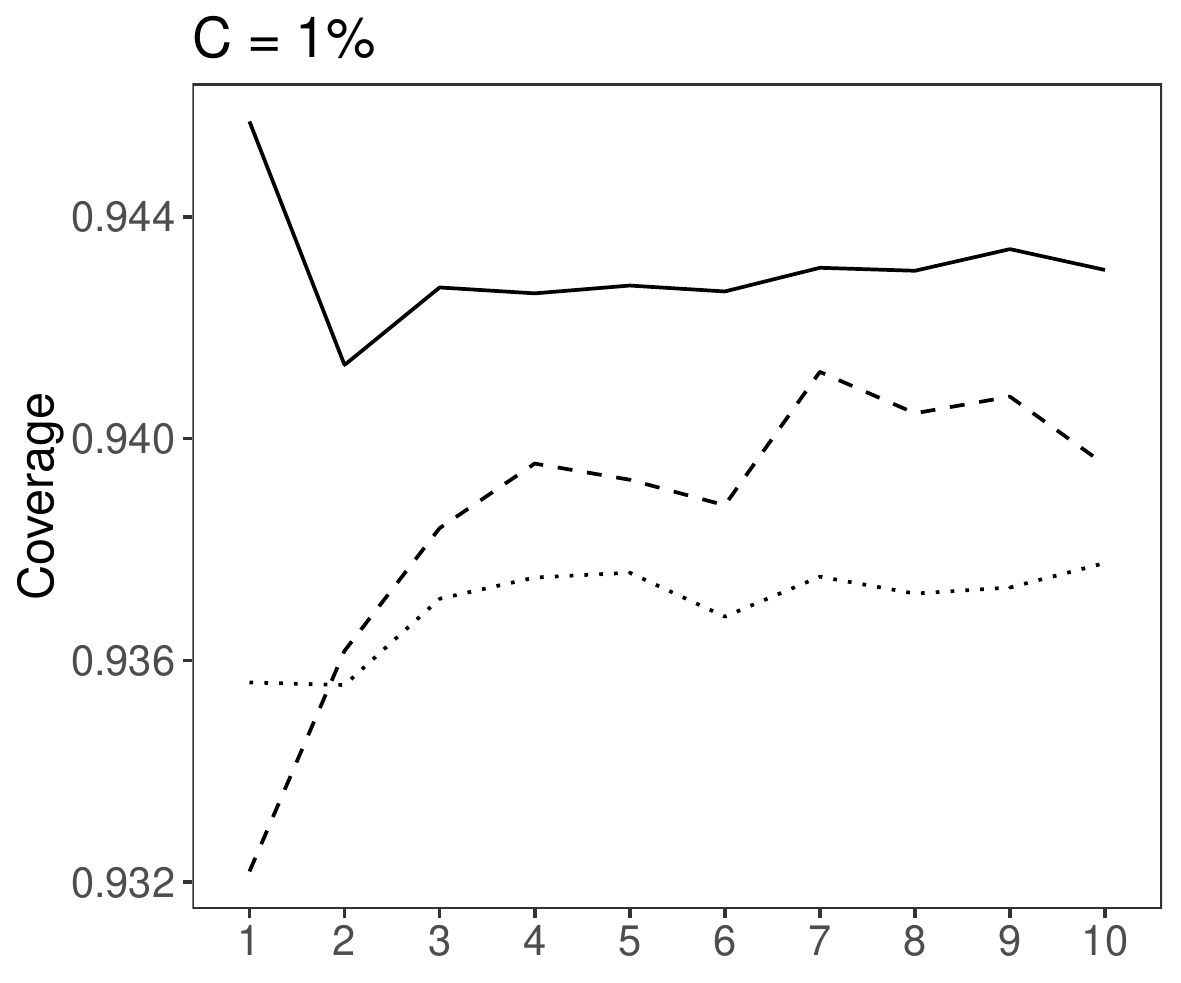}
  \includegraphics[width=5.9cm]{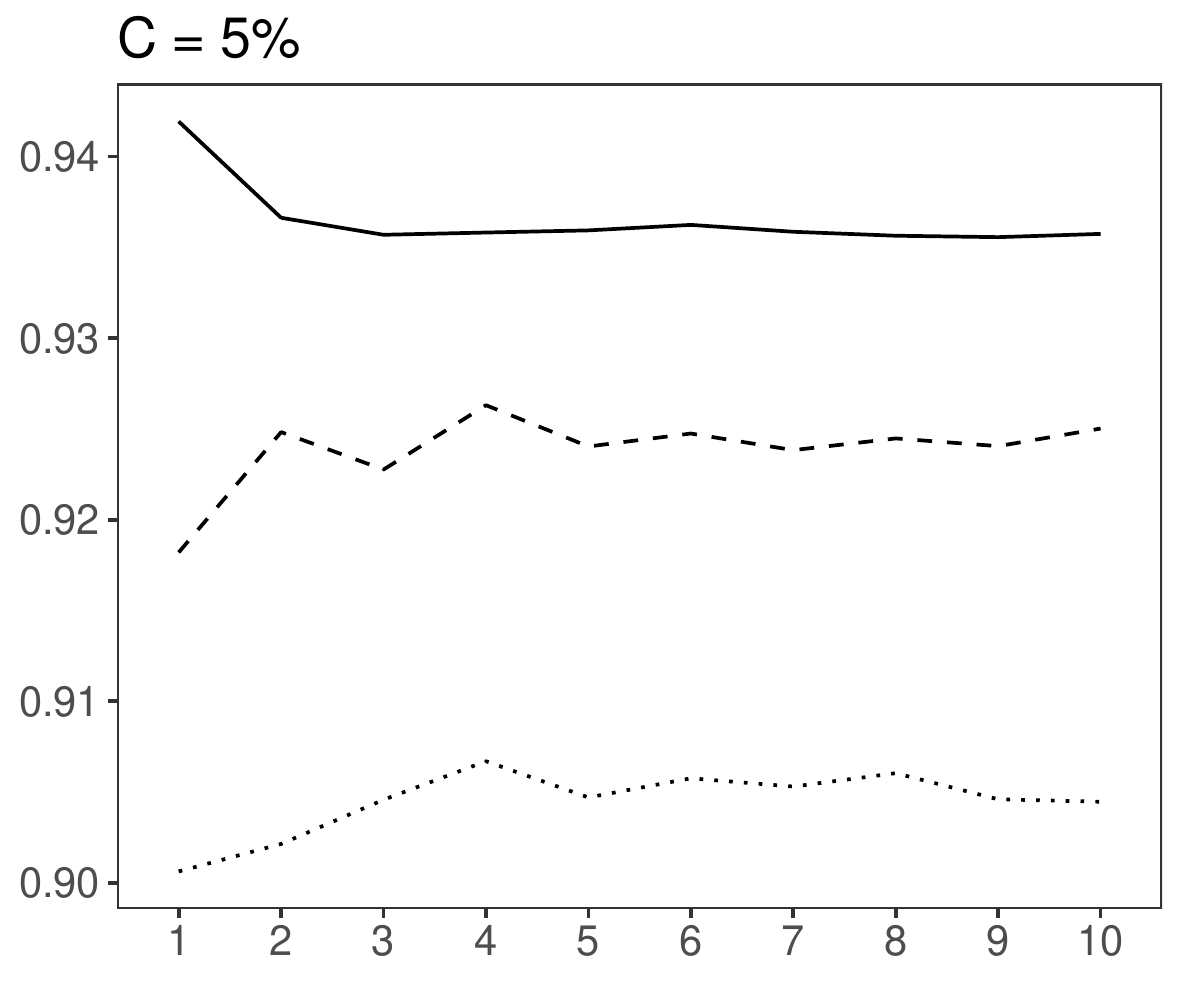}
  \includegraphics[width=5.9cm]{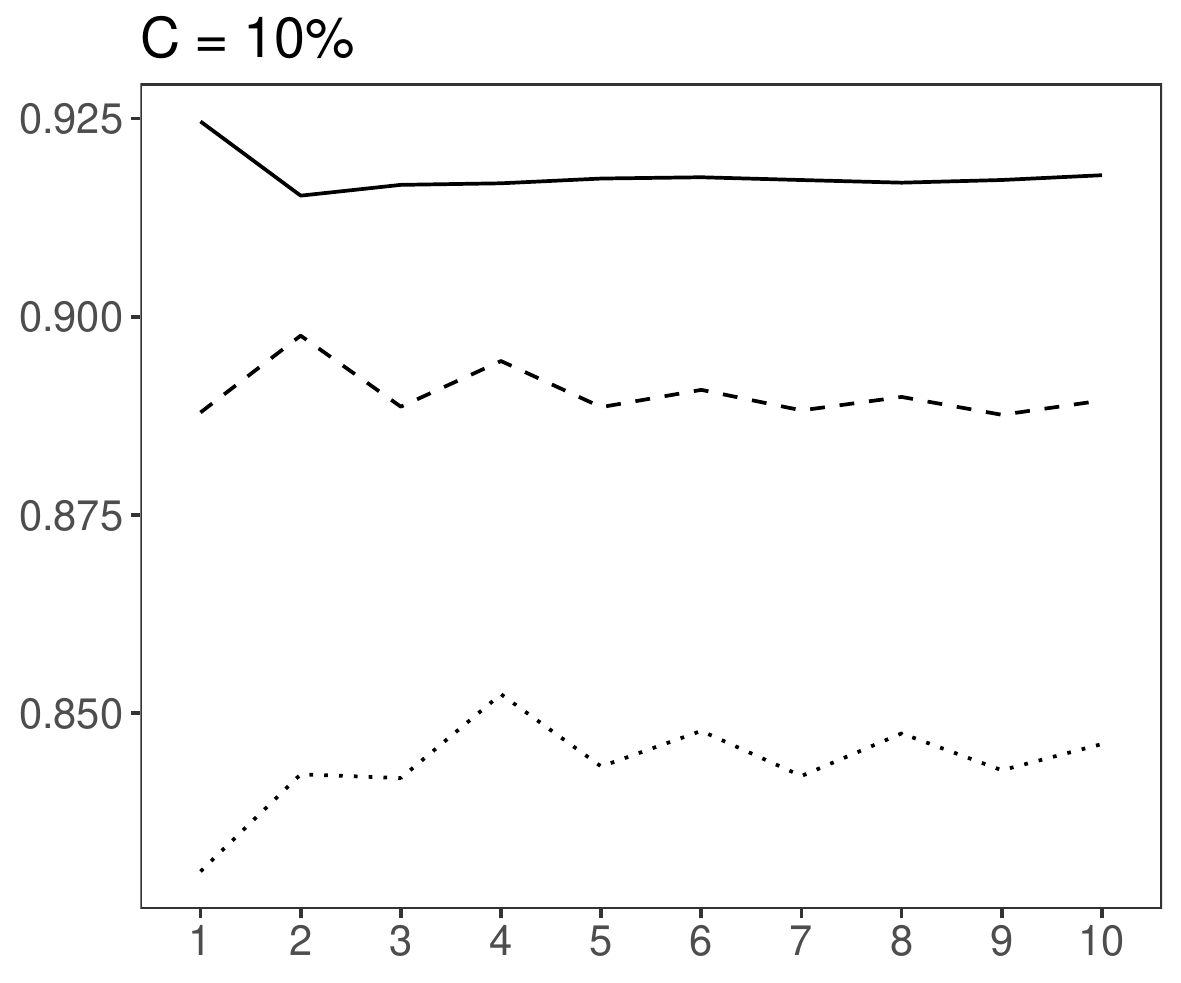}
\\
  \includegraphics[width=5.9cm]{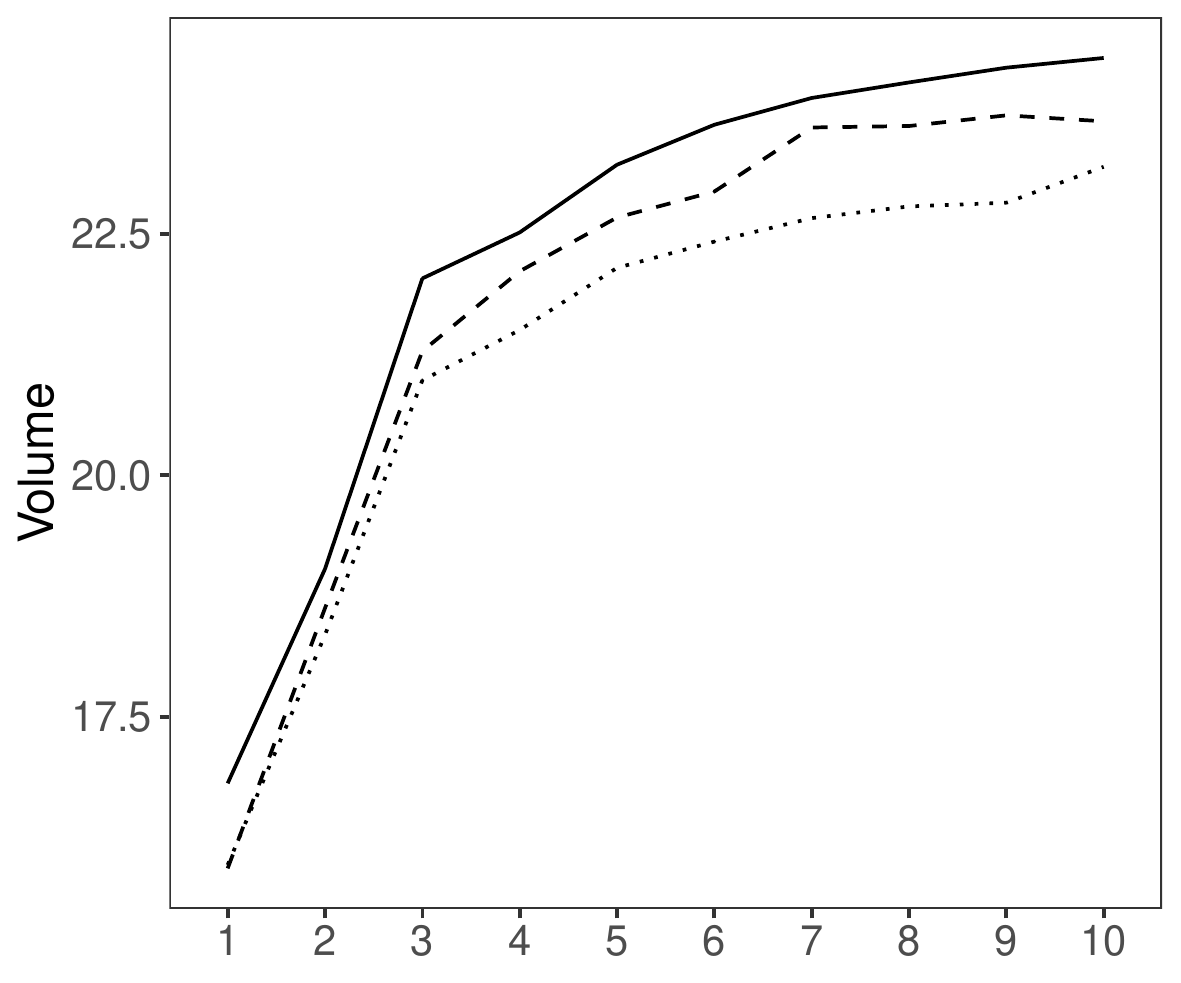}
  \includegraphics[width=5.9cm]{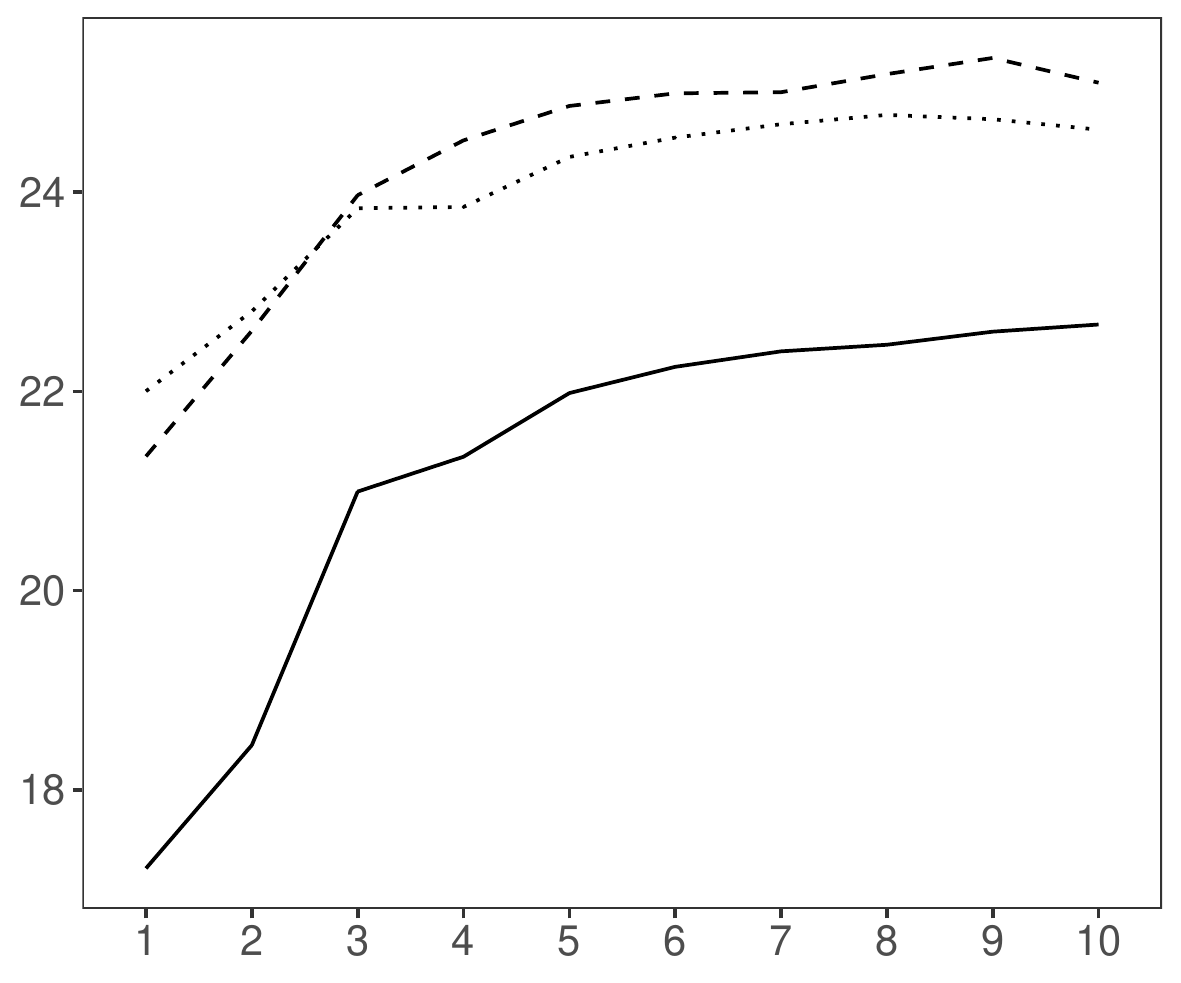}
  \includegraphics[width=5.9cm]{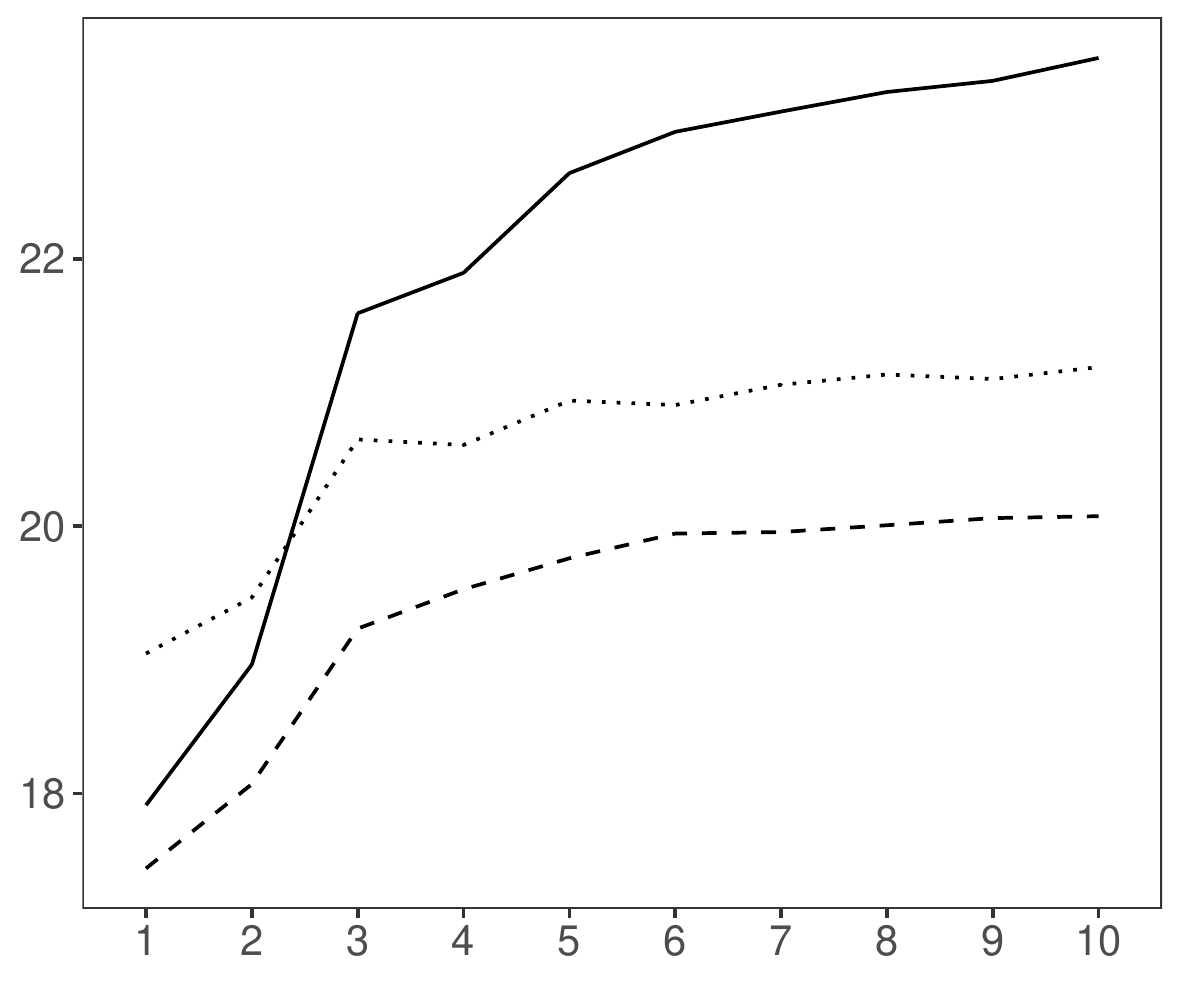}
\\
  \includegraphics[width=5.9cm]{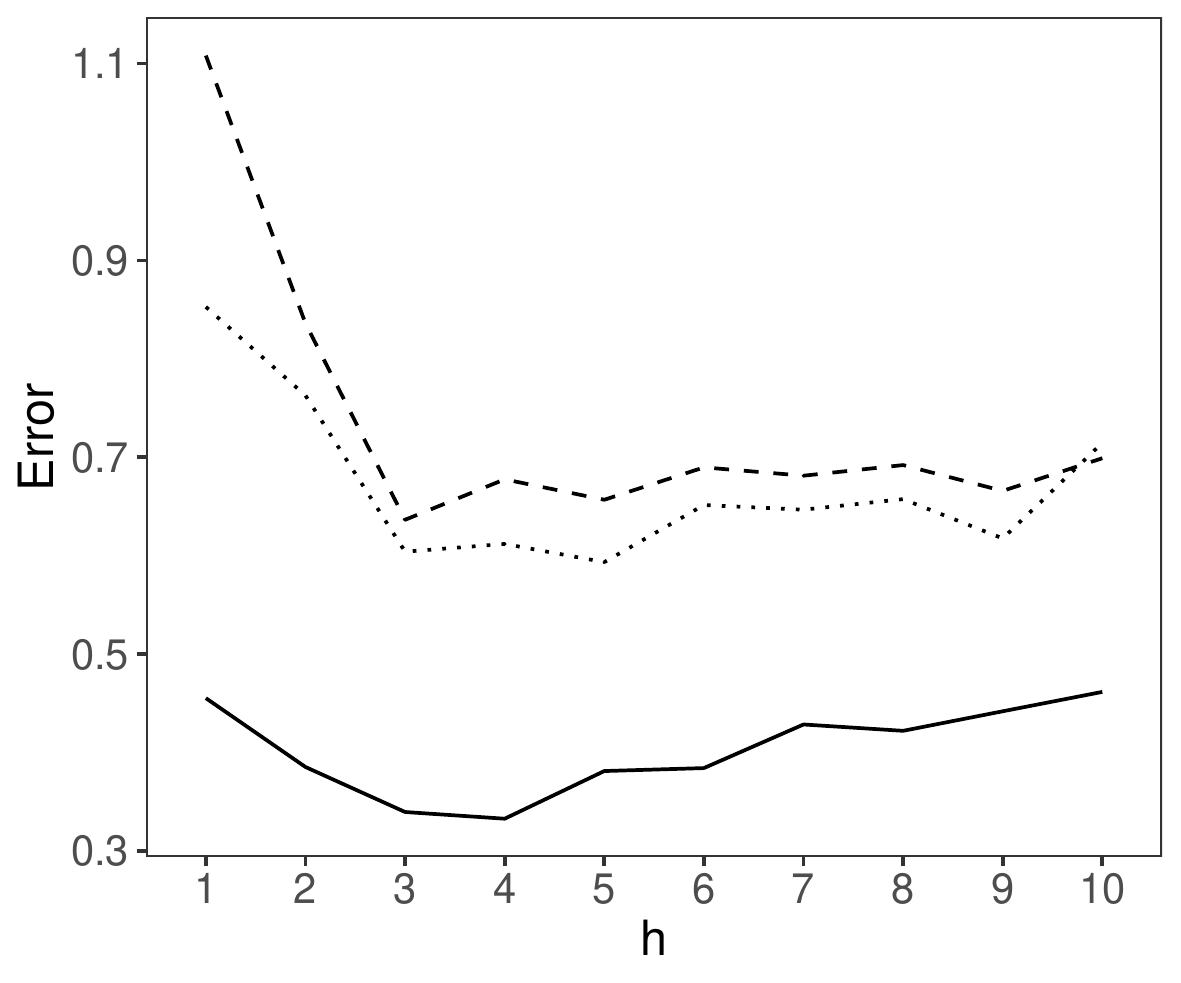}
  \includegraphics[width=5.9cm]{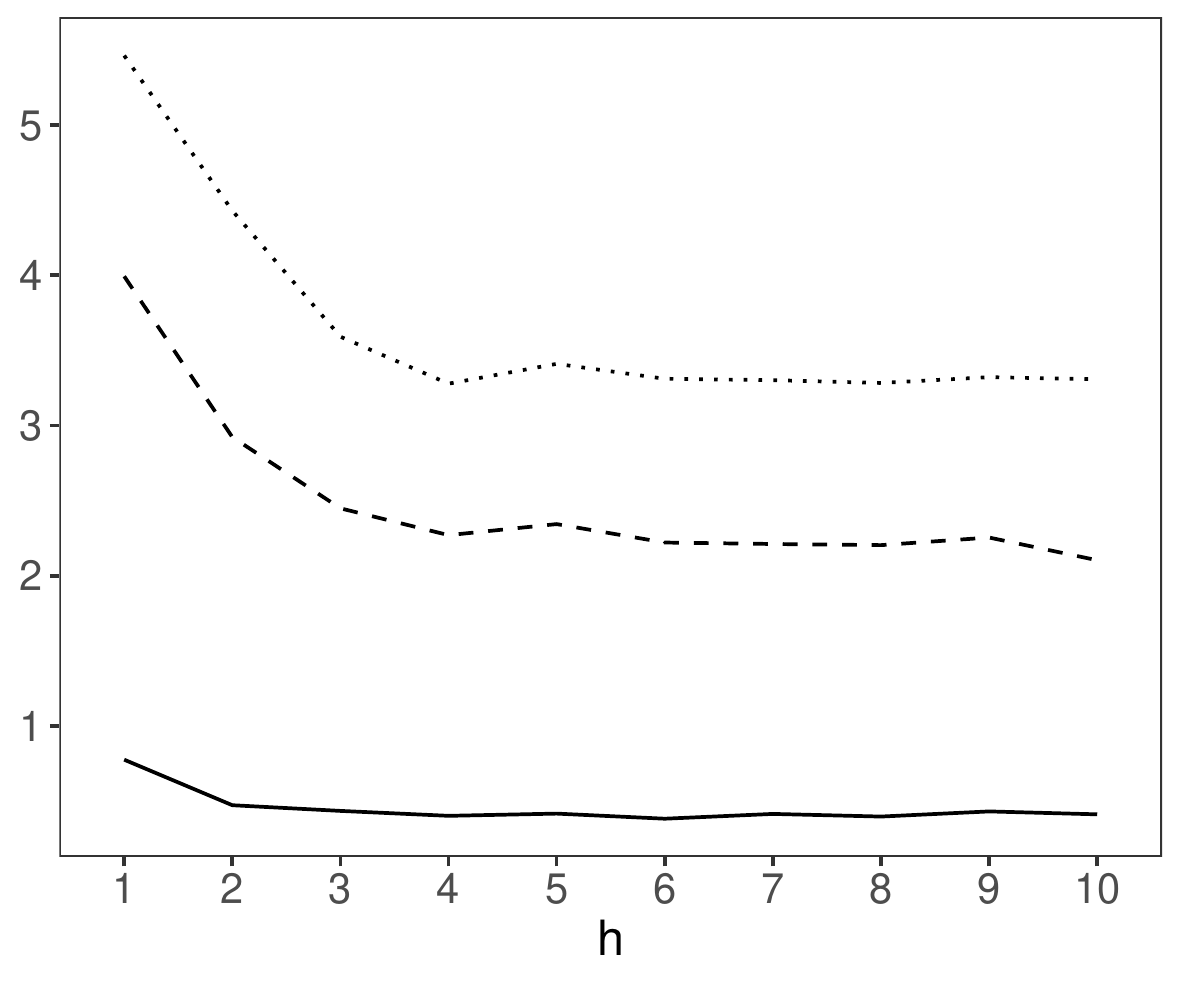}
  \includegraphics[width=5.9cm]{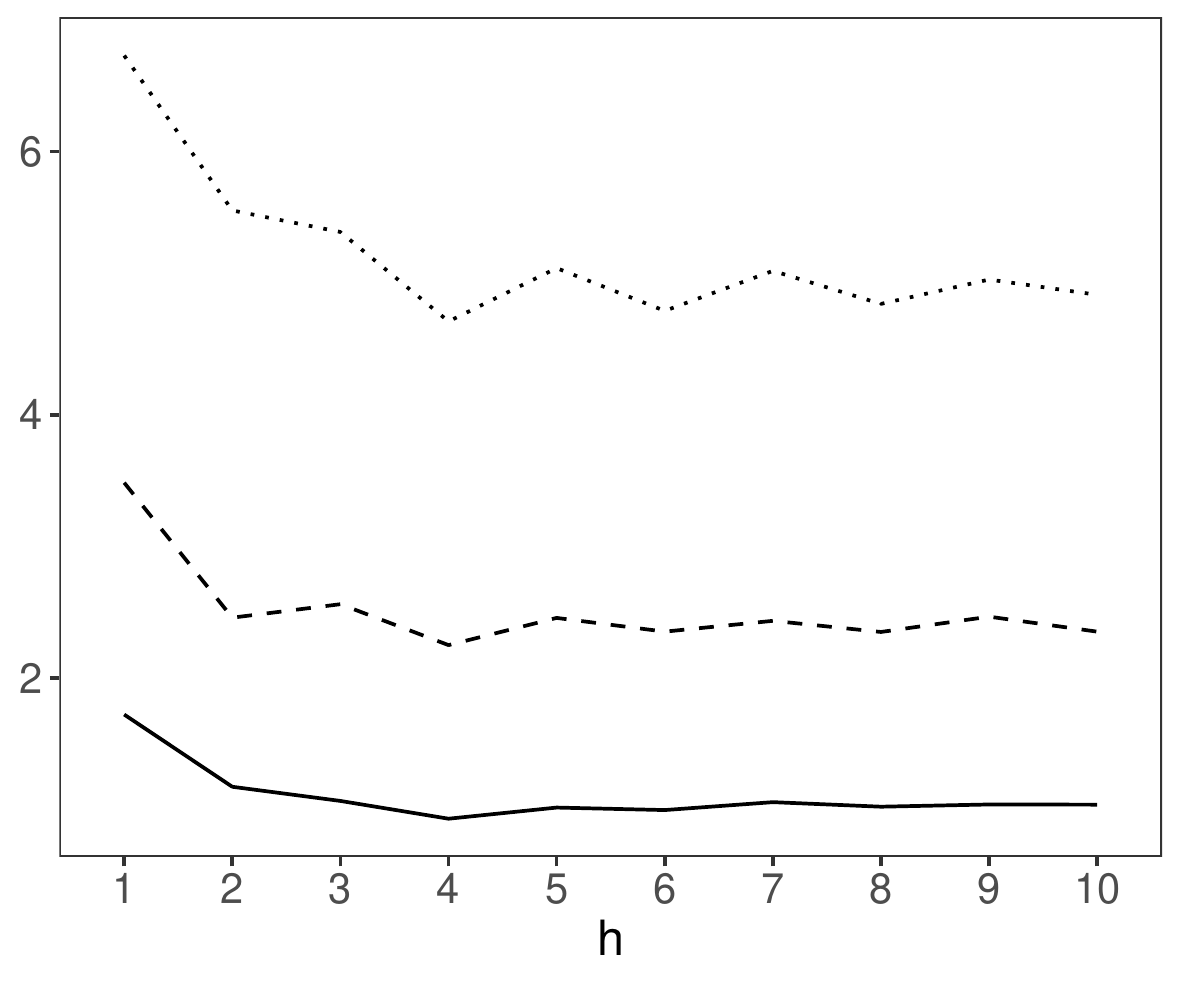}
  \caption{\small{Monte Carlo averages of the performance metrics for the VAR(2) model when $C = \left[ 1\%~\text{(first column)}, 5\%~\text{(second column)}, 10\%~\text{(third column)} \right]$ of the generated data are contaminated by deliberately inserted IOs and $\epsilon_t \sim t_5$: coverage probability (first row), volume of the Bonferroni cube (second row), and squared error between the empirical and bootstrap Bonferroni cubes (third row). Methods: OLS (dotted line), weighted likelihood (solid line), and Rob-VAR (dashed line).}}
  \label{fig:9}
\end{figure}

Our results are reported in Figures~\ref{fig:5}-\ref{fig:9}. When no outlier is present in the data, all bootstrap methods tend to produce similar coverage probabilities and volume of forecast cubes, as shown by Figure~\ref{fig:5}. Only the Rob-VAR produces slightly different coverage probabilities compared with the proposed and OLS methods when the errors follow $N(0,1)$ and $t_5$ distributions. Figure~\ref{fig:5} also shows that the proposed method has fewer errors than the other two methods when the errors are generated from $N(0,1)$ and $t_5$ distributions. However, it produces larger errors than the other methods when the errors follow a $\chi^2(4)$ distribution.

When the data have AOs/IOs and the error terms follow $N(0,1)$ distribution, the proposed method has significantly better performance compared with the other two methods. In other words, the proposed method produces better coverage probabilities, less volume, and Bonferroni forecast cubes more similar to the empirical Bonferroni forecast cube compared with other bootstrap methods. All the methods result in under-coverage when outliers contaminate the data; however, the coverage probabilities produced by the proposed method are much closer to the nominal level than those of other bootstrap methods (see Figures~\ref{fig:6} and~\ref{fig:7}). When the data have AOs/IOs but $t_5$ distributed error terms (Figures~\ref{fig:8} and~\ref{fig:9}), the proposed method still has a superior performance over other methods, but it performs slightly worse compared with the case where the errors follow $N(0,1)$ distribution.

\section{Empirical data examples} \label{sec:real}

Through two data analyses, we study the forecasting performance of the proposed method. For the univariate AR model, we consider monthly Brazil/U.S. dollar exchange rate data; quarterly GDP and immediate rate (IR) data for Turkey are considered for the VAR model. Both sets of data can be found at \url{https://stlouisfed.org}.

\subsection{Brazil/U.S. dollar exchange rate data}

Original monthly Brazil/U.S. dollar exchange rate data were obtained starting from February 1, 1999 to August 1, 2018 (235 observations), and a logarithmic return series generated via 
\begin{equation*}
P_t = \ln(y_t/y_{t-1})*100, 
\end{equation*}
where $y_t$ represents the observed exchange rate at the $t$\textsuperscript{th} month. Time series plots of the original and logarithmically transformed exchange rate data are presented in Appendix, which shows several clear outlying observations. 

The $p$-values obtained by the Ljung-Box (LB) and augmented Dickey-Fuller (ADF) $t$-statistics tests ($p$-value $\leq 0.01$) suggest that the return series is a stationary process with mean zero. The autocorrelation and partial-autocorrelation plots of the return series (not presented here) show that an AR model may be appropriate to model the return series. The optimal lag order $p$ for this series is obtained as 3, AR(3), via the AIC. In addition, the structural break test based on $F$ statistics in the \texttt{R} package \texttt{strucchange} of \cite{Zeileis} is applied to the returns series to check if there is a structural break in the series. In this test, a $F$ statistic is computed for every predetermined potential change point and the structural break test is performed using the calculated $F$ statistics. For the return series of  monthly Brazil/U.S. dollar exchange rate data, this test procedure produce the $p$-value$ = 0.188$, which indicates that there is no clear structural break in the data.

\begin{figure}[!htbp]
\centering
\includegraphics[width=10cm]{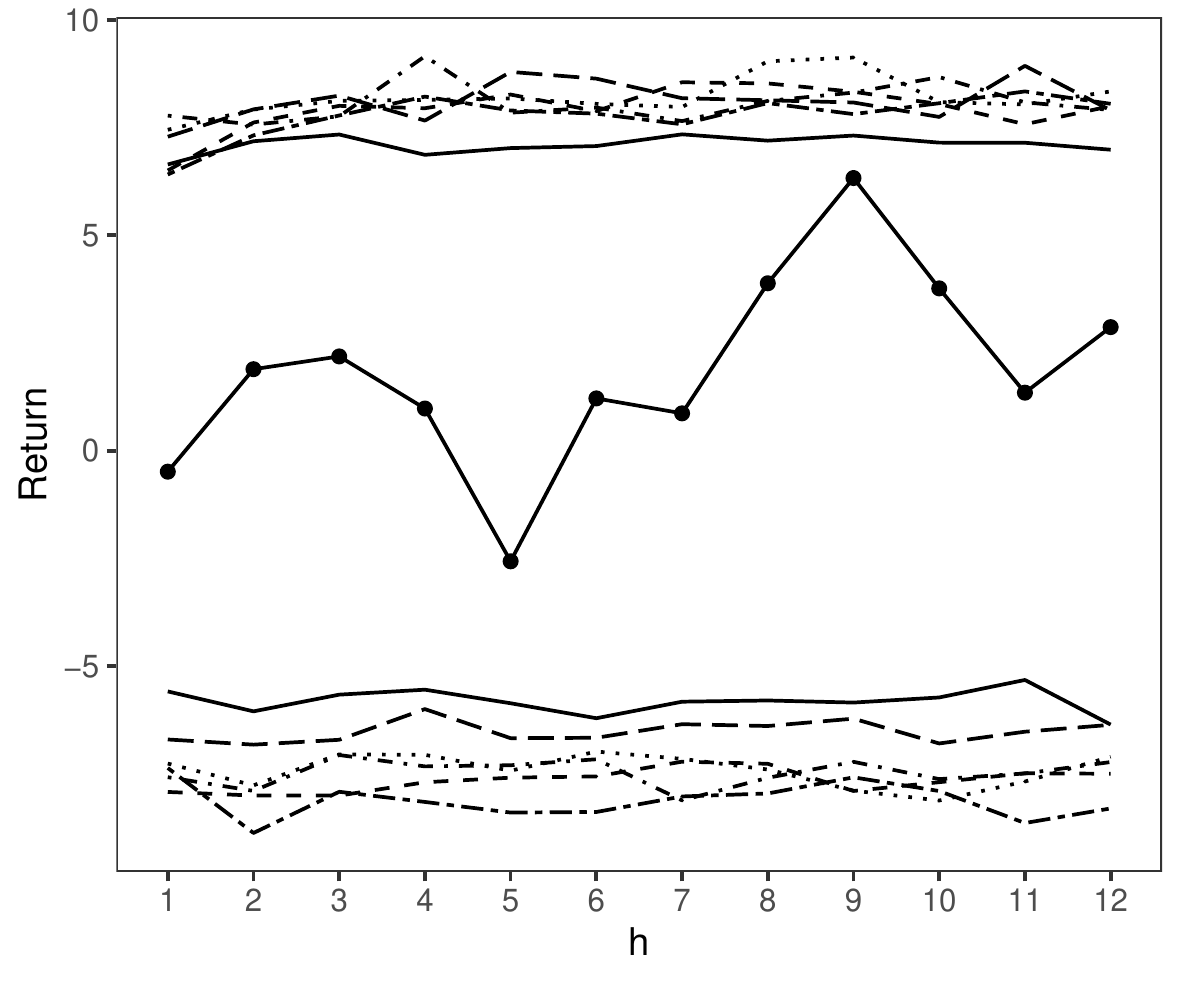}
\caption{\small{Computing $h = 12$ steps ahead prediction intervals for the Brazil/U.S. dollar exchange rate data. Methods: OLS (dotted line), weighted likelihood (solid line), Rob-YW (dashed line), Rob-Reg (dot-dashed line), Rob-Flt (long dashed line), and Rob-GM (two dashed line). Note: points with solid line represent the observed returns.}}
\label{fig:11}
\end{figure}

To obtain out-of-sample bootstrap prediction intervals for the return series, we divide the data into two parts; we estimate the model based on observations from February 1, 1999, to August 1, 2017, to construct $12$-step-ahead prediction intervals from September 1, 2017, to August 1, 2018. Our results are shown in Figure~\ref{fig:11}, which shows the prediction interval produced by the proposed method is narrower than those obtained by other bootstrap methods, and all prediction intervals include all the observed future values.

\subsection{Real GDP and immediate rate data for Turkey}

Quarterly GDP and IR data for Turkey, which consist of 76 observations, were obtained for 1999$Q_1$ to 2017$Q_3$. The GDP and IR data were then logarithmically transformed as $GDP_t = \ln(GDP_t/GDP_{t-1})$ and $IR_t = \ln(IR_t/IR_{t-1})$, respectively. The time series plots of the original and transformed data are given in Appendix, which shows several outliers for both series. 

For this dataset, the model is constructed based on data spanning from 1999$Q_1$ to 2016$Q_3$ to calculate Bonferroni cubes for four-step-ahead forecasts from 2016$Q_4$ to 2017$Q_3$. Table~\ref{tab:1} presents the sample statistics and Jarque-Bera (JB) normality and ADF stationarity test results. The results indicate that: 
\begin{inparaenum}
\item[1)] the data do not follow a Gaussian process, either individually or jointly, and
\item[2)] both GDP and IR series are stationary processes according to ADF test results. 
\end{inparaenum}
Moreover, the results of the LB test (not presented here) indicate that there is a dynamic dependence on the conditional mean for this dataset. The optimal lag-order is determined by AIC, with the results showing that a VAR(3) model is optimal.

\begin{table}[!htbp]
\centering
\tabcolsep 0.25in
\caption{\small{Sample statistics, JB and ADF test results of $GDP$ and $IR$ ($p$-values in brackets).}}\label{tab:1}
\begin{tabular}{@{}lcccccccc@{}}
\toprule
Series & Mean & Sd & Skewness & Kurtosis & JB & ADF \\
\midrule
$GDP$ & 1.22 & 2.37 & 2.22 & 0.73 &14.85 & -4.27\\
& & & & & (0.00) & ($< 0.01$) \\
$IR$ & -3.32 & 32.51 & 3.87 & 9.84 & 382.83 & -4.33 \\
& & & & & (0.00) & ($< 0.01$) \\
($GDP,IR$) & & & 1.76 & 18.78 & 179.54 \\
& & & & &(0.00) \\
\bottomrule
\end{tabular}
\end{table}

\begin{figure}[!ht]
\centering
\includegraphics[width=16cm]{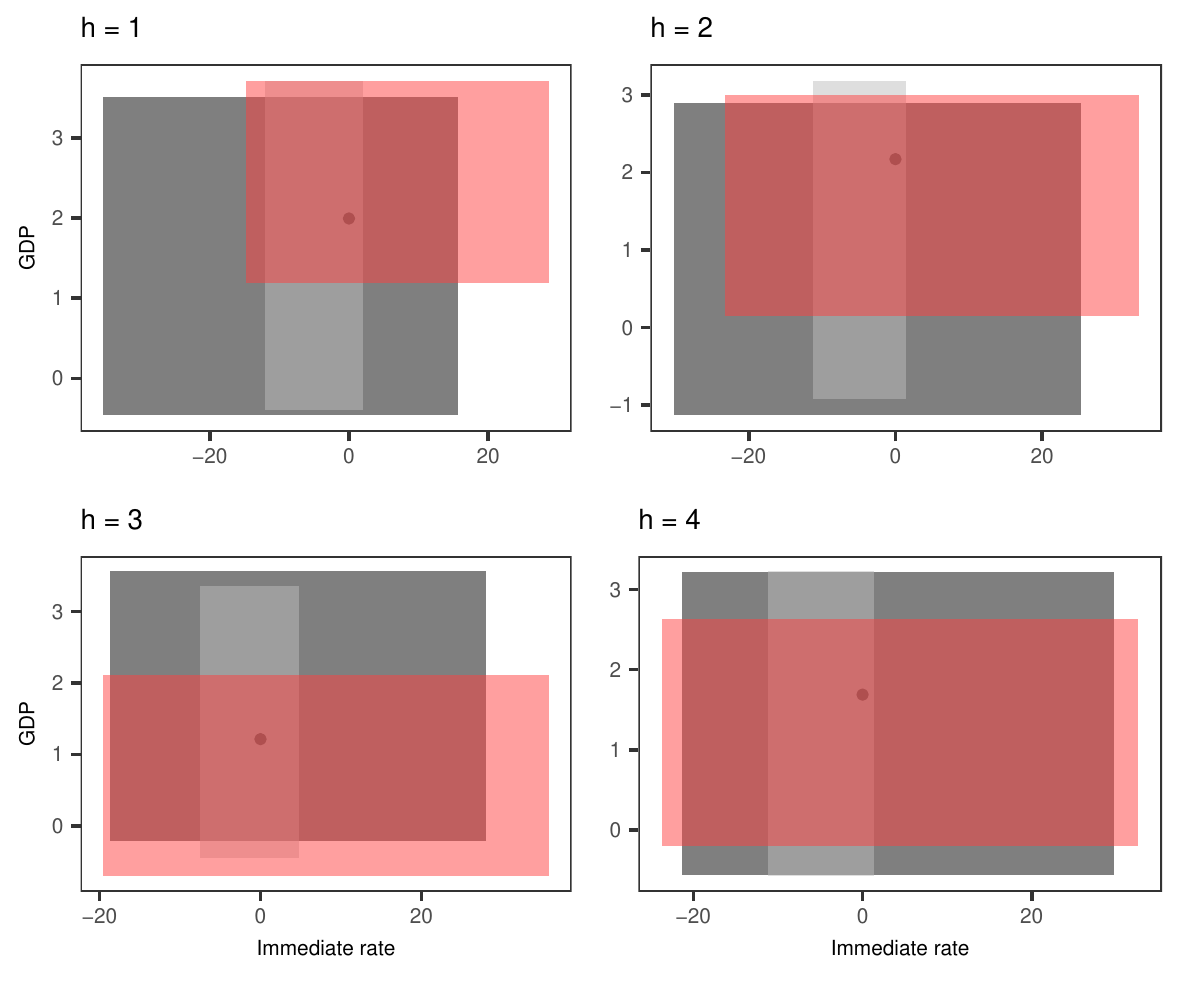}
\caption{\small{95\% Bonferroni cubes of OLS (dark gray area), Rob-VAR (red area), and weighted likelihood (light grey area) for $h$-step-ahead forecasts of GDP and immediate rate data, where $h=1, \cdots, 4$.}}
\label{fig:13}
\end{figure}

The constructed Bonferroni forecast cubes and observed out-of-sample values are presented in Figure~\ref{fig:13}. The bootstrap forecast cubes have considerably less volume than the Bonferroni cubes obtained by other bootstrap methods that use the OLS and multivariate least trimmed squares. Because of the high forecast errors caused by the presence of outliers, the weighted likelihood-based bootstrap method is less affected by the outliers and it thus provides Bonferroni cubes with less volume than the other two bootstrap methods.

\section{Conclusion} \label{sec:conclusion}

We propose a robust procedure to obtain prediction intervals and multivariate forecast regions for future values in the AR and VAR models. The proposed idea is based on using the weighted likelihood estimators and weighted residuals, which are calculated under the assumption of normally distributed error terms, in the bootstrap method. The finite sample properties of the proposed method are examined via several Monte Carlo studies and two empirical data examples, and results are compared with classical and other existing robust methods. Our records show that the proposed procedure has similar coverage performance with generally smaller interval length/volume of forecast cube compared with existing bootstrap methods, when the data have no outliers. In addition, it has superior performance in terms of coverage and interval length/volume than the other bootstrap methods, when outliers are present in the data and the error terms follow a normal distribution. 

The proposed method may lose some efficiency in its coverage performance when the errors differ from normality (such as when the errors follow $t_5$ distribution). This is because the weights $\omega_t$ do not converge to 1, when the errors follow a non-normal distribution. However, our results have demonstrated that the proposed method is also flexible enough to treat deviances from the normality and produces superior performance over the existing methods. For future research, the proposed method could be used to construct prediction intervals for financial returns and volatilities, in autoregressive conditional heteroskedasticity models \citep[see, e.g.,][]{PRR, chen, BeyaztasRevstat, BeyaztasCompE, BeyaztasShankya}.



\newpage
\section*{Appendix: Additional figures for the empirical data analyses}\label{appx}

\begin{figure}[!htbp]
\centering
\includegraphics[width=12.3cm]{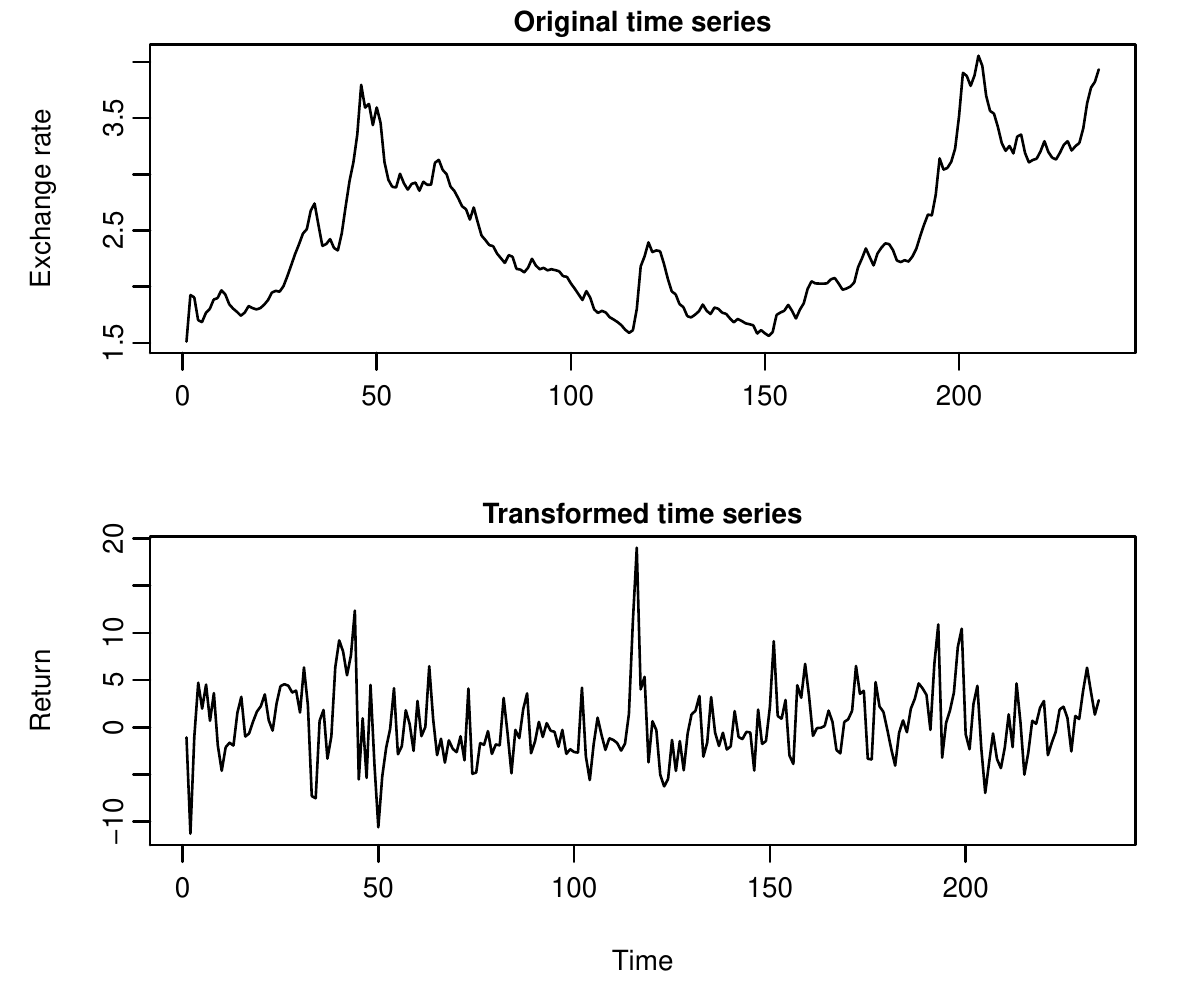}
\caption{\small{Original and transformed time series plots of Brazil/U.S. dollar exchange rates.}}
\label{fig:10}
\end{figure}

\begin{figure}[!htbp]
\centering
\includegraphics[width=16cm]{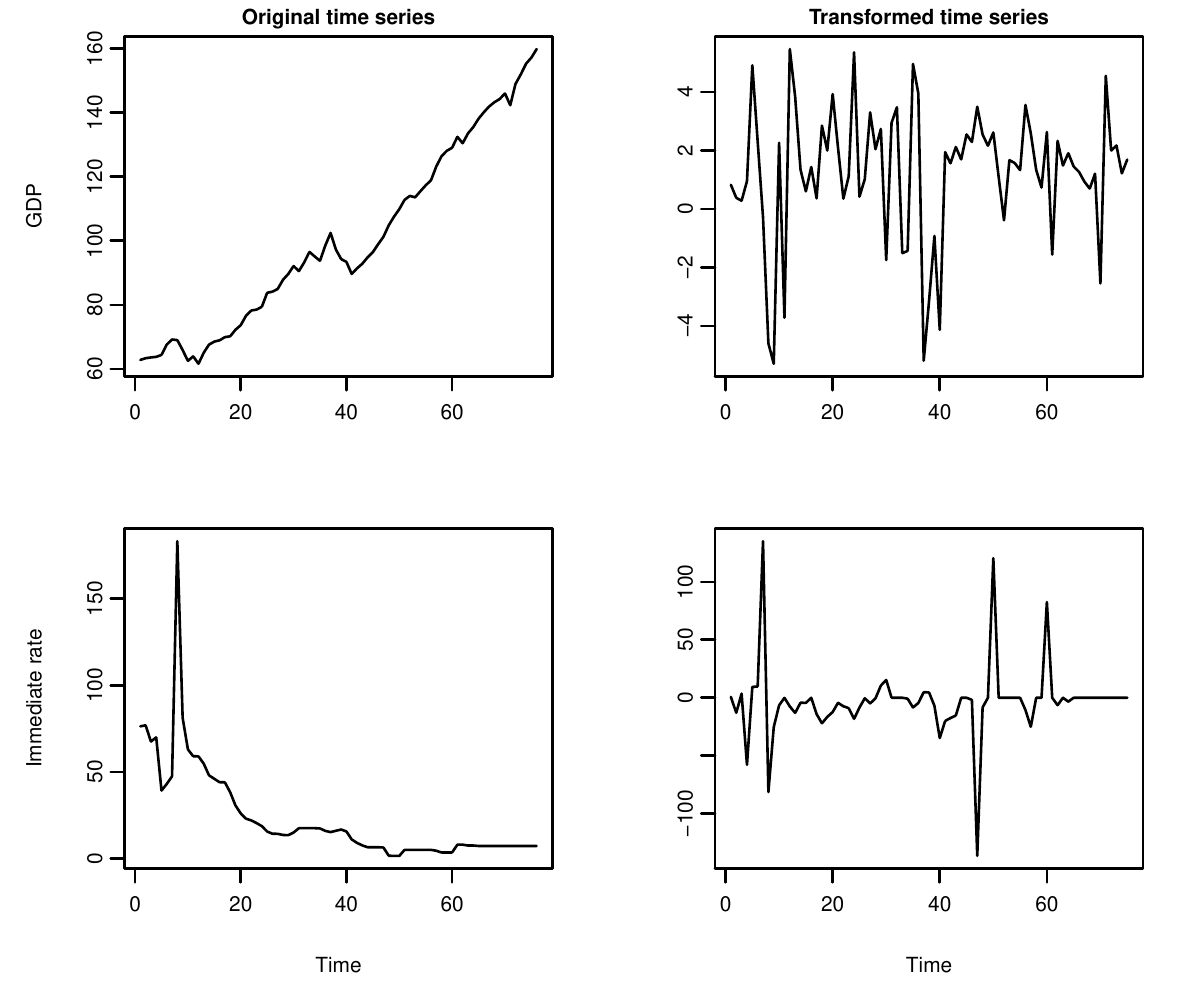}
\caption{\small{Original and transformed time series plots of GDP and immediate rate data.}}
\label{fig:12}
\end{figure}

\clearpage
\bibliographystyle{agsm}

\end{document}